\documentclass[dvips]{aa} % for the abstract without structuration 
\usepackage{graphicx,amssymb,amsmath,hyperref,multirow}
\usepackage[usenames,dvipsnames]{xcolor}
\usepackage{rotating}
\usepackage{txfonts}
\usepackage{lscape}
\usepackage{amsmath}
\usepackage{natbib,twoopt}
\usepackage{longtable}
\usepackage{multicol}
\bibpunct{(}{)}{;}{a}{}{,} %% natbib format like A&A and ApJ
\newcommandtwoopt{\citeyearads}[3][][]%
{\href{http://adsabs.harvard.edu/abs/#3}{\citeyear[#1][#2]{#3}}}
\hypersetup{colorlinks,linkcolor=red,citecolor=blue,urlcolor=blue}

\newcommand{\bs}{GBS}
\newcommand{\feh}{[Fe/H]}
\newcommand{\teff}{$\mathrm{T}_{\mathrm{eff}}$}
\newcommand{\logg}{$\log g$}
\newcommand{\vmic}{$v_{\mathrm{mic}}$}
\newcommand{\vmac}{$v_{\mathrm{mac}}$}
\newcommand{\vsini}{$v{\sin{i}}$}
\newcommand{\alfCenA}{$\alpha$~Cen~A}
\newcommand{\alfCenB}{$\alpha$~Cen~B}
\newcommand{\alfCet}{$\alpha$~Cet}
\newcommand{\alfTau}{\object{$\alpha$~Tau}}
\newcommand{\betAra}{\object{$\beta$~Ara}}
\newcommand{\betGem}{\object{$\beta$~Gem}}
\newcommand{\betHyi}{\object{$\beta$~Hyi}}
\newcommand{\betVir}{\object{$\beta$~Vir}}
\newcommand{\delEri}{\object{$\delta$~Eri}}
\newcommand{\epsEri}{\object{$\epsilon$~Eri}}
\newcommand{\epsFor}{\object{$\epsilon$~For}}
\newcommand{\epsVir}{\object{$\epsilon$~Vir}}
\newcommand{\etaBoo}{\object{$\eta$~Boo}}
\newcommand{\gamSge}{\object{$\gamma$~Sge}}
\newcommand{\ksiHya}{\object{$\xi$~Hya}}
\newcommand{\muAra}{\object{$\mu$~Ara}}
\newcommand{\muCas}{\object{$\mu$~Cas}}
\newcommand{\muLeo}{\object{$\mu$~Leo}}
\newcommand{\psiPhe}{\object{$\psi$~Phe}}
\newcommand{\tauCet}{\object{$\tau$~Cet}}
\newcommand{\cygA}{\object{61~Cyg~A}}
\newcommand{\cygB}{\object{61~Cyg~B}}

\newcommand{\tab}[1]{Table~\ref{#1}}
\newcommand{\fig}[1]{Fig.~\ref{#1}}
\newcommand{\sect}[1]{Sect.~\ref{#1}}
\newcommand{\FeI}{$\ion{Fe}{i}$}

\newcommand{\mgiant}{$M-$giants}
\newcommand{\kdwarf}{$K-$dwarfs}
\newcommand{\fgkgiant}{$FGK-$giants}
\newcommand{\fgdwarf}{$FG-$dwarfs}
\newcommand{\metalpoor}{$metal-poor$}

\defcitealias{2014A&A...564A.133J}{Paper~III}	

\begin{document}

\title{{Gaia FGK benchmark stars: abundances of $\alpha$ and iron-peak elements}
\thanks{Based on NARVAL and HARPS data obtained within the Gaia DPAC (Data Processing and
Analysis Consortium) and coordinated by the GBOG (Ground-Based Observations for Gaia) working group and on data retrieved from the ESO-ADP database.}
\thanks{Tables XX are only available in electronic form 
at the CDS via anonymous ftp to cdsarc.u-strasbg.fr (130.79.128.5) or via http://cdsweb.u-strasbg.fr/cgi-bin/qcat?J/A+A/}
}
\author{
	P. Jofr\'e \inst{\ref{ioa}}
	\and U. Heiter \inst{\ref{uppsala}} 
	\and C. Soubiran \inst{\ref{LAB}}
	\and S. Blanco-Cuaresma \inst{{\ref{LAB},\ref{geneva}}}
	\and T. Masseron \inst{\ref{ioa}}
	\and T. Nordlander \inst{\ref{uppsala}}
	\and L. Chemin \inst{\ref{LAB}}
	\and C. C. Worley \inst{\ref{ioa}}
	\and S. Van Eck \inst{\ref{ULB}}
	\and A. Hourihane \inst{\ref{ioa}}
	\and G. Gilmore \inst{\ref{ioa}}
	%here alphabetical
	\and V. Adibekyan \inst{\ref{uporto}}
	\and M. Bergemann \inst{\ref{ioa},\ref{mpia}}
	\and T. Cantat-Gaudin \inst{\ref{pad1}}
	\and E. Delgado-Mena \inst{\ref{uporto}}
	\and J. I. Gonz\'alez Hern\'andez \inst{\ref{UCM},\ref{tenerife}}
	\and G. Guiglion \inst{\ref{OCA}}
	\and C. Lardo \inst{\ref{liverpool}}
	\and P. de Laverny \inst{\ref{OCA}}
	\and K. Lind \inst{\ref{uppsala}}
	\and L. Magrini \inst{\ref{Arcetri}}
	\and S. Mikolaitis \inst{\ref{OCA}, \ref{vilnus}}
	\and D. Montes \inst{\ref{UCM}}
	\and E. Pancino \inst{\ref{bol1}, \ref{bol2}}
	\and A. Recio-Blanco \inst{\ref{OCA}}
	\and R. Sordo \inst{\ref{pad1}}
	\and S. Sousa \inst{\ref{uporto}}
	\and H. M. Tabernero  \inst{\ref{UCM}}
	\and A. Vallenari \inst{\ref{pad1}}
	}

\offprints{ \\ 
P. Jofr\'e, \email{pjofre@ast.cam.ac.uk}}

\institute{
	Institute of Astronomy, University of Cambridge, Madingley Road, Cambridge CB3 0HA, United Kingdom \label{ioa}
	\and Department of Physics and Astronomy,  Uppsala University, Box 516, 75120 Uppsala, Sweden \label{uppsala}
	\and   Univ. Bordeaux, CNRS, Laboratoire d'Astrophysique de Bordeaux (UMR 5804), F-33270, Floirac, France \label{LAB} 
	\and {Observatoire de Gen\'eve, Universit\'e de Gen\'eve, CH-1290 Versoix, Switzerland}\label{geneva}
	\and {Institut d'Astronomie et d'Astrophysique, U. Libre de Bruxelles, CP 226, Boulevard du Triomphe, 1050 Bruxelles, Belgium}\label{ULB}
	\and {Instituto de Astrof\'isica e Ci\^encias do Espa\c{c}o, Universidade do Porto, CAUP, Rua das Estrelas, 4150-762 Porto, Portugal} \label{uporto}
	\and {Max-Planck Institute for Astronomy, 69117, Heidelberg, Germany}\label{mpia}
	\and {INAF, Osservatorio  di Padova,  Universit\`a di Padova, Vicolo Osservatorio 5, Padova 35122, Italy} \label{pad1}
	%\and {Dipartimento di Fisica e Astronomia, Universit\`a di Padova Padova, vicolo Osservatorio 3, 35122 Padova, Italy} \label{pad2}
	\and {Laboratoire Lagrange (UMR7293), Univ. Nice Sophia Antipolis, CNRS, Observatoire de la C\^ote d'Azur,  06304 Nice, France} \label{OCA}
	\and{Astrophysics Research Institute, Liverpool John Moores University, 146 Brownlow Hill, Liverpool L3 5RF, UK}\label{liverpool}
	\and {INAF/Osservatorio Astrofisico di Arcetri, Largo Enrico Fermi 5 50125 Firenze, Italy}\label{Arcetri} 
	\and {Institute of Theoretical Physics and Astronomy, Vilnius University, A. Go\v{s}tauto 12, LT-01108 Vilnius, Lithuania} \label{vilnus}
	\and INAF-Osservatorio Astronomico di Bologna, Via Ranzani 1, 40127 Bologna, Italy  \label{bol1} 
	\and {ASI Science Data Center, via del Politecnico s/n, 00133 Roma, Italy} \label{bol2}
	\and {Dpto. Astrof\' isica, Facultad de CC. F\' isicas, Universidad Complutense de Madrid, E-28040 Madrid, Spain} \label{UCM}
	\and {Instituto de Astrof\'i sica de Canarias, E-38205 La Laguna, Tenerife, Spain}\label{tenerife}
	%\and {Dipartimento di Fisica \& Astronomia, Universit\'a  degli Studi di Bologna, Viale Berti Pichat 6/2, 40127 Bologna, Italy}\label{bol3}
	}

\authorrunning{Jofr\'e et al. }
\titlerunning{Gaia benchmark stars $\alpha$ and iron abundances}
   \date{}%Recived, acceptet

\abstract
 {In the current era of large spectroscopic surveys of the Milky Way, reference stars for calibrating astrophysical parameters and chemical abundances are of paramount importance. }{We determine elemental abundances of Mg, Si, Ca, Sc, Ti, V, Cr, Mn, Co and Ni for our predefined set of {\it Gaia} FGK benchmark stars. }{By analysing high-resolution and high-signal to noise spectra taken from several archive datasets, we combined results of eight different methods to determine abundances on a line-by-line basis. We perform a detailed homogeneous analysis of the systematic uncertainties, such as differential versus absolute abundance analysis, as well as we assess errors due to NLTE and the stellar parameters in our final abundances. }{Our results are provided  by listing final abundances and the different sources of uncertainties, as well as line-by-line and method-by-method abundances. }{The {\it Gaia} FGK benchmark stars atmospheric parameters are already being widely used for calibration of several pipelines applied to different surveys. With the added reference abundances of 10 elements this set is very suitable to calibrate the chemical abundances obtained by these pipelines. }
 
 \maketitle
 
 \section{Introduction}
 
 Much of our understanding on the structure and evolution of  the Milky Way comes nowadays from the analysis of large stellar spectroscopic surveys.  After the revolution of the way to pursue Galactic science with the low-resolution spectra from SDSS data \citep[see ][for a review]{2012ARA&A..50..251I}, new surveys are on-going. These  have much higher resolution than SDSS allowing not only to determine the stellar parameters of the stars more precisely  but also the chemical abundances of several individual elements. Examples of such projects are the {\it Gaia}-ESO Survey \citep[hereafter GES]{ 2012Msngr.147...25G, 2013Msngr.154...47R}, RAVE \citep{2006AJ....132.1645S}, APOGEE \citep{2008AN....329.1018A},  GALAH \citep{2015MNRAS.449.2604D} and the future billion of stars from the Radial Velocity Spectrograph  (RVS) from {\it Gaia}.  Furthermore, several 
  groups have collected over the years large samples of stars, creating ``independent" surveys for the same purpose of unraveling the structure and chemical enrichment history of our Galaxy \citep[e.g.][and references therein]{2011MNRAS.414.2893F, 2012AA...545A..32A, 2013ApJ...764...78R, 2014AA...562A..71B}.

 To parametrise these data properly in an automatic way, and furthermore,  to link the data between the different surveys in a consistent way,  good standard calibrators are needed.  To this aim, we have defined a sample, the {\it Gaia FGK benchmark stars} (hereafter GBS), which includes 34 FGK stars of a wide range of metallicities and gravities. These stars should be representative of the different FGK stellar populations of the Galaxy. The sample is presented in \cite[][hereafter Paper~I]{2015arXiv150606095H} describing the determination of effective temperature and surface gravity.  Briefly, the GBS were chosen such that the angular diameter, bolometric flux and distance of the stars are known.  { Angular diameters are known from interferometric observations for most of the stars with accuracies better than 1\%; bolometric fluxes are known from integrations of the observed spectral energy distribution for most of the stars with accuracies better than  5\%; and distances are known from parallaxes with accuracies better than 2\%.  The source and value for each star can be found in Paper~I.}  This  information allowed us to determine the temperature directly from the Stefan-Boltzmann relation.    With \teff\  and luminosity in hand, the mass could be determined homogeneously from stellar evolution models, and then the surface gravity using the Newton's law of gravity (see Paper~I for details).

 The third main atmospheric parameter for the characterisation of stellar spectra is the metallicity, [Fe/H], which was determined from a spectroscopic analysis. Since the GBS are located in the Northern and Southern hemispheres, we built a spectral library collecting high resolution and high signal-to-noise  (SNR) spectra \citep[hereafter Paper~II]{2014A&A...566A..98B}. Using this spectral library, we determined the metallicity  from iron lines \cite[hereafter Paper~III]{2014A&A...564A.133J}.  In that work we combined the results of six different methods, which used the same input atmosphere models and line list.  There are several studies in the literature reporting metallicities for the GBS, but as pointed out in Paper~III, they have a large scatter due to the different methods and input data employed in the analysis.  We determined the metallicity homogeneously, such that the [Fe/H] values for all stars can be used as reference in the same way.    In addition to a final [Fe/H] value, we provided  the results of each method for each star and spectral line.  This allows the GBS to be excellent material of reference when particular methods or spectral regions are being investigated.  A summary of this series of papers and the parameters of the GBS can be also found in \cite{2014ASInC..11..159J}.

 The material of the GBS are already being used to evaluate and calibrate several methods to determine parameters. One example is the GES pipeline \citep[Recio-Blanco et al in prep]{2014A&A...570A.122S}, where the spectra of the GBS have been observed  by the survey for this purpose \citep[see also Pancino et al. subm. and ][for calibration strategy of GES]{2013Msngr.154...47R}.  In addition, recently in \cite{2014A&A...570A..68D}, as part of the AMBRE project \citep{2013Msngr.153...18L} consisting in determining stellar parameters of the ESO archive spectra, the GBS parameters are used to show consistency. Furthermore, with the tools described in Paper~II we have  created GBS spectral libraries to have them in an SDSS-like data format which were analysed by \cite{2014MNRAS.443..698S}. We also created libraries to reproduce RAVE-like data which helps to improve the analysis of metal-rich stars of the RAVE sample \citep{2014sf2a.conf..431K} and GALAH-like spectra, which were initially used to develop its pipeline.  Recently, some GBS have started being observed for GALAH with its own instrument \citep{2015MNRAS.449.2604D}. Indeed, the GBS are showing the potential to be excellent stars to cross-calibrate different survey data.

 In this paper we present the further step in our analysis which is the determination of individual abundances.  The motivation for this is that high-resolution spectroscopic surveys determine not only the main stellar parameters automatically, but also individual abundances. Thus, a reference value for these abundances is needed.   Since the GBS are well known, there is an extensive list of previous works that have measured individual abundances, but none of them have done it for the whole sample. Under the same argument as in Paper~III (inhomogeneity in the literature), we determined the abundances in an homogeneous way for all the GBS. 
    
    We focus in this article on the abundance determination of the $\alpha$  elements Mg, Si, Ca and Ti  and the iron-peak elements Sc, V, Cr, Mn, Co and Ni. There are two main reasons for starting with these elements. The first one is a practical reason: the data contains at least 12 spectral lines for each of the elements, which allow us to follow a similar procedure as in Paper~III for deriving the iron abundances. The second reason is that $\alpha$ and iron-peak element are widely used for Galactic chemo-dynamical studies \citep[see e.g. ][and references therein]{2014AA...562A..71B, 2014A&A...568A..71B, 2014A&A...571L...5J, 2014A&A...572A..33M, 2014ApJ...796...38N}. %Furthermore, although the main atmospheric parameter on stellar spectra is [Fe/H], the $\alpha$ abundances, usually denoted as [$\alpha$/Fe], change the opacities affecting the determination of [Fe/H], reason why several pipelines doing matching of large spectral regions determine [Fe/H] together with [$\alpha$/Fe] \citep{2006MNRAS.370..141R, }  Furtherthe first scientific results on the Milky Way field chemistry from the analysis of GES for example have been focused mainly in $\alpha$ and iron-peak element analyses \citep[e.g.][]{2014A&A...571L...5J, 2014arXiv1408.6687M, 2014A&A...567A...5R, 2014A&A...565A..89B}. 

This article is organised as follows. In \sect{data} we describe the data used in this work, which includes a brief description of the updates of our library as well as the atomic data considered for our analysis. In \sect{strategy} we explain the methods and strategy employed in our work, that is, we describe the different methods to determine abundances considered here, as well as the analysis procedure employed by the methods.  The analysis of our results and the determination of abundances is explained in \sect{abundances} while  the several sources of systematic errors are described in \sect{errors}, such as NLTE departures and uncertainties of the atmospheric parameters. We proceed in the article with a detailed discussion of our results for each element individually in \sect{discussion_elements}. In \sect{conclusions} we summarise and conclude  this work.

 \section{Spectroscopic data and input material}\label{data}
  In this section we describe the data we employed in this analysis.  By data we refer to the spectra (described in \sect{lib}), list of spectral lines  (described in \sect{lines}) and the atomic data and atmospheric models (described in \sect{input_mod}).

 \subsection{Spectral library}\label{lib}
 
  As in our previous work on the subject, we built a library of high resolution spectra of the GBS, using our own observations on the NARVAL spectrograph at Pic du Midi in addition to archived data.    The different spectra were processed with the tools described in Paper~II\footnote{The spectral library can downloaded from \href{http://www.blancocuaresma.com/s/}{http://www.blancocuaresma.com/s/benchmarkstars/} } and in \cite{2014A&A...569A.111B}. Briefly, the spectra were   normalised, convolved to a common resolution, radial velocity corrected and re-sampled. The final library employed here differs from the 70~k library used in Paper~III in the following aspects:
  
 \begin{itemize}
 \item {\it A new source of spectra:}   ESPaDOnS spectra were retrieved from the PolarBase \citep{2014PASP..126..469P}. They were ingested in our library in the same fashion as the standard spectra from HARPS, UVES and NARVAL. The advantage of ESPaDOnS spectra is that the original spectra cover a very large wavelength range   like those of NARVAL,  (they are the same spectrographs)  and have high resolution and high SNR.  In addition to ESPaDOnS spectra, we added for the Sun and Arcturus the spectra from the Atlas of \cite{2000vnia.book.....H}. Although the atlases were part of our library published in Paper~II, they became available after the analysis of Paper~III was carried out. 
 \item {\it New processed spectra:} Since the analysis of Paper~III, the UVES advanced data archive has provided newly reduced spectra of all archive data in an homogenous fashion. We have updated our spectra considering this. In addition, over the past year new spectra of \bs\ have been taken with UVES, which we ingested in our library. In particular, the spectrum of \alfCenA\ was kindly provided by Svetlana Hubrig before being public in the ESO archives. 
 \item {\it Wavelength coverage:} The spectral range of the HR21 Giraffe setup   ($\sim$848~--~875~nm), was included in addition to the standard UVES 580 ($\sim$480~--~680~nm). The reason was to provide reference spectra and abundances in the wavelength range cover by the Gaia-RVS spectrograph and with Milky Way field targets of GES observed with Giraffe.   
\item {\it Telluric free:} The telluric lines in the HR21 range were removed from the spectra (Sordo, R. priv. comm). 
 \item {\it Resolution:} The data for this study at all wavelength ranges have a resolving power of R = 65,000. This limit was set according to the ESPaDONs spectra, which have that resolution.
 \item {\it Normalisation:} The spectra are normalised using the newest normalisation routines of iSpec as described in \cite{2014A&A...569A.111B}. 
 \end{itemize}
 The source of the spectrum used for each star is summarised in \tab{tab:spectra}. Note that for some stars we could not find a spectrum in the HR21 range with high resolution.  
As in Paper~III, based on visual criteria, we selected ``our favourite'' spectra for each star, which were based on continuum placement and telluric contamination.  For the UVES-580 wavelength range, we took two spectra (UVES1 and UVES2 in \tab{tab:spectra}) except those cases where we had only one spectrum per star. There were three main reasons for choosing two spectra: (1)  to cover the wavelength gap of the red and blue CCD of the UVES-580 setup with data of other spectrographs, (2) to determine abundances in telluric regions with more confidence, (3) as validation check for repeated lines, which must give same abundances regardless of instrument.   In the red RVS wavelength range (HR21 in \tab{tab:spectra}) we chose only one spectrum per star, due to the fewer spectra available at high resolution in this wavelength range.   Furthermore, note the Sun as observed by HARPS has no date of observation. This spectrum is the co-addition of the three spectra of asteroids in the HARPS archive (see Paper~II for details). Note that for the star HD84937 we used the UVES and the UVES-POP spectra, which were taken the same night. This means that  we analysed the same spectrum reduced with two different pipelines.

 \begin{table}
 \caption{Spectral source used for each star. In parenthesis the date of observation of the spectra is indicated, except if the spectrum was the product of stacked spectra taken in different nights. N: Narval, U: UVES, E: ESPaDOnS, H: HARPS, P: UVES.POP, A: Atlas.}
 \label{tab:spectra}
\tiny
\begin{tabular}{c | c c c}
\hline
% Paula Jofre May 2014 spectra information for analysis
star & UVES1 & UVES2 & HR21\\
\hline
18Sco& E (2005-06-20)  &N (2012-03-10)  & E (2005-06-20)  \\
\cygA&N (2009-10-16)  &-  &N (2009-10-16)   \\
\cygB&N (2009-10-13) &-  &N (2009-10-13) \\
\alfCenA&H (2005-04-19) &U (2000-04-11)  &U (2012-01-20) \\
\alfCenB&H (2005-04-08)  &-  & -  \\
\alfCet&U (2003-08-11) &N (2009-12-09) &N (2009-12-09) \\
\alfTau&U (2004-09-24) &H  (2007-10-22)&N (2009-10-26) \\
Arcturus&N (2009-12-11) &A (2000-01-01) &A (2000-01-01)  \\
\betAra&H (2007-09-29)  &-  &-  \\
\betGem&H (2007-11-06)  &U (2008-02-25) &E (2007-12-29)  \\
\betHyi&P (2001-07-25) &H (2005-11-13)  &P (2001-07-25)  \\
\betVir&E  (2005-12-15) &H (2009-04-10) &E  (2005-12-15)  \\
\delEri&P (2001-11-28) &-  &P (2001-11-28) \\
\epsEri&H  (2005-12-28) &P (2002-10-11) &P (2002-10-11)  \\
\epsFor&H (2007-10-22) &-  &-  \\
\epsVir&E (2996-02-15) &H (2008-02-24)  &N (2009-11-27) \\
\etaBoo&N (2009-12-11) &H (2008-02-24) &N (2009-12-11) \\
\gamSge&N (2011-09-30) &-  &N( 2011-09-30)  \\
Gmb~1830&N (2012-01-09) &-  &N (2012-01-09)  \\
HD107328&H  (2007-10-22) &N (2009-11-26)  &N (2009-11-26)  \\
HD122563&E (2006-02-16) &U (2002-02-19)  &E (2006-02-16)  \\
HD140283&E (2011-06-12) &N  (2012-01-09) &E  (2011-06-12) \\
HD220009&N  (2009-10-16) &H (2007-10-22)  &N (2009-10-16)  \\
HD22879&H (2007-10-22)  &N (2009-11-27) &N (2009-11-27) \\
HD49933&E (2005-12-18) &H (2011-01-05) &E  (2005-12-18)\\
HD84937&U  (2002-11-28) &P (2002-11-28) &N (2012-01-08) \\
\ksiHya&E (2005-09-21)  &H (2008-02-24) &E (2005-09-21) \\
\muAra&H  (2004-06-08) &U (2003-09-05)  &U (2011-04-12)  \\
\muCas&N (2009-11-26)  &-  &N (2009-11-26)  \\
\muLeo&E (2006-02-17) &N (2011-12-10) &E (2006-02-17) \\
Procyon&E (2005-12-14) &H  (2007-11-06) &E (2005-12-14) \\
\psiPhe&H (2007-09-30) &U (2003-02-08) & -  \\
Sun & H (-) &A  (2000-01-01) &A (2000-01-01)  \\
\tauCet&E (2005-09-21) &H (2008-09-09)  &E (2005-09-21) \\

\hline
 \end{tabular}
 \end{table}
 
  \normalsize
 \subsection{Line list}\label{lines}
 The elements to analyse were selected by the Porto and the Epinarbo methods (see \sect{porto} and \sect{epi}, respectively), as described below. From the lines of the GES v4 line list \citep{1402-4896-90-5-054010}, we rejected all those lines whose flag related to the atomic data quality was ``N'' {  (meaning that the transition probabilities are expected to have low accuracy and the usage of these lines is not recommended). However, we allowed for lines for which the synthesis profile in the Sun and Arcturus were flagged with ``N'' (meaning that the line is strongly blended with line(s) of different species in both stars)} since we work with stars that are different form the Sun and Arcturus, for which this line could have a better synthesis profile. For details of such flags, see \cite{1402-4896-90-5-054010}.  In this article we focused on the 10 $\alpha$ and iron-peak elements, which have at least 12 spectral lines. The elements and the number of initial lines are indicated in \tab{tab:lines}. Additionally, as part of the online material, the table (INFO\_LINES) contains the wavelength and atomic data of these lines. 

 \subsubsection{Selection of lines in the  480~--~680~nm UVES  range}

The lines in this range were selected by the Epinarbo method (see \sect{epi}) mainly  on the basis of a statistical analysis of the DR1 UVES sample of the Gaia-ESO Survey \citep[e.g.][]{2014A&A...563A..44M}, which included 421 stars with recommended parameters \citep[see][]{2014A&A...570A.122S}. Equivalent widths (EWs) were measured in a homogeneous way with an automatic version of Daospec \citep[DOOp, ][]{2014A&A...562A..10C}, and the abundances were determined with the method FAMA \citep{2013A&A...558A..38M} for lines in the EW range of 15-100 m\AA. This range helps to avoid saturated lines and faint lines affected by noise.  Then,  the distribution of the deviations  from the averaged abundance of each element were computed in all 421 stars. Finally, lines with  standard deviation within the corresponding 68.2\% percentile were selected.

 \subsubsection{Selection of lines in the  848~--~875~nm HR21  range}
This selection was done by the Porto method (see \sect{porto}) where  a selection of strong lines ($\log gf > - 4$) was taken. Only those unblended lines whose EW  were potentially measurable by the ARES code \citep{sousa_ares} were considered. Then, abundances were determined for those lines using a subset of GBS spectra in the HR21 wavelength range using the method described in \sect{porto} and the stellar parameters indicated in Paper~III. { The deviation for all the lines was calculated by comparing with the mean abundance. The lines which on average (for all the stars) gave abundances  that differed from the mean abundance (derived by all the lines) by $\pm$0.3~dex were rejected. }

\begin{table}
\begin{center}
\caption{Number of initially selected lines for each element. } \label{tab:lines}
\begin{tabular}{c|c|c}
\hline
element & atom & N lines \\
\hline
%Na&      11&           5\\
Mg&      12&          12\\
%Al&      13&           3\\
Si&      14&          15\\
Ca&      20&          25\\
Sc&      21&           17\\
Ti&      22&          68\\
V&      23&          30\\
Cr&      24&          25\\
Mn&      25&          13\\
Co&      27&          22\\
Ni&      28&          25\\
%Sm&      62&           4\\
%Eu&      63&           3\\
\hline 
\end{tabular}
\end{center}
\end{table}%

\subsection{Atomic data and atmospheric models}\label{input_mod}
  
The atomic data was taken from the fourth version of the  line list created for the Gaia-ESO survey \citep{1402-4896-90-5-054010}.   Likewise, the atmospheric models are those employed by the analysis of the spectra in the Gaia-ESO survey. These are the MARCS models \citep{2008A&A...486..951G}, which are computed under the 1D-LTE assumption { and assume the standard composition for $\alpha-$enhancement respect to iron abundance}. 
  
 \section{Analysis strategy}\label{strategy}
 To determine individual abundances we employed a similar strategy as that one of Paper~III, namely fixing the stellar parameters, and using a pre-selection of lines that were analysed by different methods determining the abundances. The results were then combined at a line-by-line basis and finally NLTE departures were computed. 
 
 \subsection{Stellar parameters}
 The idea is to use the effective temperature and surface gravity from Paper~I and the metallicity  from Paper~III as well as the averaged value for micro turbulence obtained by the different methods. The initial value of macro turbulence was set to zero.  The rotational velocity is the same value employed in Paper~III, which comes from the literature.   To assess systematic errors we ran several times the same procedure, considering the uncertainties on the stellar parameters. These parameters are indicated in \tab{tab:params}.  
 
 It is important to mention here that \feh, \teff\ and \logg\ are not completely consistent between Paper~I, Paper~III, and this work, as we are continuously improving them. The metallicity was determined in Paper~III using a previous line list version for GES than here.  This should not affect significantly our results as changes between v3 and v4 of the GES linelist have not been done for atomic data of iron.  Also, for \betAra, HD140283, and HD220009, new angular diameter measurements became available \citep[by ][for HD140283, and by Th\'evenin et al in prep for \betAra\ and HD220009]{2015A&A...575A..26C} after the results of abundances by the different methods (see \sect{methods}) were provided, which explains the the differences in \teff\ for these stars. However,  as discussed in Paper~I, the resulting \teff\ values are still considered uncertain and were not recommended as reference values. Further stars with uncertain parameters were \muAra, \psiPhe, and Gmb~1830.  For HD84937, a new parallax was published \citep{2014ApJ...792..110V} since the results of abundances by the different methods were provided, explaining the difference in \logg. For Arcturus, the recommended \logg\ is $1.6 \pm 0.2$, that is, the \logg\ uncertainty is twice as large as what was considered here. Stars with uncertain \logg\ values are \epsFor, \muCas, \tauCet, HD 220009, \betAra, \psiPhe\ (see discussions in Paper I).  Although the parameters slightly evolve throughout Papers~I, II and III, the values employed are still within the errors, which is taken into account in our spectral analyses (see below).  In this paper, for completeness with Paper~I, II and III,  we analyse the whole initial GBS sample, regardless of how uncertain the stellar parameters are and our suggestions made in Paper~I to which stars should be treated as reference and which should not.

\subsection{Runs}\label{runs}

The different analysis runs were identical except of the input parameters. For each run we fixed all parameters (\teff, \logg, \feh, \vmic, \vsini), as indicated in \tab{tab:params}. Macroturbulence was determined together with the abundances for those analysis making synthesis on-the-fly.  Some methods re-normalised and shifted in radial velocity the spectra to improve their results. The different analysis runs are described below: 

\begin{itemize}
\item {\it Run - all}: Main run: determination of individual abundances of all lines and all spectra using the main stellar parameters of the input table.  
\item {\it Run - LTE}:  like before but using the metallicity value obtained before NLTE corrections (i.e. the input of [Fe/H] - LTE, see \tab{tab:params}). This run allowed us to quantify this effect in the abundances. 
\item {\it Run-errors}:   As {\it Run-all} but considering error on the stellar parameters as determined for the metallicity in Paper~III. 
\end{itemize}

%\noindent All our abundance values correspond to the logarithmic number densities relative to hydrogen, where H=12: 
%$\log\epsilon = \log(\mathrm{N_X})-\log(\mathrm{N_H})+12$.

 \begin{table*}
 \caption{Values of stellar and broadening parameters considered for the abundance determination. Effective temperature and surface gravity is derived from fundamental laws (see Paper~I for details). Metallicity and microturbulence velocity were derived consistent with these parameters in Paper~III. [Fe/H] uncertainties were obtained by quadratically summing all $\sigma$ and $\Delta$ columns in Table~3 of Paper~III. In addition we list the value of metallicity before the correction of NLTE effects, which is used in one of the analysis runs (see \sect{runs}). Rotational velocity is taken from the literature (see Paper~III for the corresponding references).}
 \label{tab:params}
\begin{tabular}{c | c c c c c c }
\hline
% Paula Jofre May 2014 intial values for analysis
star & \teff $\pm \Delta$ \teff [K] & \logg $\pm \Delta$ \logg\ (dex) & \feh $\pm \Delta$ \feh\ (dex)& \vmic $\pm \Delta$ \vmic\ [km/s] &   \feh$_{LTE}$ (dex)&  \vsini\  [km/s]\\
\hline
18~Sco&5810 $\pm$80 &4.44 $\pm$0.03 &0.03 $\pm$0.03 &1.07 $\pm$0.20 &0.01 &2.2  \\
\cygA&4374 $\pm$22 &4.63 $\pm$0.04 &-0.33 $\pm$0.38 &1.07 $\pm$0.04 &-0.33 &0.0  \\
\cygB&4044 $\pm$32 &4.67 $\pm$0.04 &-0.38 $\pm$0.03 &1.27 $\pm$0.36 &-0.38 &1.7  \\
\alfCenA&5792 $\pm$16 &4.30 $\pm$0.01 &0.26 $\pm$0.08 &1.20 $\pm$0.07 &0.24 &1.9  \\
\alfCenB&5231 $\pm$20 &4.53 $\pm$0.03 &0.22 $\pm$0.10 &0.99 $\pm$0.31 &0.22 &1.0  \\
\alfCet&3796 $\pm$65 &0.68 $\pm$0.29 &-0.45 $\pm$0.47 &1.77 $\pm$0.40 &-0.45 &3.0  \\
\alfTau&3927 $\pm$40 &1.11 $\pm$0.15 &-0.37 $\pm$0.17 &1.63 $\pm$0.30 &-0.37 &5.0  \\
Arcturus&4286 $\pm$35 &1.64 $\pm$0.06 &-0.52 $\pm$0.08 &1.58 $\pm$0.12 &-0.53 &3.8  \\
\betAra&4173 $\pm$64 &1.04 $\pm$0.15 &-0.05 $\pm$0.39 &1.88 $\pm$0.46 &-0.05 &5.4  \\
\betGem&4858 $\pm$60 &2.90 $\pm$0.06 &0.13 $\pm$0.16 &1.28 $\pm$0.21 &0.12 &2.0  \\
\betHyi&5873 $\pm$45 &3.98 $\pm$0.02 &-0.04 $\pm$0.06 &1.26 $\pm$0.05 &-0.07 &3.3  \\
\betVir&6083 $\pm$41 &4.10 $\pm$0.02 &0.24 $\pm$0.07 &1.33 $\pm$0.09 &0.21 &2.0  \\
\delEri&4954 $\pm$26 &3.75 $\pm$0.02 &0.06 $\pm$0.05 &1.10 $\pm$0.22 &0.06 &0.7  \\
\epsEri&5076 $\pm$30 &4.60 $\pm$0.03 &-0.09 $\pm$0.06 &1.14 $\pm$0.05 &-0.10 &2.4  \\
\epsFor&5123 $\pm$78 &3.52 $\pm$0.07 &-0.60 $\pm$0.10 &1.04 $\pm$0.13 &-0.62 &4.2  \\
\epsVir&4983 $\pm$61 &2.77 $\pm$0.02 &0.15 $\pm$0.16 &1.39 $\pm$0.25 &0.13 &2.0  \\
\etaBoo&6099 $\pm$28 &3.80 $\pm$0.02 &0.32 $\pm$0.08 &1.52 $\pm$0.19 &0.30 &12.7  \\
\gamSge&3807 $\pm$49 &1.05 $\pm$0.34 &-0.17 $\pm$0.39 &1.67 $\pm$0.34 &-0.16 &6.0  \\
Gmb~1830&4827 $\pm$55 &4.60 $\pm$0.03 &-1.46 $\pm$0.39 &1.11 $\pm$0.57 &-1.46 &0.5  \\
HD107328&4496 $\pm$59 &2.09 $\pm$0.14 &-0.33 $\pm$0.16 &1.65 $\pm$0.26 &-0.34 &1.9  \\
HD122563&4587 $\pm$60 &1.61 $\pm$0.07 &-2.64 $\pm$0.22 &1.92 $\pm$0.11 &-2.74 &5.0  \\
HD140283&5514 $\pm$120 &3.57 $\pm$0.12 &-2.36 $\pm$0.10 &1.56 $\pm$0.20 &-2.43 &5.0  \\
HD220009&4275 $\pm$54 &1.47 $\pm$0.14 &-0.74 $\pm$0.13 &1.49 $\pm$0.14 &-0.75 &1.0  \\
HD22879&5868 $\pm$89 &4.27 $\pm$0.03 &-0.86 $\pm$0.05 &1.05 $\pm$0.19 &-0.88 &4.4  \\
HD49933&6635 $\pm$91 &4.20 $\pm$0.03 &-0.41 $\pm$0.08 &1.46 $\pm$0.35 &-0.46 &10.0  \\
HD84937&6356 $\pm$97 &4.15 $\pm$0.06 &-2.03 $\pm$0.08 &1.39 $\pm$0.24 &-2.09 &5.2  \\
\ksiHya&5044 $\pm$38 &2.87 $\pm$0.02 &0.16 $\pm$0.20 &1.40 $\pm$0.32 &0.14 &2.4  \\
\muAra&5902 $\pm$66 &4.30 $\pm$0.03 &0.35 $\pm$0.13 &1.17 $\pm$0.13 &0.33 &2.2  \\
\muCas&5308 $\pm$29 &4.41 $\pm$0.01 &-0.81 $\pm$0.03 &0.96 $\pm$0.29 &-0.82 &0.0  \\
\muLeo&4474 $\pm$60 &2.51 $\pm$0.09 &0.25 $\pm$0.15 &1.28 $\pm$0.26 &0.26 &5.1  \\
Procyon&6554 $\pm$84 &3.99 $\pm$0.02 &0.01 $\pm$0.08 &1.66 $\pm$0.11 &-0.04 &2.8  \\
\psiPhe&3472 $\pm$92 &0.51 $\pm$0.18 &-1.24 $\pm$0.39 &1.75 $\pm$0.33 &-1.23 &3.0  \\
Sun&5777 $\pm$1 &4.44 $\pm$0.00 &0.0300 $\pm$0.05 &1.06 $\pm$0.18 &0.02 &1.6  \\
\tauCet&5414 $\pm$21 &4.49 $\pm$0.01 &-0.49 $\pm$0.03 &0.89 $\pm$0.28 &-0.50 &0.4  \\

\hline
 \end{tabular}
 \end{table*}

\subsection{Methods to determine abundances}\label{methods}
Eight methods were used to determine the abundances and are described briefly in this section. Most of the methods were employed in the metallicity determination of Paper~III and in the determination of \teff, \logg\ and abundances within the Gaia-ESO Survey for the UVES data \citep[WG11 pipeline, see][for details]{2014A&A...570A.122S}.  { A summary of the methods can be found in \tab{tab:methods} and are briefly explained below. }

\begin{table}
 \caption{Summary of methods employed to the determination of abundances in this work. The name of the method, the approach (EW: equivalent width, synth: synthesis), the radiative transfer code employed and the wrapper code that uses the radiative transfer code (if applicable) are indicated. }
 \label{tab:methods}
{\small 
\begin{tabular}{c | c c c}
\hline 
name & approach  & radiative transfer code & wrapper \\
\hline 
iSpec & synth &  SPECTRUM & iSpec\\
ULB & synth/EW & Turbospectrum & BACCHUS\\
Porto & EW & MOOG & \\
Bologna & EW & SYNTHE & GALA\\
Epinarbo & EW & MOOG & FAMA\\
GAUGUIN & synth & Turbospectrum\\
Synspec & synth & Turbospectrum\\
UCM & EW & MOOG & StePar\\
\hline
 \end{tabular}
 }
 \end{table}

\subsubsection{iSpec}

iSpec \citep{2014A&A...569A.111B} is a spectroscopic framework that implements routines for the determination of chemical abundances by using the spectral fitting technique. Given a set of atmospheric parameters, atomic data and wavelength ranges, iSpec generates synthetic spectra on the fly and minimises the difference with the observed spectra by applying a least-square algorithm.

We developed a completely automatic pipeline for the analysis of the GBS. Each absorption line of each spectrum was analysed separately by the same homogeneous process. Even though iSpec includes routines to identify unreliable or doubtful solutions, we did not apply any automatic filtering to facilitate the comparison with the rest of the methods.

\subsubsection{ULB}

The BACCHUS (for Brussels Automatic Code for Characterising High
accUracy Spectra) consists in three different modules respectively
designed to derive EWs, stellar parameters and
abundances. For the purpose of this paper, only the modules for
measuring abundances and EWs have been used. The current
version relies on the radiative transfer code Turbospectrum
\citep{Alvarez1998,Plez2012}.  This method has been employed in Paper~III, as well as for all  for the WG11 pipeline.

With fixed stellar parameters, the first step consists in determining
average line broadening parameters (i.e. macroturbulence parameter in
the present case) using a selection of clean Fe lines. Then, for each
element and each line, the abundance determination module proceeds in
the following way: 
(i) a spectrum synthesis, using the full set of (atomic and molecular)
lines, is used for local continuum level finding (correcting for a
possible spectrum slope);
(ii) cosmic and telluric rejections are performed;
(iii) local signal-to-noise is estimated;
(iv) a series of flux points contributing to a given
absorption line is selected.
Abundances are then derived by comparing the observed spectrum
with a set of convolved synthetic spectra characterised by different
abundances.  Four different diagnostics are used:  line-profile fitting, core line
intensity comparison, global goodness-of-fit estimate, and EW comparison. Each diagnostic yields validation flags.
Based on those flags, a decision tree then rejects the
line, or accepts it keeping the best matching abundance.

One supplementary asset of the code is the computation of EWs. They are computed not directly on the observed
spectrum, but internally from the synthetic spectrum with the
best-matching abundance. This way, we have access to the information
about the contribution of blending lines, allowing a clean computation
of the equivalent width of the line of interest.

\subsubsection{Porto}\label{porto}
It employes  ARES \citep{sousa_ares} to measure EWs (automatically normalising the spectra) and MOOG \citep{sneden} to derive abundances. For refractory element abundances (from Na to Ni),  the fast rotator \etaBoo\ and stars with \teff\ < 4200 K (usually the EWs with ARES for these stars are not good) were rejected. This method has been employed in Paper~III, as well as   for the WG11 pipeline.%For heavier elements, all elements and lines were considered for the analysis.

\subsubsection{Bologna}
Bologna analysis is based on the same method of Paper~III. It has also been used in the WG11 pipeline. 
In particular, we ran DAOSPEC \citep{2008PASP..120.1332S} to measure EWs through DOOp \citep{2014A&A...562A..10C} until the input and output FWHM of the absorption lines agreed within 3\%.
The abundance analysis was carried out with GALA \citep{2013ApJ...766...78M}, an automatic program for 
atmospheric parameter and chemical abundance determination from atomic lines based on the SYNTHE code \citep{kurucz}. 
In order to provide measurements for all the selected lines, discrepant lines with respect to the fits of the slopes of Fe abundance versus EW, excitation potential, and wavelength were rejected with a very large  $5\sigma$ cut. All stars and elements were analysed in this fashion.

\subsubsection{Epinarbo}\label{epi}
This method is based on EWs from DOOp, which are measured in a similar way to the Bologna method (see above). The abundance determination is done with the code FAMA \citep{2013A&A...558A..38M}, which is based on MOOG. This method has been employed in Paper~III, as well as  for the WG11 pipeline.

FAMA can perform stellar parameter determinations, or work with fixed parameters and return elemental abundances. In this analysis all parameters were kept fixed, including the microturbulent velocity. We provided abundances for all the selected lines that were detected. No abundance was returned if the line was detected but its EW was smaller than 5\,{m$\AA$} or larger than 140\,{m$\AA$}, to avoid measurement errors associated with very weak or very strong lines.

\subsubsection{Nice/GAUGUIN}

For a given benchmark star, we first assumed an $[\alpha/\text{Fe}]$ enrichment consistent with the typical properties of Milky Way stars.
Then, we normalised the observed spectrum by (i) linearly interpolating a synthetic spectrum in the GES synthetic spectra grid\footnote{Synthesised with the GES v4 line list and convolved to the observed resolution and to the rotational velocity of the star, more details on the grid computation in \cite{grid_patrick}}, (ii) estimating
a ratio between the synthetic flux and the observed one over a spectral range of $20\,$\AA, centered of the interested line and, (iii) fitting this ratio by a polynomial function. Finaly, the observed spectra is divided by the polynomial fit in order to adjust its continuum.

The individual chemical abundances were then derived as follows: (i) 1-D synthetic spectra grids for each stars are built from the initial 4-D GES grid. These grids in the searched chemical abundances are cut around the analised spectral line. Practically, for the iron-peak species derivation, we
linearly interpolated the GES grid on \teff, \logg\  and $[\alpha/\text{Fe}]$. For the $\alpha$
element cases, we linearly interpolated this grid on \teff, \logg\ and [M/H]. (ii) Then, we looked for the minimal difference between the observed
spectrum and the 1-D synthetic spectra grids. 
(iii) The solution was finally refined with the
Gauss-Newton algorithm GAUGUIN \citep{gauguin}.

\subsubsection{Nice/Synspec}
 
  \begin{figure*}
  \resizebox{\hsize}{!}{\includegraphics{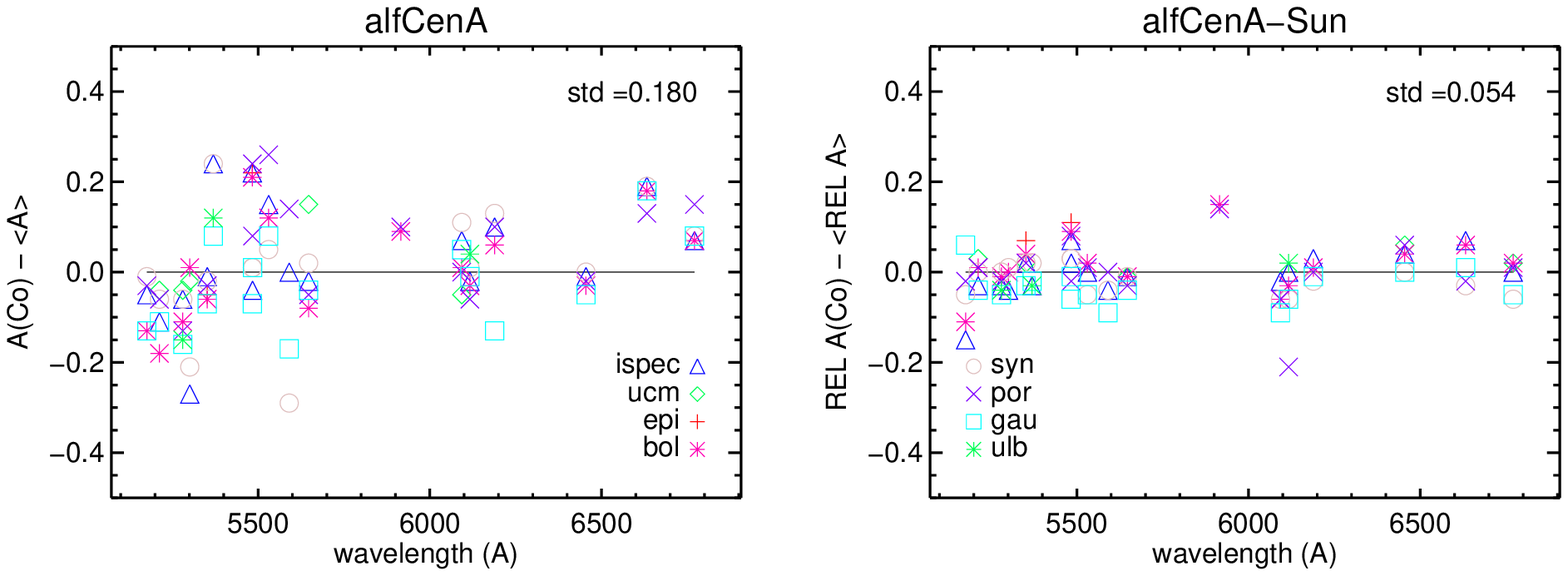}}
  \resizebox{\hsize}{!}{\includegraphics{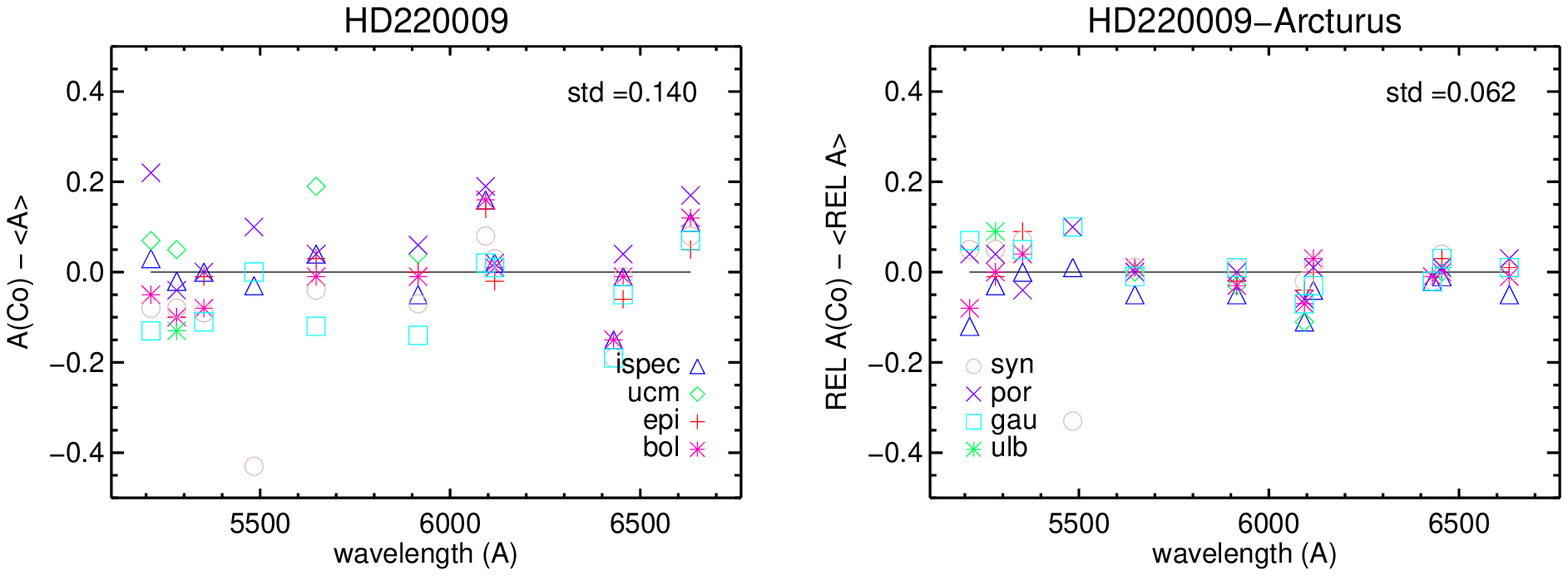}}
  \caption{Abundances of Co at a line by line and method by method basis as a function of wavelength. Colours and symbols represent the different methods, which are indicated in the legend. Top panels:  absolute (left) and relative to the Sun (right) abundances of the star \alfCenA. Bottom panels: absolute (left) and relative to Arcturus (right) abundances of the star HD220009. The horizontal line represents the mean of all abundances with its standard deviation indicated at the top right of each panel. }
  \label{fig:nodes-abu}
\end{figure*}

We adopted the recent version (v12.1.1) of the spectrum synthesis code Turbospectrum \citep{Plez2012}. This pipeline determines the continuum in two steps. First, it takes the normalised spectra from the library. Second, it adjusts the continuum locally in the region ($\pm 5\AA$) around every line of interest. It was done by selecting the possible line-free zones of the synthetic spectrum, defined as regions where the intensity of the synthetic spectrum is depressed by less than 0.02. If the possible line-free zones were too narrow or did not exist, we iteratively searched for the possible less contaminated zones in the synthetic spectrum.
Finally we determined abundances with the method described in \citet{2014A&A...572A..33M}. This method has been employed in the WG11 pipeline.

\subsubsection{UCM}
This method is based on EWs and has been used in Paper III and is part of the WG11 GES pipeline. Different line selections were considered for different stars.  The division of stars was based on metallicity and  surface gravity.  Metallicity was divided in metal-rich ([Fe/H] $\geq -0.30$),  metal-poor ($-0.30 <$ [Fe/H] $\geq -1.50$), and very metal-poor ([Fe/H] $< -1.50$). Surface gravity was divided in: giants (\logg $<$ 4.00) and dwarfs (\logg $\geq$ 4.00). However, we decided to merge into one single region the very metal poor stars ([Fe/H] $<$ -1.50).
The EWs were  measured using TAME \citep{2012MNRAS.425.3162K}. We followed the approach of \citet{2012MNRAS.425.3162K} to adjust the $rejt$ parameter of TAME according to the SNR of each spectrum.  The abundance analysis was carried out with  using a wrapper program for MOOG, in order to take care of the elemental abundances
automatically, based on {\scshape StePar}  \citep[see][]{2012A&A...547A..13T}. We also made a rejection of outliers for those lines that deviate more than three of the standard deviation.

     \begin{figure*}
 \resizebox{\hsize}{!}{\includegraphics[angle=90]{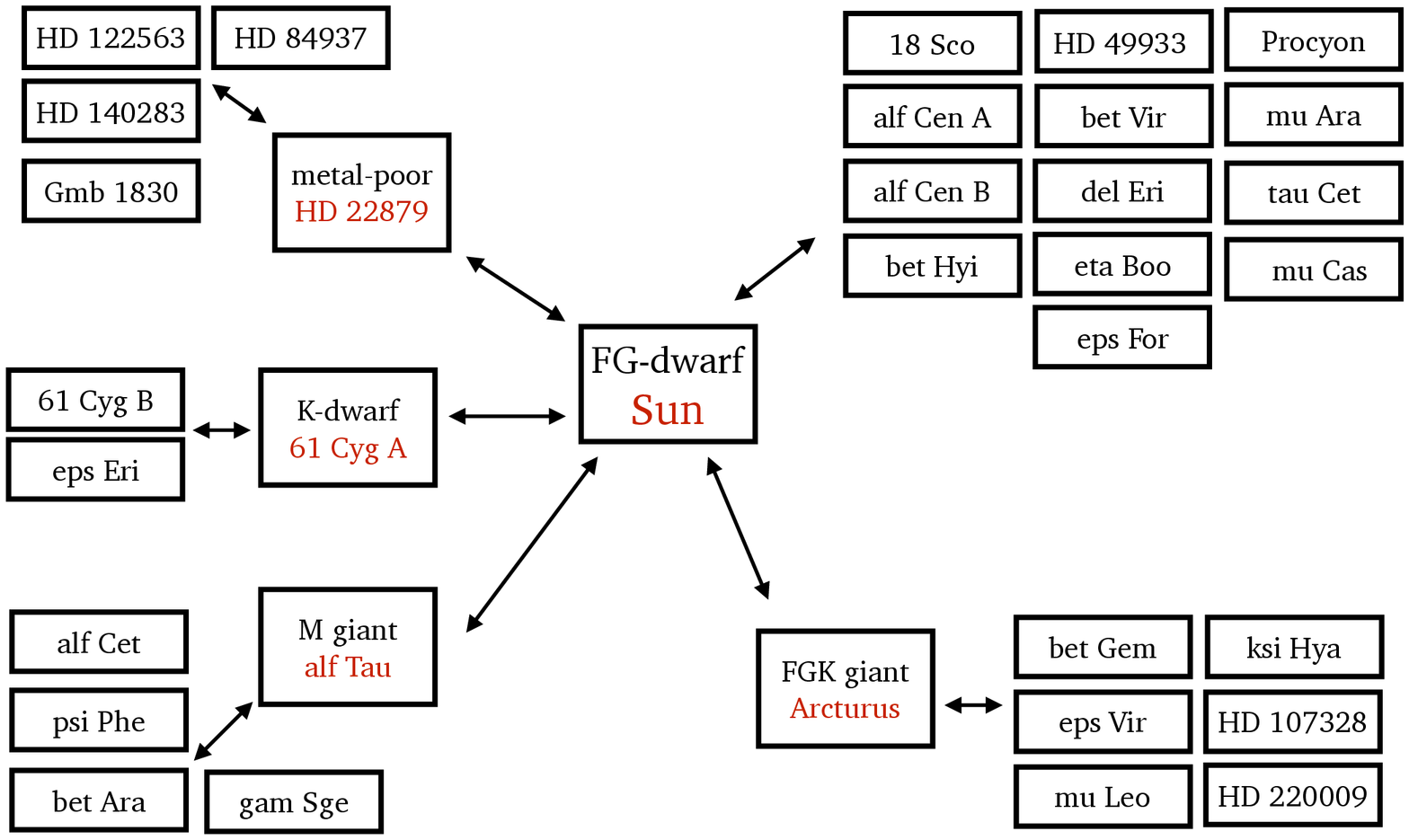}}
   \vspace{-3cm}
   \caption{Schematic picture of how the GBS are differentiated against each other.  Stars are associated in five different groups according to their spectral type. One star is chosen as a reference for each group (in red). The rest of the stars in that group is analysed with respect to the reference star, and are connected with arrows. The reference stars are finally analysed with respect to the Sun, which is  the zero point.}
  \label{fig:scheme}
\end{figure*}

 \begin{table}
 \begin{center}
\caption{Final absolute abundances for the Sun obtained in this work (here), where the standard deviation at a line-by-line basis is indicated as $\sigma$.  For comparison, we list the abundances of \cite[G07]{2007SSRv..130..105G}  with their reported error in the last two columns. -- here we need new table with all odd-Z with hfs only}
 \label{tab:sun}
\begin{tabular}{c | c c  | c c  }
\hline
Element & $\log \epsilon_{\mathrm{here}}$ & $\sigma_{\mathrm{here}}$  & $\log \epsilon_{\mathrm{G07}}$ & $\sigma_{\mathrm{G07}}$  \\
\hline 
%Final abundances for the Sun
% Cambridge, P Jofre Mon Jun 22 12:47:09 2015
% element logeps sigma  logepsG sigmaG
Mg &  7.65 &  0.08&  7.53 &  0.09 \\
Si &  7.49 &  0.08&  7.51 &  0.04 \\
Ca &  6.32 &  0.09&  6.31 &  0.04 \\
Ti &  4.90 &  0.07&  4.90 &  0.06 \\
\hline 
Sc &  3.22 &  0.14&  3.17 &  0.10 \\
V &  3.93 &  0.04&  4.00 &  0.04 \\
Cr &  5.58 &  0.06&  5.64 &  0.10 \\
Mn &  5.30 &  0.09&  5.39 &  0.04 \\
Co &  4.89 &  0.09&  4.92 &  0.08 \\
Ni &  6.18 &  0.10&  6.23 &  0.04 \\

\hline
 \end{tabular}
 \end{center}
 \end{table}

 \section{Determination of Elemental Abundances}\label{abundances}

%The goal of this work is to determine in a consistent way as much  accurate abundances as possible for all stars.  This is not possible to be done in an automatic fashion because the stars and elements are very different from each other. %While we have more that 60 Ti, we have only 1 line for Ru for example. In addition, many of the lines are not detectable in metal-poor stars and many are blended or saturated in cool giants.  Thus, significant individual inspection had to be carried out to define the final sample of lines that provide an accurate and reliable abundance. 

Following Paper III,  we firstly selected only lines with $-6.0 < \log ({\mathrm{EW}/\lambda}) < -4.8$ (which helps to avoid very weak lines or saturated lines) and grouped the stars into {\it metal-poor}, {\it FG-dwarfs}\footnote{The subgiants (cf. Paper~I) are included in the group of \fgdwarf\ in this work.}, {\it FGK giants}, {\it M giants} and {\it K dwarfs}.  Furthermore, to avoid effects due to normalisation or bad employment of atomic data, we performed a differential abundance analysis. For that we chose one reference star in each of the groups, being HD22879, the Sun, Arcturus, \alfTau\ and \cygA, respectively. We looked for the lines in the allowed EW range analysed by each method for the reference star and then looked for common lines for that method in the rest of the stars in that group. Like this we could have differential abundances for each individual method, which we then could combine with a much lower dispersion at a line-by-line and method-by-method basis than using absolute abundances.  The advantage of using differential abundances for Milky Way studies with elemental abundances has been discussed in e.g. \cite{2007AA...468..679S, 2011ApJ...743..135R} and \cite{2013NewAR..57...80F}.

One example of differential abundance results on line-by-line and method-by-method basis is shown in \fig{fig:nodes-abu}.  In the figure, we plotted the results of individual line abundances of Co as function of wavelength for all methods in different colours and symbols. For better visualisation of the symbol definition in the figure, the legend is split in the two panels.   The top panels illustrate an example of an {\it FG-dwarf} star, \alfCenA, which has as reference star the Sun. The bottom panels illustrate an example of a {\it FGK giant} star, HD220009, which has Arcturus as reference star.  The left panels show the absolute abundances minus the mean of all abundances,  while the right panels show the relative abundance with respect to the reference minus its mean.   The standard deviation of this mean is indicated in the top right side of each panel.  We plot these relative abundances with respect to the mean only for illustration purposes, aiming at keeping the same scale in both cases. This  allows us to focus on the dispersion of each case. Note that the scatter of different methods for individual lines is considerably decreased from absolute to relative abundances. This mostly reflects on the removal of method-to-method systematic errors such as the approaches to normalise the data. In addition,  note that some absolute abundances agree well between methods, but deviate significantly from the mean. One example is the reddest Co line, which for both stars yields absolute abundances higher than the mean. This would suggest  a revision of the atomic data, in particular of  $\log gf$. When using differential abundances, one can see that this line yields abundances that agree better with the mean.

Since our abundances are relative to a reference star in each group, we needed to determine the abundances of those reference stars separately. This is also done in a differential way with respect to the Sun. This implies that not all available lines were used but only those for which reliable EWs could be measured in the spectra of both the reference star and the Sun.  An extensive discussion of this strategy can be found in the following section.

A scheme of the differential analysis employed in this work is shown in \fig{fig:scheme}. The zero point is the Sun, for which we determine absolute abundances. The lines for the Sun were carefully inspected as well as its atomic data as discussed in \cite{1402-4896-90-5-054010}.  The first group of {\it FG-dwarfs} is analysed with respect to the Sun by differentiating the abundances obtained by each method for each line.  This group contains the stars indicated at the top right of the scheme.  The other reference stars, which are in red colour { in the large boxes}  in the figure, are also analysed by differentiating between common lines with respect to the Sun in the same way as all the stars from the {\it FG-dwarfs} group.   Finally, the rest of the stars are analysed with respect to the reference stars of each of the groups.  In summary, all groups of stars except the {\it FG-dwarfs} are analysed in a two-step approach, differentiating the stars with respect to a representative reference star, which is then analysed with respect to the Sun.

   \begin{figure}
  \resizebox{\hsize}{!}{\includegraphics{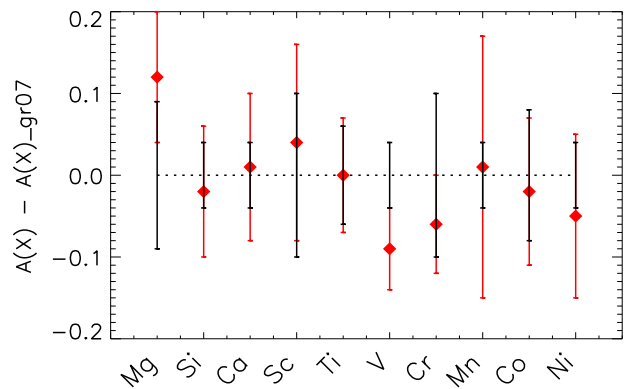}}
  \caption{The difference between abundances of the Sun obtained by  us and by \cite{2007SSRv..130..105G} is displayed with a red diamond. The red error bars correspond to our line-to-line scatter while the black error bars are the uncertainties listed in Table 1 of  \cite{2007SSRv..130..105G}.}
  \label{fig:sun}
\end{figure}

\subsection{Analysis of reference stars}\label{ref_gbs}

In the  case of reference stars, the Sun and the rest of the  stars are very different from each other, which meant that common lines were in some cases very few.  This issue is of crucial importance for our final results. In the spirit of homogeneity, having the Sun for zero point and employing a differential analysis in steps for the rest of the stars is the best way to proceed.  In this section we compare this approach with direct determination of absolute abundances for the reference stars. We show that similar mean abundances are derived and that the homogeneous/differential approach, although with significant loss of lines in some cases,  is the best possible one for our purpose.

\subsubsection{Solar abundances}\label{solar_abundances}

 We defined the Sun as our zero point, for which we needed to determine the abundances in an absolute way. For that, we used all lines 
 with $-6.0 < \log ({\mathrm{EW}/\lambda}) < -4.8$  from all methods and defined the final abundance to be the median of all abundances, after 1.5 $\sigma-$clipping of all abundances.   { This rejects in most of the cases less than 10\% of the total measurements}.
 
  The comparison of our abundances including the line-to-line standard deviation and the solar abundance of  \cite{2007SSRv..130..105G} is displayed in \fig{fig:sun}. We compare with the solar abundances of  \cite{2007SSRv..130..105G} because these are the solar abundances employed for the chemical analyses of the Gaia-ESO Survey \citep[see e.g.][]{2014A&A...570A.122S}.  In the figure, we plotted with red diamonds the difference of our results from the values of  \cite{2007SSRv..130..105G}, while in black we plotted only the errors of  \cite{2007SSRv..130..105G}. Note that our values are obtained under LTE, which might cause some of the slight discrepancies seen in the figure.  In any case,  within the errors, our abundances agree well with  those of  \cite{2007SSRv..130..105G} except for vanadium.  Nonetheless, our results for V agree well with \cite{2015AA...577A...9B}.   As extensively discussed in e.g. \cite{2014ApJS..215...20L}, optical lines of $\ion{V}{i}$ are among the weakest ones produced by Fe-peak elements in the Sun (most of them with $ \log ({\mathrm{EW}/\lambda}) < -6$) partly because of the slight under abundance of V with respect to other Fe-peak elements in the Sun.

 { Note that the line-to-line scatter of  some elements is quite large.  Indeed, the odd-Z elements V, Sc, Mn and Co are affected by  hyperfine structure splitting \citep[hfs, for a recent discussion see][]{2015AA...577A...9B}. None of the EW methods considered hfs in the determination of abundances, which could be  translated to greater  abundances from the derived EW in some lines. The synthesis methods ULB, GAUGUIN  and Synspec considered hfs in the line modelling, while iSpec did not.  If only methods that consider hfs are taken into account, the Mn abundance of the Sun is brought down from 5.43 to 5.30.  The latter value has a more significant difference with respect to  \cite{2007SSRv..130..105G}   and subsequent papers on solar abundances such as \cite{2009ARA&A..47..481A} and \cite{2015A&A...573A..26S} of 0.09~dex, but agree well with \cite{2015AA...577A...9B}.   Our values and those of  \cite{2015AA...577A...9B} are performed under LTE considerations, whereas   \cite{2007SSRv..130..105G}, \cite{2009ARA&A..47..481A} and \cite{2015A&A...573A..26S}  consider NLTE.   { An extensive discussion on the effects of hfs is found in \sect{hfs}}.

   \begin{figure*}
  \resizebox{\hsize}{!}{\includegraphics{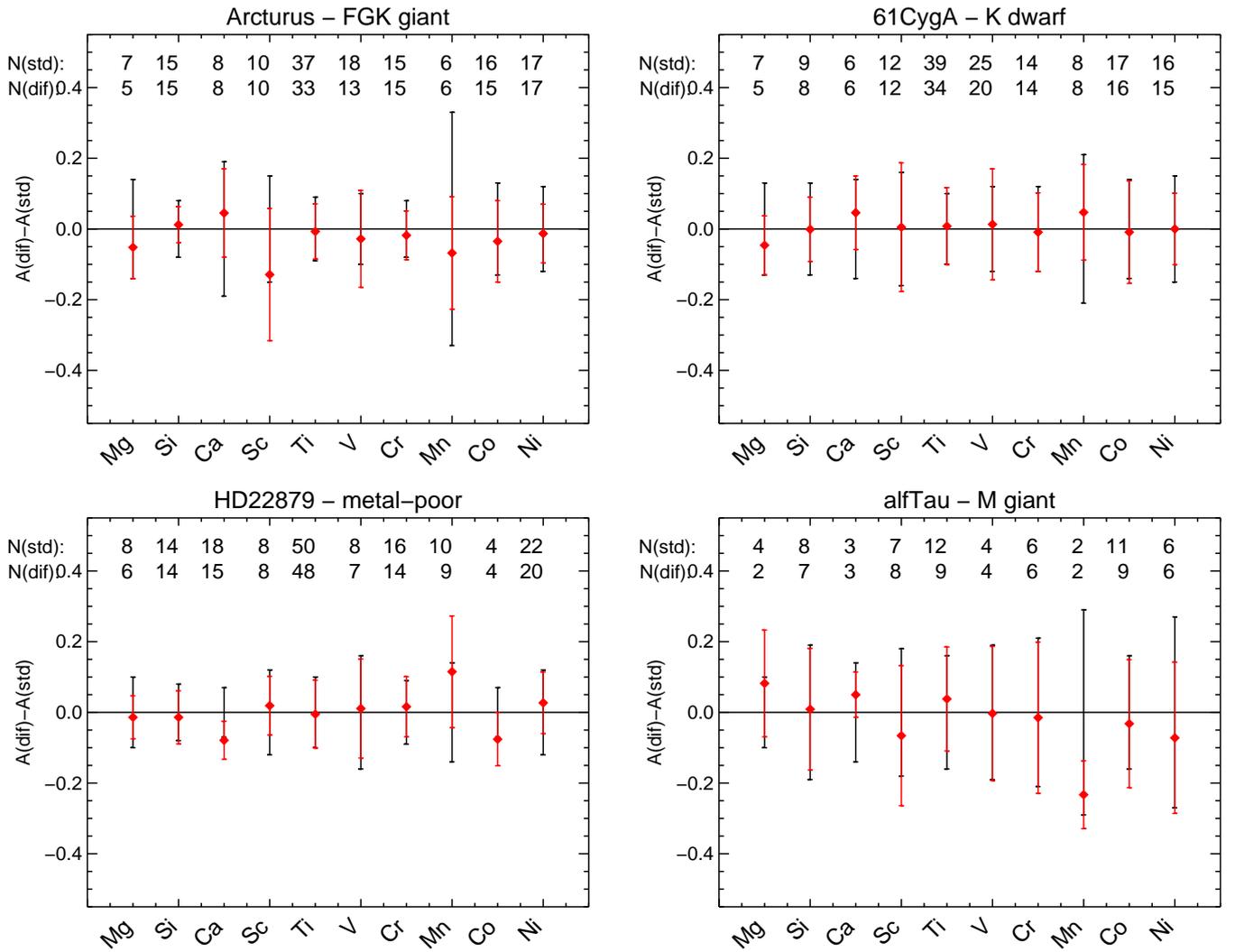}}
  \caption{The line-to-line scatter of abundances of the reference stars  determined directly from the absolute values obtained from all methods is indicated with black error bars (std). Red diamonds represent the difference of these abundances with respect to the final abundances obtained by performing a differential analysis with respect to the Sun (dif), with the error bar representing the line-to-line scatter.  At the top of each panel the number of lines used for the determination of the abundances with both approaches. }
  \label{fig:absrel}
\end{figure*}

  % Mention that the rest of the Co lines have a very well defined continuum in the vicinity, while the Mn do not, being another source of scatter for this element in the Sun. 

  Note also from  \fig{fig:sun}  the large scatter of Sc. From \sect{hfs} we can not attribute it to a hfs effect.  The scatter in this case could rather come from a NLTE effect}, which can produce differences of up to 0.2~dex in the abundances obtained for the Sun from neutral and ionised lines \citep{2008A&A...481..489Z}. We use both ionisation stages to determine the abundances of Sc.    A extensive discussion for each element can be found in \sect{discussion_elements}.
  
  {  The final absolute abundances for the Sun are indicated in \tab{tab:sun}. The horizontal line divides the $\alpha$ elements (top) and the iron-peak elements (bottom).  The measurements of individual lines can be found as part of the online material in the table (SUN) for the Sun.  Note tat the values listed for Sc, V, Mn and Co consider only the results from ULB, GAUGUIN and Synspec, which is diffferent than the values plotted in \fig{fig:sun} for the discussion, which consider the measurements of all methods.}
 
 %{ From \fig{fig:hfs} we conclude that the solar absolute abundances of V, Sc and Co can be determined safely from the combinations of all 8 methods, but the abundances of Mn for the Sun need to be determined from only the ULB, GAUGIN and Synspec methods, who performed a synthesis approach considering hfs. In Table (SUN) all values are provided, but only those mentioned here are used to compute the values indicated in \tab{tab:sun}.}

\subsubsection{Differential vs absolute approach of reference stars}

  \begin{figure*}
 \vspace{-0.3cm}
  \resizebox{\hsize}{!}{\includegraphics{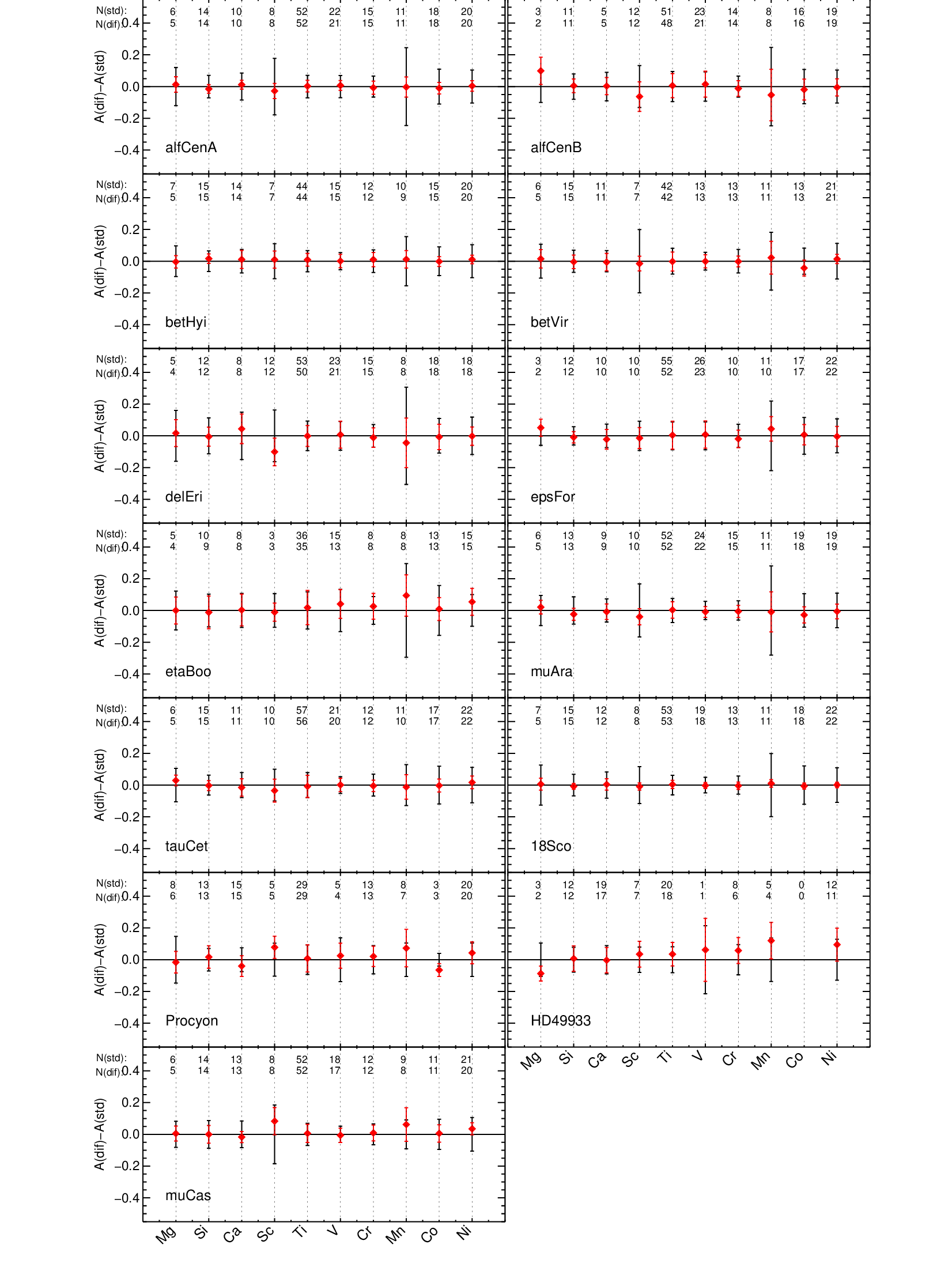}}
 \vspace{-0.5cm}
  \caption{Comparison of the abundances of \fgdwarf\  determined using the standard approach and by performing a differential analysis with respect to the Sun. Same information as \fig{fig:absrel}}
  \label{fig:absrel-fdwarfs}
\end{figure*}

 Employing the differential strategy of reference stars with respect to the Sun meant that a considerable number of lines had to be discarded in some cases.  The few lines left were carefully checked in order to have few, but trustable lines for the differential abundance of the reference stars.  We studied the effects of our differential analysis of stars that have very different spectra such as the reference stars with respect to what would be the ``standard" analysis, namely the determination of abundances considering the direct measurements of all methods.  In this case, we relaxed our line-strength criterion to enhance the number of overlapping lines between the reference stars. We selected the lines with reduced equivalent width of $-6.5 < \log ({\mathrm{EW}/\lambda}) < -4.7$, that is, we allowed for slightly weaker and slightly stronger lines than for the differential analysis of one group of stars. For Mn we even allowed for stronger lines ( $\log ({\mathrm{EW}/\lambda}) < -4.6$) to have more lines to analyse. The standard abundances were calculated using 1.5$\sigma$ clipping of all measurements of all methods, in the same fashion than for the Sun. 
 
 The comparison of abundances for both approaches (standard v/s differential) is displayed in \fig{fig:absrel}, for all reference stars and elements. To obtain the absolute abundances with the differential approach, we added to the final differential abundances the results obtained for the Sun listed in \tab{tab:sun}.  Around the zero line the error bars in black represent the standard deviation of the line-by-line scatter of the standard measurements of all methods (std in the figure). The difference between the standard and the differential (dif in the figure) final abundances is indicated with red diamonds in the figure, with the error bar corresponding to the line-by-line scatter. Each element is indicated in the bottom of the panel, and each panel represents one reference star, with its name and its group indicated as title. On the top of each panel two sequences with numbers can be seen. They correspond to the number of lines used for the determination of abundances in each approach. The upper sequence indicates the number of lines used for the standard approach while the lower sequence indicates the number of lines used for the differential approach. 
 
 From \fig{fig:absrel} we can see that the line-by-line dispersion is significantly decreased when the differential approach is used for the determination of the final abundances. The differences of the final values in both approaches is also within the errors.  This suggests that the differential approach provides robust final abundances while improving the internal precision due to systematic uncertainties in the methods and the atomic data. 
 
 The number of lines used in the differential approach drops in almost every case,  as expected because the allowed strength of the line needs to be satisfied in both, the Sun and the reference star. For Arcturus, \cygA\ and HD22879 in most of the cases the number of lines loss with the differential analysis is minimal, whereas for  \alfTau\ the lost can be significant. This is not surprising as spectra of  cool giants are extremely different to the solar spectrum.

\subsubsection{Summary}
 
 To summarise, the differential approach in two steps for \fgkgiant, \kdwarf\ and \metalpoor,  namely one differential step with respect to Arcturus, \cygA\ and HD22879,  respectively, and a second one with respect to the Sun, is better than the standard approach of taking all absolute abundances and performing $\sigma$ clipping. This is because without loosing too many lines, we are able to retrieve abundances with the same absolute value yet better precision.
 
  The two-step differential approach for \mgiant\  is less obvious because strong lines for the Sun might be saturated or blended for \alfTau\ due to the difference in effective temperature of about 2000~K and gravity of 3.5~dex. The  lines used for the determination of abundances were carefully inspected for blends and normalisation problems.  These few overlapping lines between the Sun and \alfTau\  are able to give us more accurate abundances without affecting the final absolute abundance significantly.  There is one exceptional case where no overlapping lines were found: V. For V, four clean lines could be used, which are not detectable in the Sun, yielding relatively consistent results between different methods. The Mn line at $\lambda 5004$~\AA\ lacks a good continuum in its vicinity. Thus, the absolute abundance obtained for this line gave very different results for each method, which can be noted from the large error bar, and should  be treated with care.

  It is important to discuss here that the reference star of the \metalpoor\ group was chosen to be the most metal-rich star because it provided a better link between the Sun and the rest of the metal-poor stars. We performed similar differential tests with the star HD140283 ([Fe/H] $\sim -2.5$) and the Sun finding that most of the lines were lost, either because they were too weak in the metal-poor star or saturated in the Sun. Furthermore, no V and Co lines were visible in our spectral range for HD140283. This implies that when using HD140283 as reference, no V and Co abundances could be provided for any star in the \metalpoor\ group.  The few lines left for the rest of the elements (varying normally from 1 to 3) were so weak that only synthesis methods could provide abundances, which were very uncertain, mostly due to different normalisation placements.

To conclude, the absolute abundances of the Sun provide the zero point for all abundances of the reference benchmark stars of all groups, where a reference star was used to differentiate with respect to the Sun, except for the \fgdwarf, for which the reference star was directly the Sun. The abundance of V for \mgiant\ needs a special treatment as no common good lines of V in the reference star and the Sun could be found. In this case, the zero point for V was the abundance of \alfTau\ (see below).

\subsubsection{Vanadium for $\alpha$~Tau}

The lines  $\lambda 5592$, $\lambda 5632$, $\lambda 6002$ and $\lambda 6565$~\AA\ are clean lines (not blended by molecules) in this cool giant which can be used to measure V abundances. %Equivalent widths methods such as Porto and UCM had troubles measuring the EW of these lines but EPINARBO and Bologna were able to determine V abundances that are consistent with the synthesis methods.  We comment that DAOSPEC  \citep{2008PASP..120.1332S}  normalises the continuum to an "effective continuum" level, that is, it takes statistically into account the contamination by small unresolved (and numerous) lines in the region of each line. It was thus expected that DAOSPEC performs well in line-crowded regions, because it refits the continuum in several loops after removing the fitted lines.  The fact that Bologna and EPINARBO are able to obtain results for \alfTau\ is a proof of this expectation. The individual results of each method and line can be found in the online material for the star \alfTau.  
The continuum normalisation is  difficult for this star,  being probably the cause of large discrepancies among the different methods seen in some  extreme cases, which can be even more than 0.5~dex (see line  $\lambda 5592$~\AA\ in table (ALF TAU).  It is impressive to realise that even when using the same atomic data and atmospheric models, as well as the same very high SNR and resolution spectra, different methods can obtain very different abundances for a given line. This particular case is a strong argument of how better is to employ differential approaches because this cancel some of the systematic errors of a given method. Like this, one is able not only to achieve a higher precision of a measured line, but also allows for better comparison of the results with another independent method.

\subsection{Differential vs absolute approach of group stars}
It is instructive to visualise the global effect of the abundances obtained in the standard and the differential way for the stars of the same groups. Figures~\ref{fig:absrel-fdwarfs}, \ref{fig:absrel-redgiants}, \ref{fig:absrel-kwarfs}, \ref{fig:absrel-metalpoor} and \ref{fig:absrel-mgiants} show the comparison of differential versus standard approach  for the groups of \fgdwarf, \fgkgiant, \kdwarf, \metalpoor\ and \mgiant, respectively.  As for the reference stars, in black colour the scatter of the line-to-line and method-to-method for the standard approach is plotted with black error bars while the same scatter but for the differential approach is plotted with red error bars. The red diamond shows the difference in the final value obtained with both approaches. The numbers on the top indicate the number of lines used in the standard and the differential approach.  Each panel shows one star, which is indicated in the bottom left part of it.

    \begin{figure}
  \resizebox{\hsize}{!}{\includegraphics{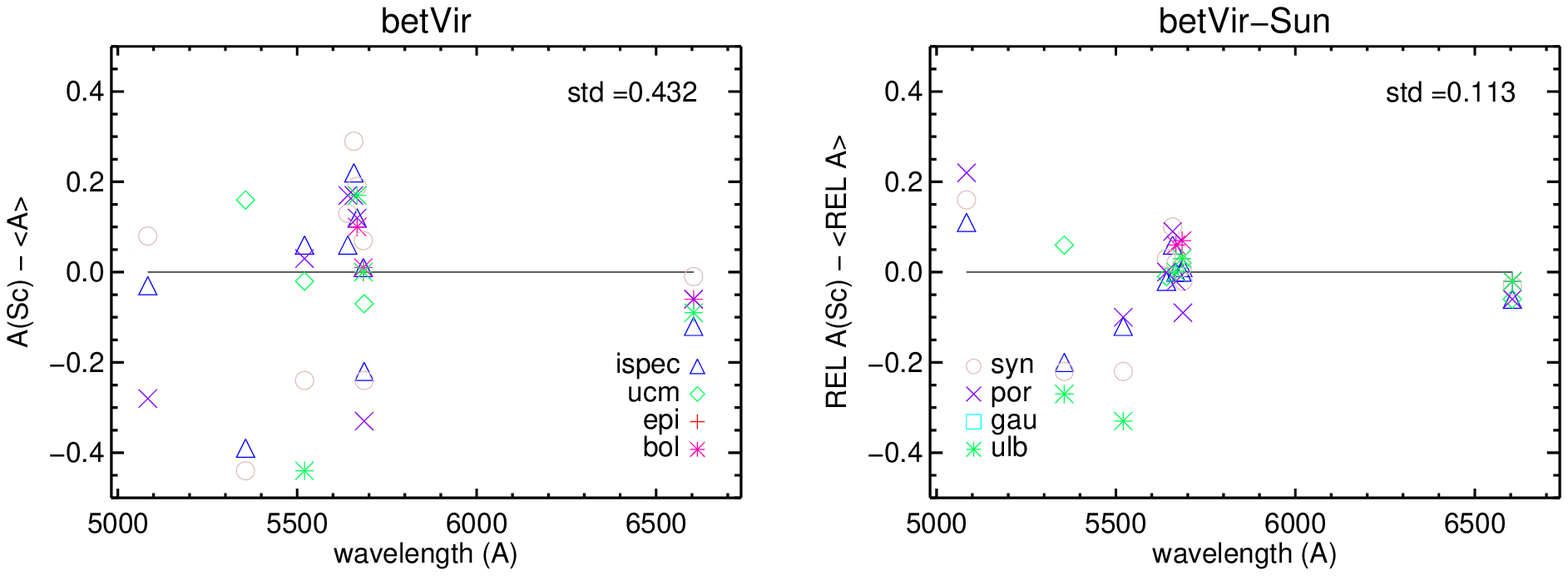}}
  \resizebox{\hsize}{!}{\includegraphics{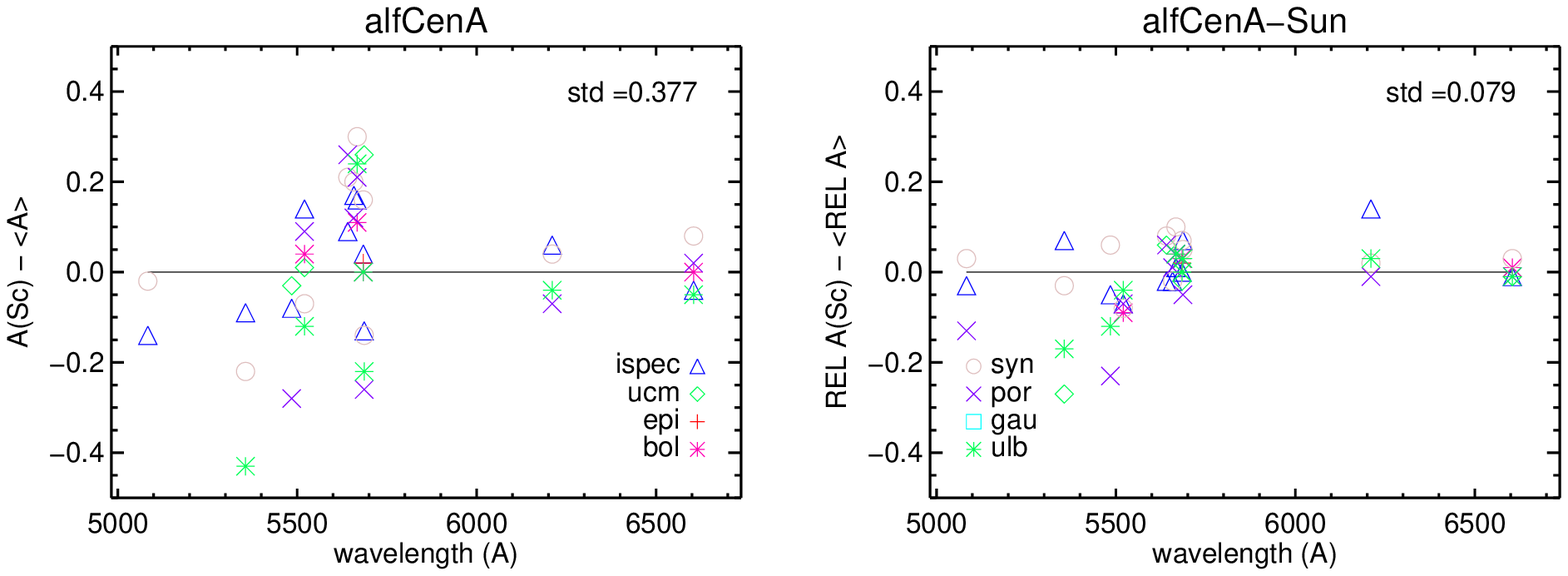}}
  \caption{Line-by-line abundances in the standard approach (left) and differential approach (right) of scandium for two \fgdwarf. The different symbols represent different methods}
  \label{fig:Sc-fgdwarfs}
\end{figure}

 The \fgdwarf\ have in general a large number of lines, which remains the same when differentiating with the Sun in most of the cases (see \fig{fig:absrel-fdwarfs}). It is expected that this group uses a large number of lines as these lines were selected from the analysis of the Gaia-ESO Survey, which contains mostly solar-type stars \citep[e.g.][]{2014A&A...570A.122S}. The lines are clean and numerous, making the standard deviation in general very small, even in the standard approach. The scatter of the differential approach is, however, still considerably decreased to values of the order of 0.01~dex.  Recent works on main-sequence stars perform differential analysis of chemical abundances with respect to the Sun \citep[e.g.][]{2012A&A...543A..29M, 2014AA...562A..71B, 2015A&A...575A..51L, 2015AA...577A...9B}, which allows to do a very precise analysis of relative differences in Galactic stellar populations. Note that the dispersion of Mn is systematically higher than the rest of the elements.  { As mentioned in \sect{solar_abundances}, the lines are affected by hfs}, adding a source of error in the EW measurement and line modelling.   Indeed, several Mn lines are very strong, partly due to effects of hfs, so we allowed for stronger lines as in the reference stars ( $\log ({\mathrm{EW}/\lambda}) < -4.6$)  to have more lines to analyse. { We can see here that, like for the reference stars, the effects of hfs in the line-to-line scatter is significantly cancelled when differential abundances are determined. Further discussions on this regards is found in \sect{hfs}}.

   \begin{figure}
  \resizebox{\hsize}{!}{\includegraphics{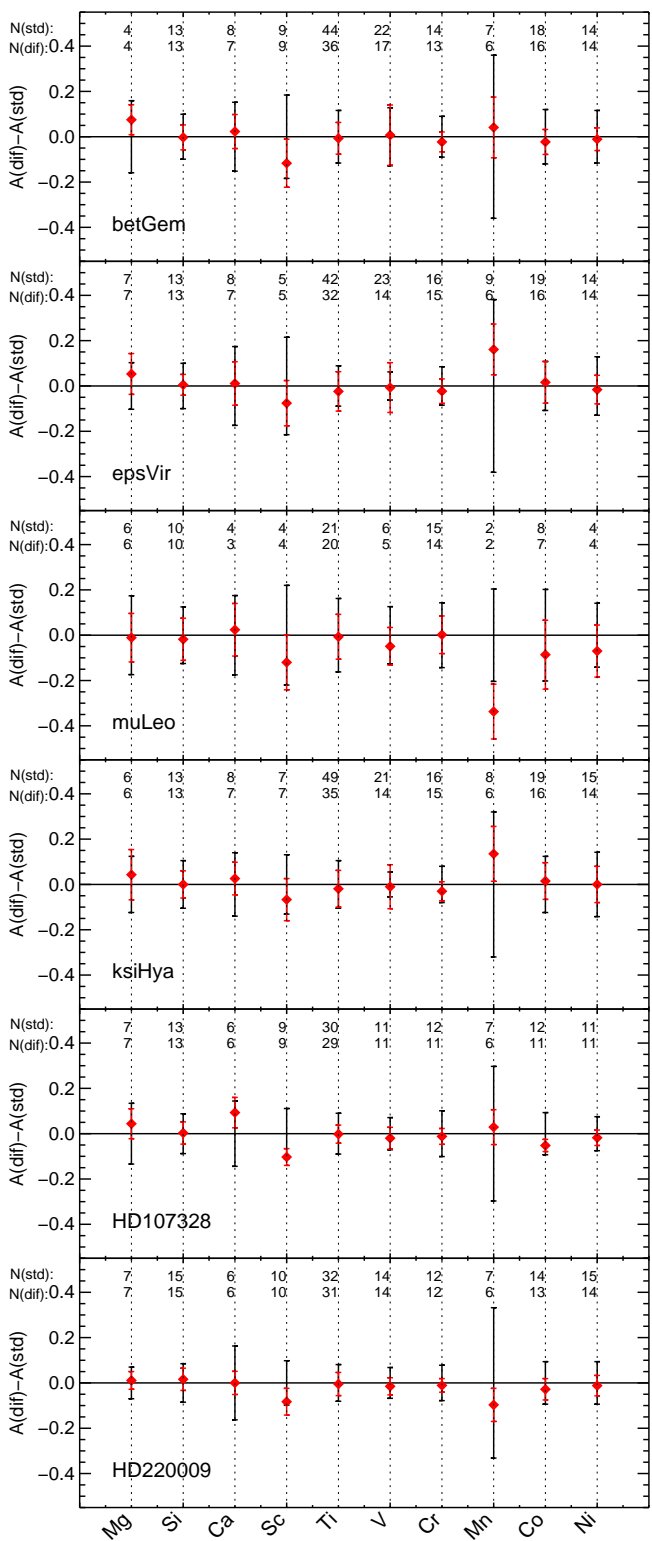}}
  \caption{Comparison of the abundances of \fgkgiant\  determined using the standard approach and by performing a differential analysis with respect to Arcturus. Same information as \fig{fig:absrel}}
  \label{fig:absrel-redgiants}
\end{figure}

  The case of  Sc is worth commenting as it also has a systematic high scatter in the standard approach.  The differential analysis however yields a scatter that compares to the rest of the elements. The Sc lines are clean for these kind of stars, and abundances agree between methods for a given line quite well. Two examples are shown in \fig{fig:Sc-fgdwarfs}, which is similar to \fig{fig:nodes-abu}.  We show abundances of Sc at line-by-line and method-by-method of the stars \alfCenA\ and \betVir\, which presented significant larger scatter from the standard analysis with respect to the rest of the elements in \fig{fig:absrel-fdwarfs}.  One can see in the left panels of \fig{fig:Sc-fgdwarfs} that the methods obtain more or less consistent results for the same lines around $\lambda 5400$~\AA, but they are very different for the different lines.  The Sc abundances derived from the neutral line $\lambda 5356$~\AA, for example, are systematically lower for all methods, suggesting either that the atomic data of this line could be revisited, or that the NLTE effects of this line are rather strong. NLTE corrections enhances the abundances of Sc of neutral lines in the Sun \citep{2008A&A...481..489Z}.  We recall that we could see in \fig{fig:sun} that for the Sun we obtained a large line-by-line dispersion.  If the dispersion is caused by NLTE effects, then a differential analysis would remove part of this effect, at least for the stars that are being differentiated with respect to the Sun. 
  
 It is instructive to discuss the case of 18~Sco, a classical solar twin. The same lines are used for the differential and standard approach except for Mg  and V. The Mg line at $\lambda 6319$~\AA\ and the V line at $\lambda 6296$~\AA\ had a gap in the solar atlas due to a blend from a telluric feature.  The line-to-line scatter is greatly decreased in the differential approach, which is expected because the spectra of these twins are almost identical.  For this reason, chemical analyses of solar twins are commonly performed differentially \citep[e.g.][ and references therein]{2014ApJ...791...14M, 2012A&A...543A..29M, 2015arXiv150407598N}. This allows to detect at great accuracy slight differences in their chemical pattern that otherwise would be undetectable.  \\

\noindent  The \fgkgiant\ (\fig{fig:absrel-redgiants}) also have a relatively large number of lines analysed, although slightly less than the \fgdwarf.  In our reduced EW cut, several lines are rejected because they saturate in giants.  The final value using both approaches remains unchanged within the errors, and the scatter systematically decreases.  Only Mn for \muLeo\ has a different final abundance in the standard and the differential approach. We inspected the two lines used ($\lambda 5004$ and $\lambda 5117$~\AA), both having a continuum difficult to identify in their vicinity for both stars, \muLeo\ and Arcturus. The abundances of Mn for \muLeo\ which rely only on those two lines, should be treated with care.  As for the \fgdwarf, number of lines remains very similar between the standard and the differential approach, meaning that Arcturus is a good reference star for the \fgkgiant\ group. \\
 
   \begin{figure}
  \resizebox{\hsize}{!}{\includegraphics{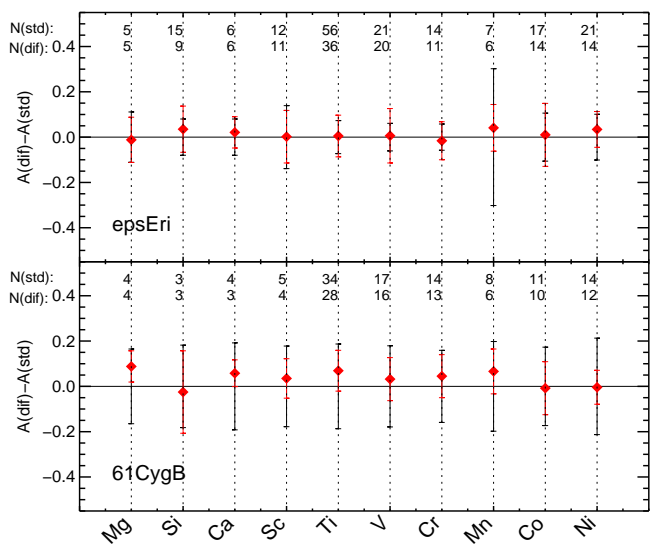}}
  \caption{Comparison of the abundances of \kdwarf\  determined using the standard approach and by performing a differential analysis with respect to \cygA. Same information as \fig{fig:absrel}}
  \label{fig:absrel-kwarfs}
\end{figure}

   \begin{figure}
  \resizebox{\hsize}{!}{\includegraphics{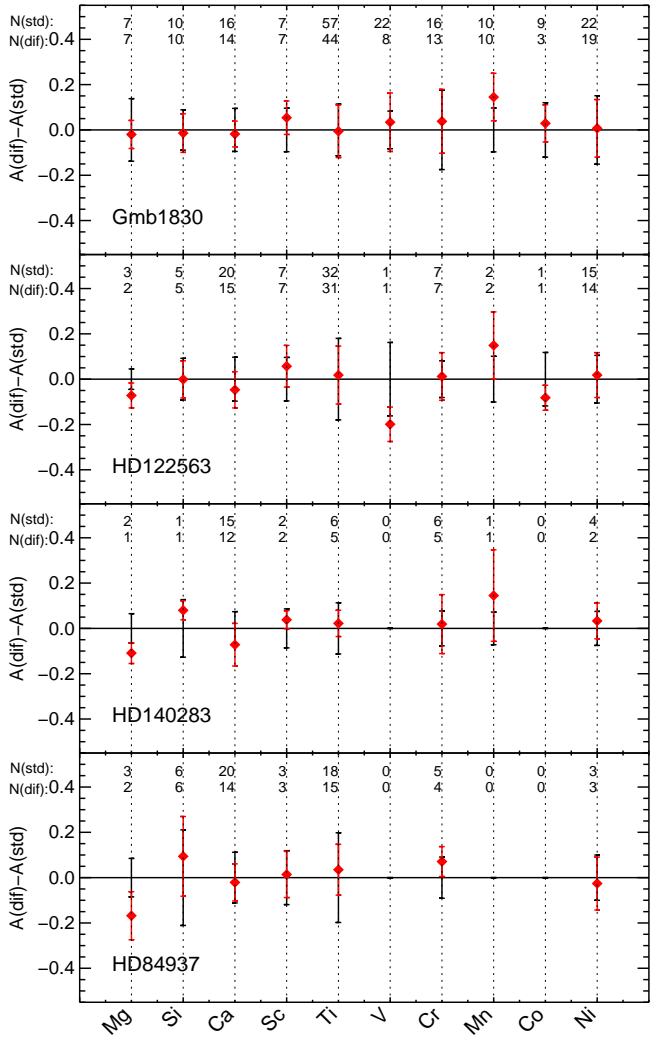}}
  \caption{Comparison of the abundances of \metalpoor\ stars determined using the standard approach and by performing a differential analysis with respect to HD22879. Same information as \fig{fig:absrel}}
  \label{fig:absrel-metalpoor}
\end{figure}

 \noindent Regarding the \kdwarf\ (see \fig{fig:absrel-kwarfs}), while a large number of lines was used for the determination of elemental abundances of \delEri, for \cygB\  fewer lines were used. This star is very cold, meaning that most of the lines are blended with molecules and can not be used. For those fewer selected lines, almost all overlap with \cygA, allowing us to perform a differential analysis in this case improving the precision of our measurements in every case while keeping the final abundance unchanged within the errors.  We point out here that the Si line at $\lambda 6371$~\AA\ was removed from the analysis of \cygB\ because it was contaminated with a telluric line. \\
 
\noindent  Metal-poor stars are more difficult to analyse in a differential approach because they are all very different from each other spanning a metallicity range of 1.5~dex or more. Moreover, \teff\ and \logg\ of the \metalpoor\ group also cover a wide range.  One can see in \fig{fig:absrel-metalpoor} that the number of lines decreases in every case with respect to the \fgdwarf, which is expected because of the low metallicities. The case of Gmb~1830 still preserves most of the lines after performing differential analysis, with the finally abundances practically unchanged.  Interestingly, HD122563 also preserves most of the lines after differentiating, with results notably better in every case. This case is an example of the degeneracies of stellar parameters: a very metal-poor giant has most of its lines of the same size than a more metal-rich dwarf.   We are able to measure V for HD122563 from one line, whose final value differs by $\sim$0.2~dex when using standard or differential approach. This line, however, should be treated with care as it is on the wing of H$\beta$, making the continuum more difficult to set. 
 
  Regarding HD140283, as previously discussed, very few lines are detected (Mg, Si, Mn), or no line at all (V and Co).  It is unfortunate that two initially selected silicon lines ($\lambda 5701$ and $\lambda 5948$~\AA) had to be removed because they were blended by telluric features in every spectrum of our library.  The right wing of the $\lambda 5948$~\AA\ line can still be used for synthesis methods, yielding an abundance that is consistent with the only weak but clean line at $\lambda 5708$~\AA\ we have left.   The abundance of Mn is less precise when the differential approach is employed. Although the final value agrees within the errors, it is worth commenting that the only line used ($\lambda 4823$~\AA) has a slight blend in the left wing in the spectrum of HD22879, the reference star. The Mn abundance obtained from this line for the reference star varies between EW and synthesis method by 0.1~dex { probably due to hfs (see \sect{hfs})}.  Finally, HD84937 has few lines in general, but they are not lost after performing differential analysis. Some elements can not be measured in the standard approach (V, Mn, Co), and thus can not be measured in the differential approach either.\\

\noindent  The last group of \mgiant\ is the most difficult one. These cool giants have few clean and unsaturated lines in general, especially from our initial selection of lines which was done based on the Gaia-ESO data, which contains very few of such cool giants.  Furthermore, detecting the continuum is very challenging, as well as fitting the right profile to the lines. That makes a large dispersion of measured abundances  line-to-line and method-to-method in general, especially for the extreme cool star \psiPhe.   From \fig{fig:absrel-mgiants} we can see that the error bars are significantly larger than those of the other groups. However, the differential analysis yields better results in terms of precision in most of the cases, in particular for \alfCet\ and \gamSge. There are several cases  that lines did not provide trustable abundances because the spectrum in that region was too crowded with molecules. In these cases  the lines were rejected by hand which meant to not have abundances for some of the elements. These cases have zeroes at the top sequences of \fig{fig:absrel-mgiants}.

 \begin{figure}
  \resizebox{\hsize}{!}{\includegraphics{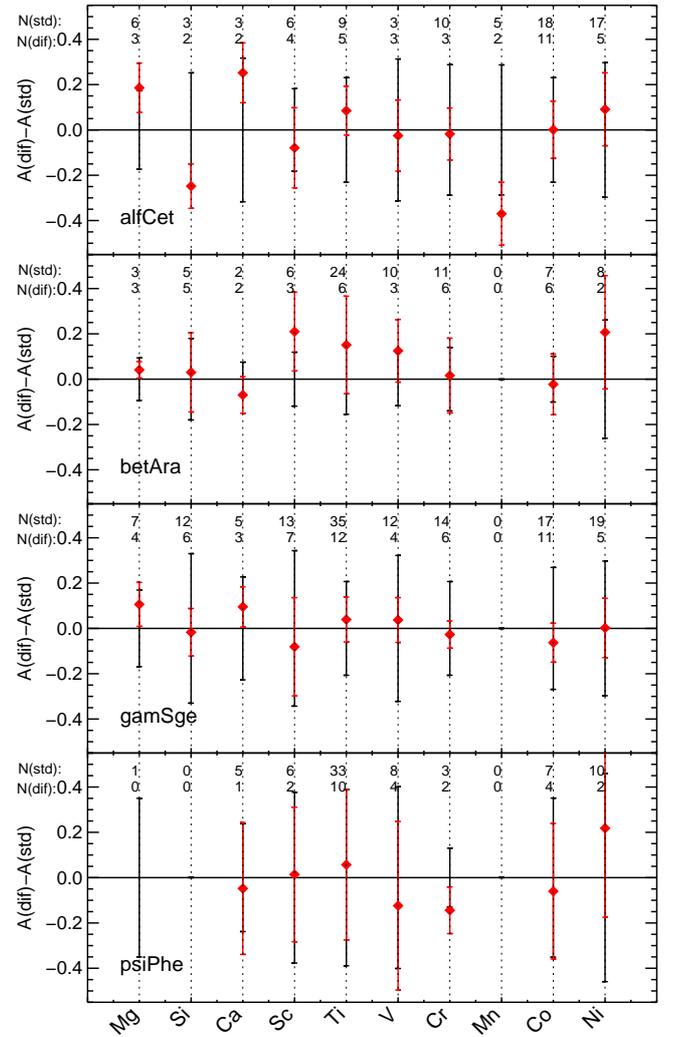}}
  \caption{Comparison of the abundances of \mgiant\  determined using the standard approach and by performing a differential analysis with respect to \alfTau. Same information as \fig{fig:absrel}}
  \label{fig:absrel-mgiants}
\end{figure}

\subsection{Hyperfine Structure splitting}\label{hfs}

{ The odd-Z elements Sc, V, Mn and Co analysed in this work are affected by hyperfine structure splitting.  Since we have methods that considered hfs and some that did not, we performed an analysis to quantify the effect of hfs in the measured abundances.   In  \fig{fig:hfs}  we plot in each panel the absolute abundances of the Sun for the four odd-Z elements for each line and method, as a function of wavelength. The open black triangles correspond to the abundances derived from the EW and iSpec methods (i.e. no hfs), while the filled red circles represent the abundances obtained by the methods considering hfs.  The dotted black line represents the standard deviation of all measurements, while the red dashed line represents the standard deviation of the methods that consider hfs only. 
  %How are the results when only ULB, gAU and syn are used? 
     \begin{figure}
  \resizebox{\hsize}{!}{\includegraphics{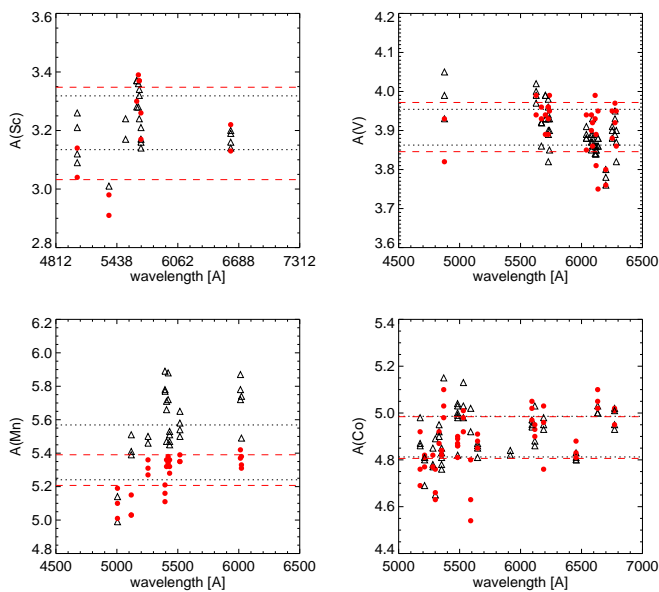}}
  \caption{Abundances of the Sun for the odd-Z elements determined by the different methods as a function of wavelength. In red filled circles we plot the methods that consider hfs in the determination of abundances. In black open triangles we plot the methods that neglect hfs. }
  \label{fig:hfs}
\end{figure}

  One can see in the figure that V, Sc, and Co, although affected by hfs, a systematic effect in the final abundance is not significant, where the averaged absolute value and the scatter at a line-by-line and method-by-method basis remain essentially the same.  AThis is not surprising since the line-profiles in the Sun for those elements are symmetric and can be well represented with a Gaussian profile.  few individual line abundances of V and Co might be affected - those where the non-hfs abundances are systematically higher than the hfs ones: for V only one line at $\lambda$4875\AA, for Co the four lines around $\lambda$5500\AA:  $\lambda$5483, $\lambda$5530 and $\lambda$5590~\AA.   The case of Mn, however, shows a strong effect due to hfs, explaining the large scatter seen in \fig{fig:sun}. On the contrary to V, Sc and Co, several strong Mn lines presented a pronounced boxy-shape, in particular, $\lambda$5407, $\lambda$5420 and $\lambda$5516~\AA. The line  profile at  $\lambda$5420~\AA\ can be seen in Fig. 1 of  \cite{2015A&A...573A..26S}.  This analysis suggests us that only Mn hfs should be taken into account, which in our analysis it means that only the abundances of Mn obtained from ULB, GAUGUIN and Synspec should be taken for the Sun.  For the rest of the odd-Z elements, all methods can be used without affecting the line-to-line precision and absolute value of the solar abundance.

     \begin{figure}
  \resizebox{\hsize}{!}{\includegraphics{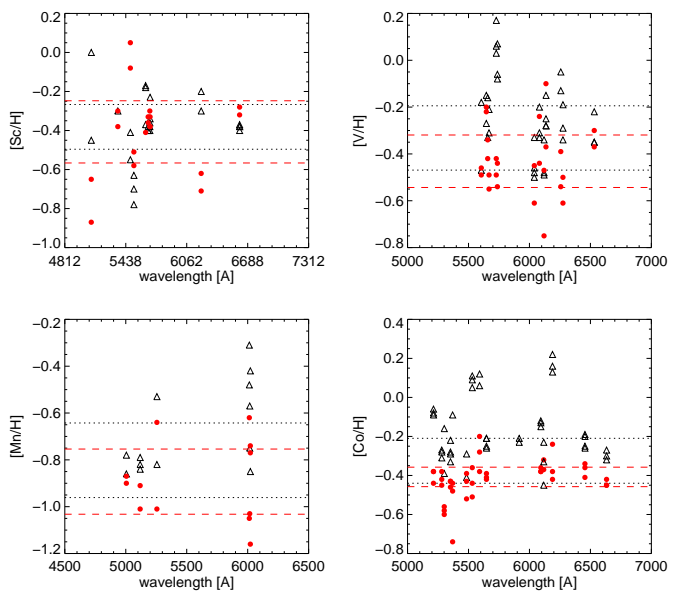}}
  \caption{Similar as \fig{fig:hfs} but for Arcturus respect to the Sun. }
  \label{fig:hfs_arcturus}
\end{figure}
  
  Next, we investigated if the differential approach cancels the effect of hfs, such that all measurements can be employed, regardless of the consideration of hfs. For that we did a similar analysis to that  shown in \fig{fig:hfs} but comparing the differential abundances obtained by the methods. We started with the reference benchmarks, which are very different from each other. In \fig{fig:hfs_arcturus} we show the case of Arcturus  with respect to the Sun.   In contrast to the Sun, we can see that the effects of hfs are important for this giant for V and Co. The lines are stronger than in the Sun, making the effect of hfs more pronounced. Since we see that hfs affects differently  the different kind of stars, when performing a differential analysis respect to stars that are different from each other, the effect of hfs can not be cancelled. Similarly, we could see that hfs affects significantly the abundances of Co and V. For Mn, it is  more difficult to say due to the few lines used. The metal-poor reference star HD22879, presents a slight offset in the abundances with and without hfs for Sc and V, but for Mn and Co the offset is not clear, again due to the few lines used. The cool dwarf \cygA\ shows significant offset for vanadium only.  This analysis suggests that although hfs is not prominent for the determination of Sc, V and Co for the Sun, it affects significantly stars different of the Sun. Thus, we confirm that to achieve more reliable results in an homogeneous manner, only methods employing hfs should be considered for all the reference stars, including the Sun.

     \begin{figure}
  \resizebox{\hsize}{!}{\includegraphics{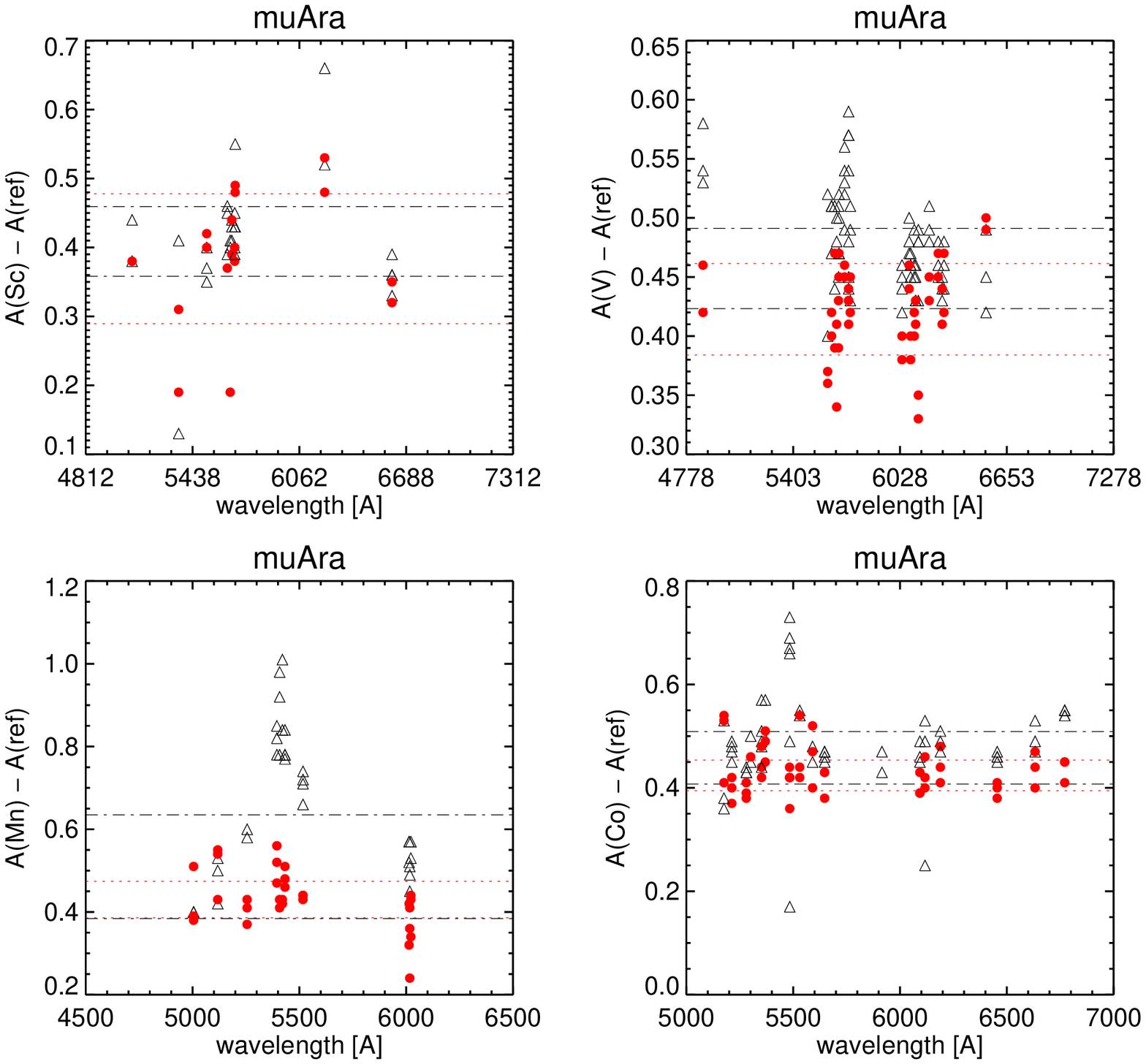}}
   \resizebox{\hsize}{!}{\includegraphics{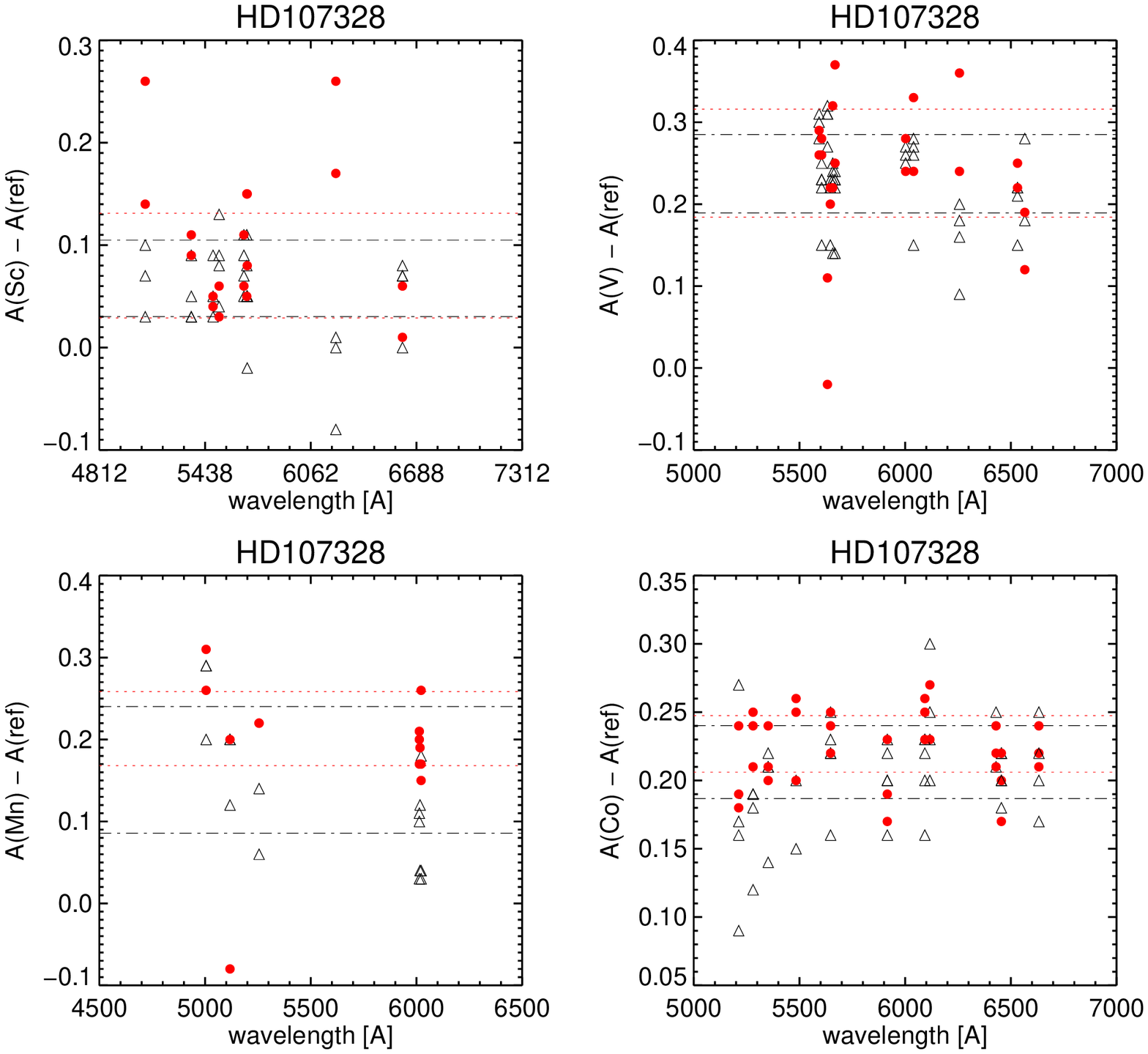}}

  \caption{Similar as \fig{fig:hfs} but for a \fgdwarf\ star (\muAra) respect to the Sun and for a \fgkgiant\ (HD107328) respect to Arcturus. The name of the star is indicated in the title of each panel. }
  \label{fig:hfs_group}
\end{figure}

  Finally, we investigated the effect of hfs in the differential analysis when stars are similar to each other. That is, for stars within their group.  Examples are shown in  \fig{fig:hfs_group} for \muAra, a star from the \fgdwarf\ group, and for HD107328, a star from the \fgkgiant\ group.  We chose these examples as one case presenting systematic differences when considering hfs (\muAra) and one case not presenting significant differences (HD107283). In the first case, the dwarf star has similar temperature and surface gravity as the Sun, but it is considerably more metal-rich than the Sun, so its lines are much stronger than the Sun. Even in the differential approach the effects of hfs in this case are not totally cancelled, where we see that V, Mn and Co show notable overabundances for methods that neglect hfs. On the other hand, HD107283 has a more  similar metallicity than Arcturus, but a slightly higher surface gravity and higher temperature. There is no evidence of any odd-Z element having particularly different abundances when methods consider hfs or not. We know, from \fig{fig:hfs_arcturus}, that V and Co are strongly affected by hfs. This suggests that in the case of HD107283 analysed differentially respect to Arcturus, the differential procedure cancels the effects of hfs.  
  
  To conclude, since the GBS are slightly different from each other,  the effects of hfs can not be cancelled via differential approach for all stars in a group in the same way. Furthermore, since our aim is to achieve the most reliable abundances in an homogeneous way for all GBS, we restrict the determination of abundances of Sc, V, Mn and Co from only the methods that consider hfs. That is, we neglect the results obtained by iSpec, Bologna, Porto, UCM and Epinarbo for those elements.

}
\subsection{Final abundances}

As a final step,  we visually inspected the profiles of all selected lines individually and removed unreliable abundances of lines with blends or potential bad profiles.  { The final selected  lines for each star can be found as part of the online material. These are 34 tables (one for each star) containing the individual abundances for each line for the 10 elements analysed in this article. The first column indicates atomic number of the element; the second column  the wavelength of the line in \AA; the third column the final abundance for that line obtained with the process described above (the differential abundances for that line were averaged between the methods analysing that line and then the absolute abundance determined for the reference benchmark star was added to the averaged differential abundance); 
%\footnote{We transformed the final relative abundance considering $\mathrm{A(X)} = \Delta \mathrm{A(X)} + \mathrm{A(X)}_{\mathrm{Ref}}$, where $\Delta \mathrm{A(X)} $ refers to the differential abundance with respect to the reference star and $\mathrm{A(X)}_{\mathrm{Ref}}$ corresponds to the final absolute abundance of the reference star.}
 the fourth column lists the NLTE correction obtained for that line (see \sect{nlte}); columns 5-10 show the different measurements of EWs of the Porto, Epinarbo, Bologna, UCM and ULB methods, respectively. The final eight columns correspond to the starting point of this work, namely the direct measurement of abundance for this line by each method.

 After selecting the good lines (those un-saturated and un-blended, in common between the reference and the star under investigation) the final value was determined with averaging all selected lines of all methods.  %weighted average according to the number of nodes that determined abundance for that line.
The final value for each star and element can be found in Tables~\ref{tab:mgh_fin}, \ref{tab:sih_fin}, \ref{tab:cah_fin} and \ref{tab:tih_fin} for Mg, Si, Ca and Ti, respectively, and in Tables ~\ref{tab:sch_fin}, \ref{tab:vh_fin}, \ref{tab:crh_fin}, \ref{tab:mnh_fin}, \ref{tab:coh_fin} and \ref{tab:nih_fin}, for Sc, V, Cr, Mn, Co and Ni, respectively. The first column in the tables indicates the final abundance in [X/H] notation, and the rest of the columns represent different sources of uncertainties, which are explained in detail in \sect{errors}.

More specifically, the final abundance was calculated in the following way:
\begin{equation}
\mathrm{[X/H]} = \langle {\mathrm{A(X)}} - \mathrm{A(X)}_{\mathrm{ref}} \rangle + \langle \mathrm{A(X)}_{\mathrm{ref}} - \mathrm{A(X)}_{\odot} \rangle,
\end{equation}

\noindent where $\langle {\mathrm{A(X)}} - \mathrm{A(X)}_{\mathrm{ref}} \rangle$ is the final  averaged differential abundance of the star with respect to the reference star; $ \langle \mathrm{A(X)}_{\mathrm{ref}} - \mathrm{A(X)}_{\odot} \rangle $ corresponds to the final  averaged differential abundance of the reference star with respect to the Sun. 

It is worth to comment here that the final abundances are under the assumption of LTE. The reason is that NLTE corrections for these 10 elements for the variety of spectral classes, gravities and metallicities of the GBS are not available. We calculated dedicated NLTE departure coefficients for some of the elements for all the stars but these calculations are not possible to do for every atom configuration (see \sect{nlte}). Since our work aims at homogeneity, we provide LTE abundances for all elements and stars as our default abundance, and provide the average NLTE departure of all the lines when available. 
{ Regarding the odd-Z elements, the final values listed in \tab{tab:sch_fin}, \tab{tab:vh_fin}, \tab{tab:mnh_fin} and \tab{tab:coh_fin},  contain only the abundances obtained by the ULB, GAUGUIN and Synspec methods.  However, we decided to keep the unused EWs and abundances in our final tables for each star, to enable further investigations on hfs effects in the future.  }

\subsection{Golden lines}

The atomic data of all the lines employed in this work can be found in the online table (FIN\_LINES), where we list the wavelength, $\log gf$, low excitation potential, the quality flags from the GES line list (see \sect{lines}) and flags representing if the line is a ``golden line'' (see also Paper III) or not. There are five columns with flags for ``golden classification'', one  for each of the groups mentioned above. In the columns two flags are indicated (Y and N).  We defined a line to be golden line if it is analysed in at least 50\% of the stars in the group, in that case the flag corresponds to the letter  ``Y'' (yes,  it is a golden line). If the line was analysed by less than 50\% of the stars in the group, then the line is not a golden line and the flag has the letter ``N'' (no, it is not a golden line). Finally, if the line was not analysed for any of the stars in that group, then there is no flag.

Note that to derive the final abundances, we used all the lines, not only the golden lines.  Nonetheless, it is useful to have a global view of which lines are used  most. For magnesium all lines except one are classified as golden for the \fgkgiant. However, there is no Mg  golden line for all groups.  For the case of silicon, the group of \fgdwarf\ all the lines are classified as golden and one line ($\lambda 5684$~\AA) is a golden line for all groups.  Among the numerous lines analysed for Ca,  golden lines  distribute between  the different groups, which tell us that Ca lines change significantly in different spectra. Interestingly, cool stars (giants and \kdwarf) use much less lines (3--7 lines) compared to metal-poor and \fgdwarf\ (18-19 lines).   The most numerous lines are those of titanium, with the \fgdwarf\ having many golden lines. Metal-poor stars also use several Ti lines, although very few were classified as golden.  Giants and \kdwarf\ generally do not use lines bluer than 5300~\AA, probably they saturate or blend with other lines due to their cooler temperatures with respect to the solar-type stars.   All scandium lines are employed by the \fgdwarf\ group, but about half of the lines were classified as golden. The line $\lambda 5686$~\AA\ is a golden line for the dwarfs and \fgkgiant, but was not used by the other two groups. Almost all of the used Sc lines except one ($\lambda 5641$~\AA) were classified as golden for the \fgkgiant. 

Vanadium is another element with numerous lines employed. For  dwarfs and \fgkgiant\ most of them were used  with several being classified as golden. Metal-poor stars employ very few (8 lines), none of them golden. \mgiant, on the other hand, employ even less lines (4 lines) but all of them are golden. Chromium behaves line Ca, in the sense that golden lines are distributed between the different groups, and very few lines are golden lines. The line $\lambda 5628$~\AA is a golden line for all groups except the metal-poor stars, for which we did not use that line.  No line redder than 5600~\AA\ was employed by the metal-poor stars.  Concerning Mn, most of the lines were used by the \fgdwarf, and were classified as golden. On the other hand, most of the lines were used by the \metalpoor, but only one line ($\lambda 4823$~\AA) was classified as golden. For \mgiant, only two lines were used, none of them golden. All lines used for \fgkgiant\ (6 lines only) are golden.

Regarding cobalt, most of the lines were analysed for the dwarfs, being most of them golden for the \fgdwarf\ and only those redder than 5500~\AA\ for the \kdwarf. Metal-poor stars employ only 3 Co lines, none of them golden. There are two lines that are golden lines for all groups except the metal-poor stars ($\lambda 5647$, $\lambda 5915$~\AA).  Finally, Ni has many golden lines for several groups. For \fgdwarf, all of them were golden except one, for \fgkgiant, all of them were golden, although fewer were used with respect to the \fgdwarf. Metal-poor stars employ numerous lines, but very few are golden. The cool stars use less lines than the previous groups, but several of these lines are golden.  
More details for each element and nature of lines can be found in \sect{discussion_elements}. 
%\subsection{Ionisation imbalance}

\section{Sources of uncertainties of derived abundances}\label{errors}
In this section we discuss the effect on the final abundances of different sources of uncertainties, most of them having a systematic origin, such as the consideration of stellar parameters. We also discuss the effects of NLTE corrections for some of the elements. 
\subsection{Line-by-line scatter}
In general, we have a fair amount of lines and methods providing abundances for each element/star, which allows us to determine a standard deviation around the mean of all those measurements. These values for each element/star are indicated as the column $\sigma (\epsilon)$ of Tables~\ref{tab:mgh_fin} -- \ref{tab:nih_fin}.

\subsection{Systematic errors due to the consideration of \feh\ in LTE}
We defined in Paper~III our metallicity as the value obtained for the iron abundance after NLTE correction. This opens the question on how robust our determination of more elements are when they are determined using material and methods that consider LTE. To assess that question, we made an extra run (see \sect{runs}) on determining abundances but considering \feh\ as the value obtained { in LTE, that is,}  before the NLTE correction (i.e. \feh\ = \feh\ - $\Delta_{\mathrm{LTE}}$ as explained in Paper~III). { In this section we aim to see the effect of NLTE corrections for iron in the resulting elemental abundances. }The resulting abundances were determined as explained in \sect{abundances}, and the error due to LTE is considered as the difference between both results.  These differences are listed as $\Delta$LTE in the final tables~\ref{tab:mgh_fin}, \ref{tab:sih_fin}, \ref{tab:cah_fin} and \ref{tab:tih_fin} for Mg, Si, Ca and Ti, respectively, and in the tables ~\ref{tab:sch_fin}, \ref{tab:vh_fin}, \ref{tab:crh_fin}, \ref{tab:mnh_fin}, \ref{tab:coh_fin} and \ref{tab:nih_fin}, for Sc, V, Cr, Mn, Co and Ni, respectively.  

 The difference in the abundances obtained considering the metallicity after and before NLTE corrections is displayed for all elements and stars in \fig{fig:delta_lte}. The error due to the NLTE corrections in the iron abundances for the determination of these $\alpha$ and iron-peak elements is in most of the cases negligible. This is not surprising  given the normally small corrections of NLTE for metallicity, which are in most of the cases less than 0.1 dex (see Paper III).  The effect of that difference causes different results in individual abundances which are usually  below 0.02 dex, although some cases such as [Si/H] for the cool giant \gamSge\ has a difference of 0.1~dex { in the resulting abundance}. Note that even for this extreme case, this difference is less than the differences in the value for Si abundance due to errors in the stellar parameters of the line-by-line scatter.  
 
Chromium abundances of the cool giants \alfTau\ and \psiPhe\ have the largest differences { of 0.19~dex} when considering { the iron abundances under LTE or after NLTE corrections}. This difference is, however, comparable to the line-by-line scatter of the abundances determined from the 6 selected Cr lines for \alfTau. The line-to-line scatter of [Cr/H] of \psiPhe\ is slightly lower but only two lines were employed in that case.  % It is interesting to see that Cr abundance differences for HD140283 is larger than more metal-rich stars .  On one hand, NLTE corrections of iron are larger for more metal-poor stars \citep[see e.g. ][or Paper III]{2013MNRAS.429..126R}. On the other hand, Cr can be very sensitive to NLTE departures for metal-poor stars (see below). %Other exception are the cases of [Ni/H] for the metal-poor dwarfs HD22879 and HD84937, which have a difference of 0.15 and 0.12~dex respectively. These differences are larger than the differences obtained in [Ni/H] when considering the error in the stellar parameters. The NLTE corrections in the metallicity in this cases are by an order of magnitude larger than the errors in the final metallicity (see \tab{tab:params} and Paper III for more details). Note that only 1 and 2 lines of Ni are used for the final abundance, which might cause part of this apparent large difference.  

  \begin{figure}
  \resizebox{\hsize}{!}{\includegraphics{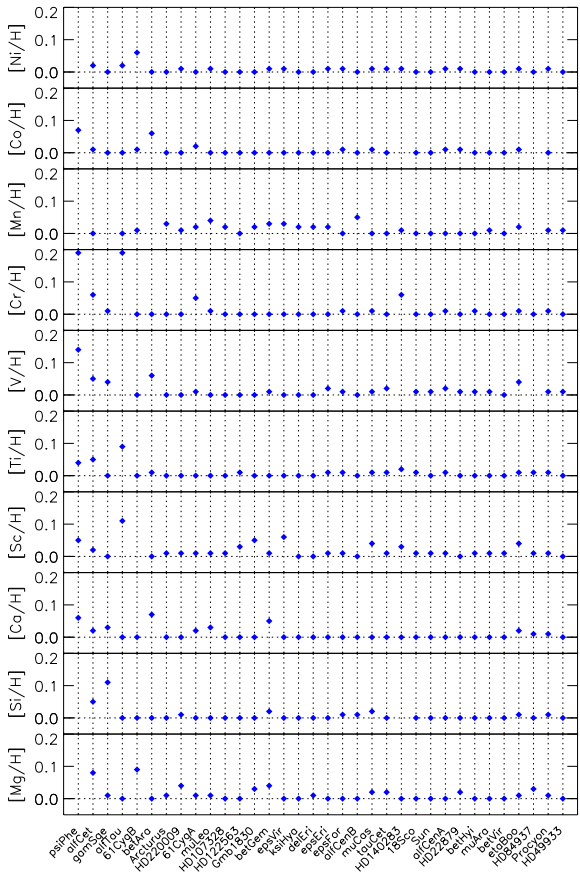}}
\vspace{-1cm}
   \caption{Differences of abundances obtained using the final \feh\ value from Paper~III compared to the \feh\ value before the NLTE corrections of iron. Stars are sorted in temperature.}
  \label{fig:delta_lte}
\end{figure}

\subsection{Errors due to uncertainties in stellar parameters}
We quantified the errors due to uncertainties in stellar parameters in the same way as in Paper~III, namely we determined abundances using the procedure explained above but changing the value of the stellar parameters considering their error (see \sect{runs}). This process was repeated eight times, two for each parameter (\teff, \logg, \feh\ and \vmic) by adding and subtracting  the error from the respective parameter. Then, the error was considered to be the difference between the values obtained from both runs of each parameter.  These values are represented as $\Delta$[Fe/H], $\Delta$\teff, $\Delta \log g$ and $\Delta $\vmic\ for the error in metallicity, temperature, surface gravity and microturbulence velocity, respectivey.  They are indicated in Tables~\ref{tab:mgh_fin}--\ref{tab:nih_fin}.  

The  differences are displayed for all elements and stars in \fig{fig:delta_param}. For this illustration we considered the total error of the stellar parameters, defined by:
\begin{eqnarray}
\Delta = \sqrt{(\Delta_{\mathrm{[Fe/H]}})^2+(\Delta_{\mathrm{Teff}})^2+(\Delta_{\log \mathrm{g}})^2+(\Delta_{v \mathrm{mic}})^2}.
\end{eqnarray}

This difference is plotted in \fig{fig:delta_param} only, it should not be treated as a total error because it would be overestimated.{ This way of treating the total error assume that the different parameters are not correlated to each other, which is not the case. If a covariance between parameters would be taken into account, the error would be smaller. In the final tables we give each error independently, here for the discussion we consider the total one as defined above. } 
 The total differences are usually comparable with the line-to-line scatter although in some cases they can be significant. This reflects on one side the some rather large error
in the stellar parameters  (see \tab{tab:params} and Paper~I and III for details) and on the other hand the dependency of the different elements on stellar parameters. For example, the cool stars  have relatively large differences in most of the determined abundances, but the uncertainties of the stellar parameters are also relatively large. Furthermore,  for the cool stars the differences are in general large for most of the elements, which might be due to the fewer lines employed  for the abundance determination. Solar-like stars, on the other hand, have in most of the cases differences below $\sim$0.05~dex except few cases such as Mn for \alfCenB, Mg for \delEri,  and Ti,  V, and Mn for \epsFor.  The 0.31~km/s error in \vmic\ of \alfCenB\  causes a difference of 0.08~dex in Mn, which is expected since \alfCenB\ is a metal-rich star and its lines are strong, i.e. more dependent on microturbulence \citep[see e.g.][]{1992oasp.book.....G}. The rest of the parameter uncertainties produce negligible differences in the abundance of Mn (see \tab{tab:mnh_fin}). In any case, the line-to-line scatter is larger than this uncertainty. This is probably because of the same reason, given that this star is metal-rich, the Mn abundances are more uncertain for strong lines that have large hfs. For the Mg abundance of \delEri\ the situation is very similar to Mn for \alfCenB, the  large error in \vmic\ causes a larger difference in [Mg/H]. This star is also slightly metal-rich, making the abundance determination of strong lines more sensitive to this parameter.  A similar behavior is seen for some elements for the metal-rich giant \muLeo. Finally, the error of $\sim$80~K  in the temperature of \epsFor\ produces a difference in [X/H] of 0.08, 0.08 and 0.09~dex for Ti, V and Mn, respectively. This agrees with what has been discussed in e.g. \cite{2004AA...425..187T} for this star.

The metal-poor stars have for most of the elements a difference below 0.1~dex  in the determined abundances. Uncertainties of that order are found for Ti, Cr and Ni, suggesting that these elements are particularly sensitive to the stellar parameters in this type of stars.  In the chemical analysis of 14 metal-poor stars of \cite{2011ApJ...742...54} an extensive discussion can be found on how the different elements are affected by stellar parameters. In their Table 7  there is a detailed description of how the abundances can change with errors in stellar parameters. If an error of 150~K is considered in the temperature (the temperature uncertainties of our metal-poor sample is of $\sim$100~K) then the abundances of Ti, Cr and Ni can result in up to $\sim$0.2~dex difference. The rest of the elements have differences of $\sim$0.15~dex for this temperature error. Similarly, if \vmic\ has a difference of 0.3~dex, then abundances of Ni, Cr and Mn can be affected by $\sim$0.15~dex as well.  As in \cite{2011ApJ...742...54}, we find that the uncertainty in surface gravity is less significant for most of the species (see Tables~\ref{tab:mgh_fin} to \ref{tab:nih_fin}).

As commented before, the error due to  \logg\ uncertainties for Arcturus has been calculated considering a value that is half large of what is reported in Paper~I. We expect the errors in the abundances to be twice as large approximately. We obtained uncertainties due to \logg\ error for Arcturus  of the order of 0.01 -- 0.02~dex (see Tables \ref{tab:mgh_fin} -- \ref{tab:nih_fin}), with the error obtained in Paper~I, the uncertainties should become 0.03 -- 0.04~dex, which is still smaller than other uncertainties listed in the tables. 
 
  \begin{figure}
  \resizebox{\hsize}{!}{\includegraphics{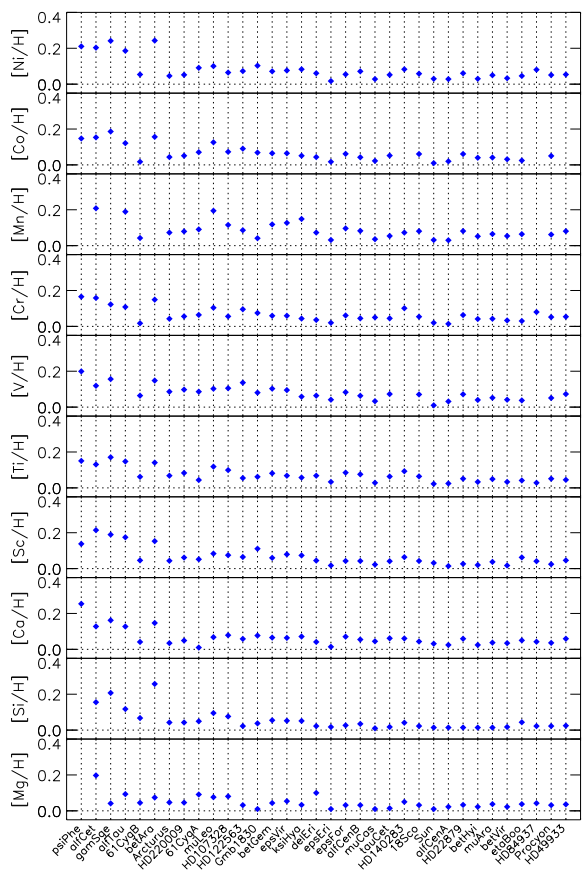}}
\vspace{-1cm}
   \caption{Differences of abundances obtained using the uncertainties of stellar parameters. The individual values shown from \tab{tab:mgh_fin} to \tab{tab:nih_fin} were summed quadratically. }
  \label{fig:delta_param}
\end{figure}

\subsection{NLTE departures}\label{nlte}
NLTE corrections were computed for selected lines of magnesium \citep{2015arXiv150407593O}, silicon \citep{Bergemann2013}, calcium \citep{Lind2013}, chromium \citep{2010AA...522A...9B} and manganese \citep{2008AA...492..823B}. The main uncertainty in these calculations is typically related to the inelastic hydrogen collision rates. In the case of magnesium, these rates have been accurately computed from quantum mechanical methods, while the other studies rely on the classical formula of \citet{Drawin1968}, rescaled by a factor $S_\text{H}$ which is calibrated empirically. Full details are given in the respective papers. Another source of uncertainty is related to the equivalent widths. The corrections were performed considering an average EW of all methods. 

The mean corrections for the elements where we calculated the NLTE departures are displayed in \fig{fig:nlte} for all stars, sorted by temperature. In general, the NLTE corrections are below 0.1~dex, which is comparable with the uncertainties discussed above. Metal-poor stars, however, have  large NLTE departure of Cr of up to 0.3~dex, which is consistent with \cite{2010AA...522A...9B}. Silicon abundances have small NLTE departures for all stars.  Mn has larger departures for the metal-poor stars, but for the rest of the stars, the departures are also normally very small, consistent with  \cite{2008AA...492..823B}.  

  \begin{figure}
  \resizebox{\hsize}{!}{\includegraphics{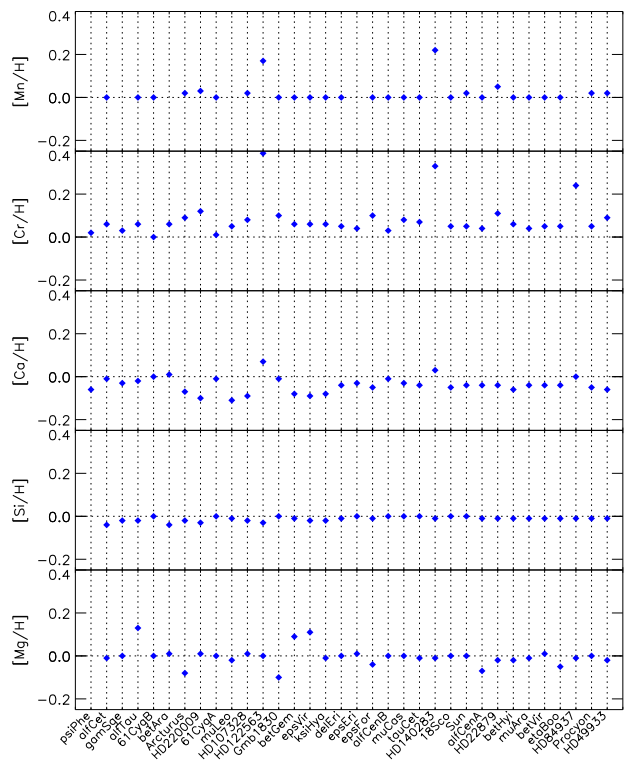}}
   \caption{NLTE departures of Mn, Cr, Ti, Ca, Si and Mg for all stars, sorted by temperature.  }
  \label{fig:nlte}
\end{figure}

  \begin{figure}
  \resizebox{\hsize}{!}{\includegraphics{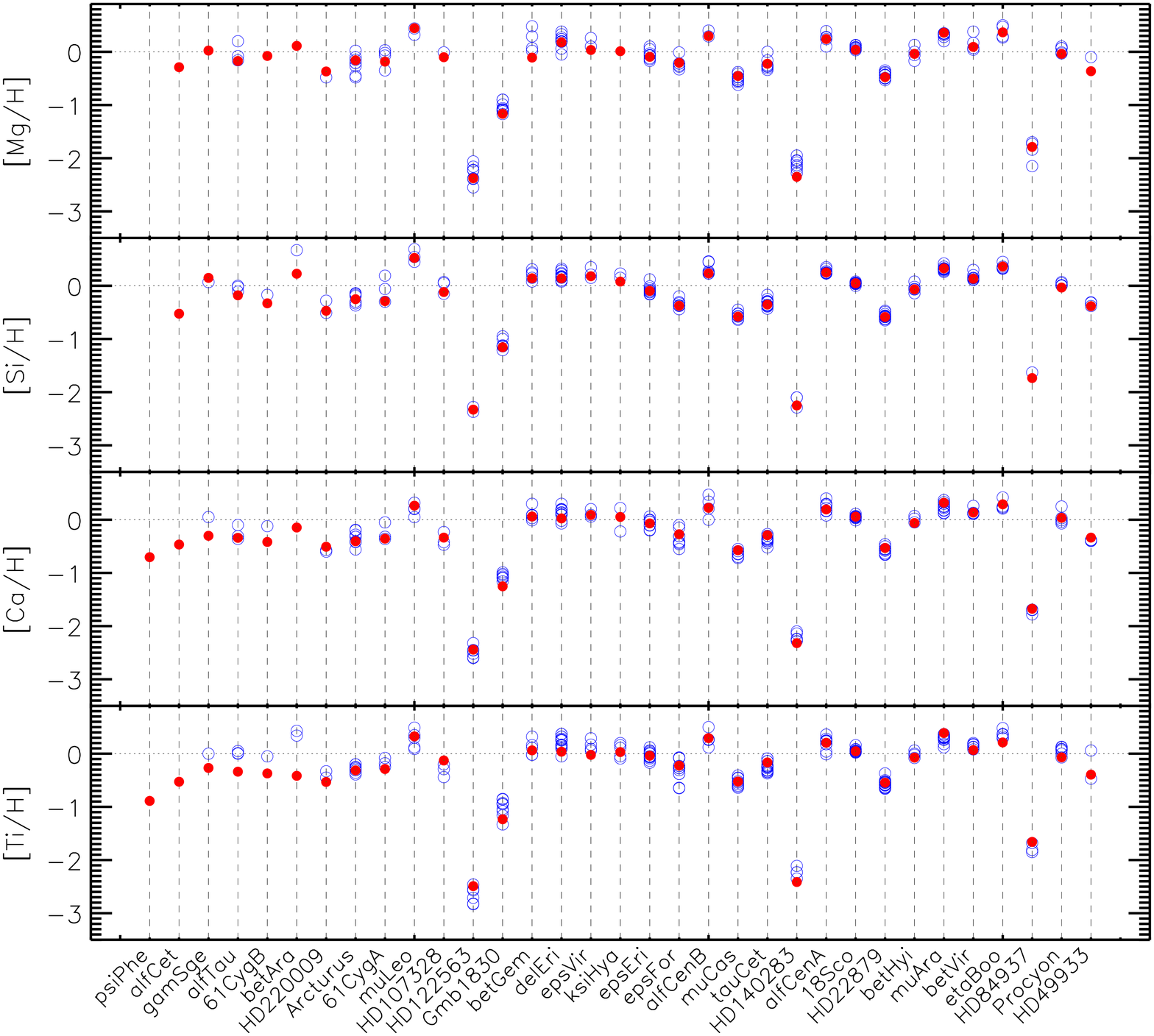}}
   \caption{Abundances of $\alpha-$elements for the GBS sample sorted by effective temperature. Filled red circles represent our final values while open blue circles represent different results from the literature (see \sect{lit}). }
  \label{fig:final_alpha}
\end{figure}

  \begin{figure}
  \resizebox{\hsize}{!}{\includegraphics{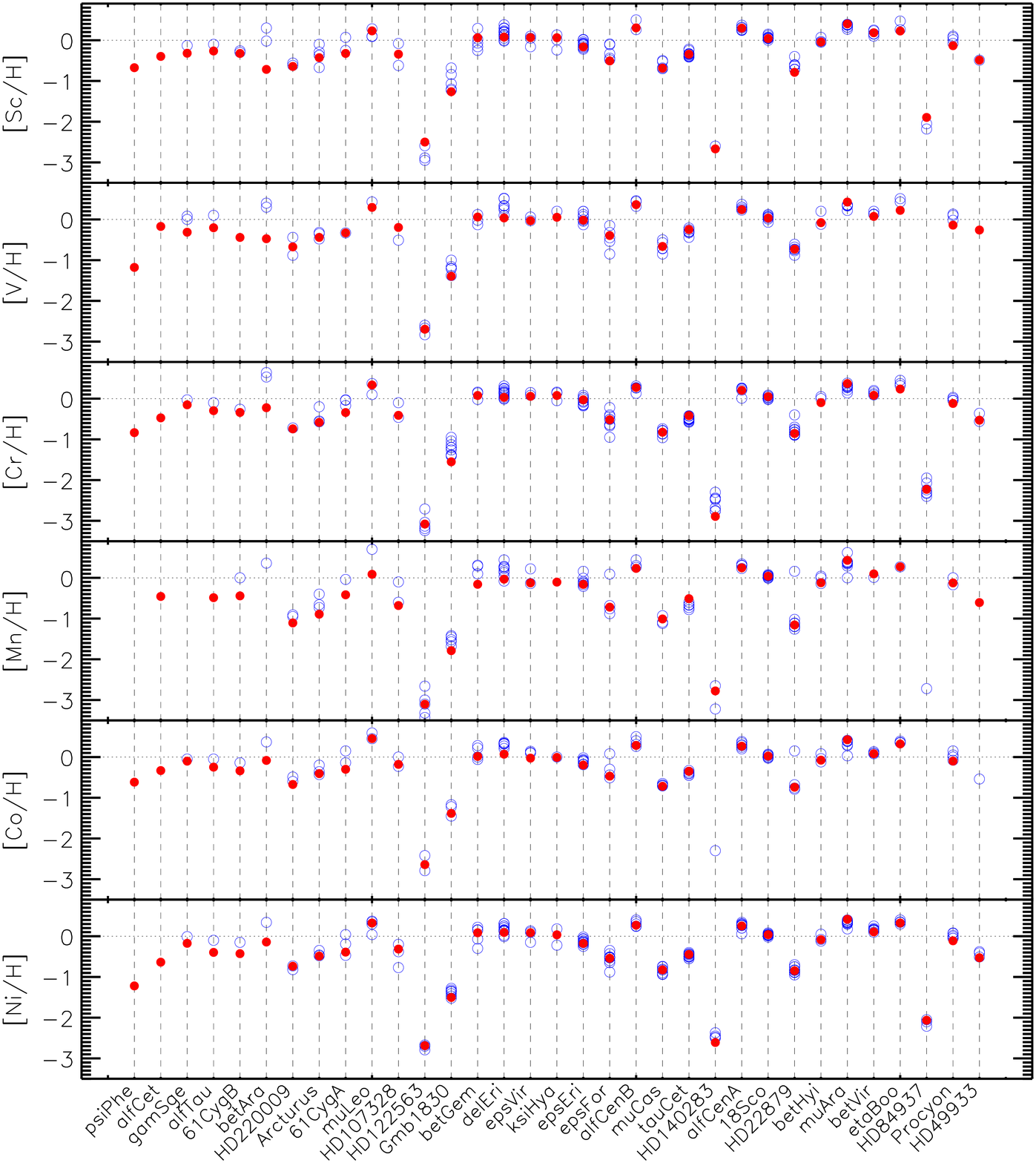}}
   \caption{Abundances of iron$-$peak elements. See caption of \fig{fig:final_alpha}}
  \label{fig:final_iron}
\end{figure}

%\section{Comparison to literature}\label{lit_comparison}

  \section{Discussion on individual abundances}\label{discussion_elements}
  
  {  In \fig{fig:final_alpha} and \fig{fig:final_iron} we show our final results, which are displayed with red filled circles. These values are listed in the second column of Tables~\ref{tab:mgh_fin}, \ref{tab:sih_fin}, \ref{tab:cah_fin} and \ref{tab:tih_fin} for Mg, Si, Ca and Ti, respectively, and of Tables ~\ref{tab:sch_fin}, \ref{tab:vh_fin}, \ref{tab:crh_fin}, \ref{tab:mnh_fin}, \ref{tab:coh_fin} and \ref{tab:nih_fin}, for Sc, V, Cr, Mn, Co and Ni, respectively.  
 The blue open circles are different values taken from an extensive search from the literature which were then transformed to [X/H] scale using the solar abundances employed by the different literature works. For a description of our literature search and a general discussion of our results in the context of previous studies, see Appendix~\ref{lit}. }  In this section we { focus on a detailed discussion of each element individually.  We point out here that this discussion is intended to be for consultation of the results of the elements for the benchmark stars focusing on the main aspects and results in a general way. Thus, this section does not contain important fundaments needed for concluding this work, and can be skipped by impatient readers.  } 
  \subsection{Magnesium}
  % why is this element interesting
  
In spite of the relatively few spectral lines of magnesium with respect to other elements analysed in this work (see \tab{tab:lines}), the abundance of this element is very important in spectroscopic studies partly because of the MgI b triplet around 5180~\AA. These lines are commonly used to derive the surface gravity of the star, especially for spectra with low resolution or low signal-to-noise \citep[see e.g.][for stellar parameter determination of SDSS or LAMOST  spectra]{2008AJ....136.2022L, 2010A&A...517A..57J, 2014ApJ...790..105L}, as many of the smaller lines vanish.  Since Mg is an $\alpha-$element, [Mg/Fe] can have different values depending on the chemical enrichment history of the environment where the star formed. Furthermore, Mg and O are the $\alpha-$elements that separate the best in the [$\alpha$/Fe] vs \feh\ diagrams used to disentangle Galactic components using chemical tagging (Hawkins et al, subm.). Because of the strength of the MgIb triplet, the abundance of this element  known for a vast amount of stars, especially distant metal-poor stars observed with low resolution spectra \citep[e.g.][]{2011AJ....141...90L, 2015arXiv150304362F}. For these reasons, a well defined scale of Mg abundances is very important for stellar spectroscopy as well as Galactic science.
  
  % how many lines do we have, how many golden lines overlap between groups, any line with gf to be corrected?
Eight of the twelve initially selected lines were used for the determination of Mg, with the triplet being excluded as its reduced equivalent width was larger than $-4.8$ for the reference stars. The atomic data of these lines can be found in the electronic table (FINAL LINES). Each line was classified as a golden line for at least one stellar group, meaning that they were used at least by 50\% of the stars in their group.  The lines at $\lambda 6318$ and $\lambda 6319$~\AA\ were golden lines for all groups except the metal-poor stars.   The line $\lambda 8712$~\AA\ has no quality flag from the GES linelist, which means its synthesis and atomic data quality have not been analysed yet. We included this line anyway because of the few lines in the { Giraffe} HR21 { (and Gaia-RVS)} range. 

%nlte and other accuracies 
For magnesium we calculated for each of the lines and stars the departures from LTE, being most of the cases very small.  The largest averaged NLTE departure of -0.1~dex is for Gmb1830 (see \tab{tab:mgh_fin}). As discussed in Appendix~\ref{lit}, the abundances of this star are determined using an effective temperature that is very different from the typical spectroscopic temperature, which might cause { that the calculations for departures from LTE in Paper~III are incorrect}.    The rest of the stars present a small NLTE correction, usually below 0.05~dex,  except for the giants \alfTau,  \betGem, \epsVir\ and Arcturus, which have an averaged departure of the order of 0.1~dex.  Note that large departures of NLTE are found for most of the lines used to determine Mg for these stars.  The lines we used for the Mg determination of giants have all an excitation potential above 5~eV (see table FINAL\_LINES).  As discussed in  \cite{1999A&A...350..955G}, for warm giants NLTE corrections can become large for high excitation Mg lines. Recently, \cite{2015arXiv150407593O} presented a detailed 3D-NLTE analysis of Mg of some of the GBS, in particular the Sun, Arcturus, Procyon and the metal-poor stars.   Although they find that NLTE departures do not significantly affect the solar abundances, for Arcturus differences of up to 0.3~dex can be reached. We found an average departure of 0.08~dex for Arcturus, with extreme cases of around 0.15~dex for the lines at $\lambda 8712$ and $\lambda 8717$~\AA.   \cite{2015arXiv150407593O} discusses further how 3D effects become important as well, making it more difficult to make a one-to-one comparison with our case.

   %literature and Sun 
   In general, our results agree well with the literature. The solar abundance of Mg is quite discrepant  with respect to the scale of \cite{2007SSRv..130..105G} among the 10 elements analysed in this work. Nevertheless,  when uncertainties are taken into account, the values agree. In our extensive literature search, we could not find Mg abundances for the cool GBS, except for \alfTau, which was analysed by \cite{2010AA...513A..35A} and by \cite{1998yCat.3193....0T}. From \cite{2010AA...513A..35A} we could derive two values for [Mg/H] as they provided two values for \feh\ using two atmosphere models. In \tab{tab:mgh_lit} the value of $\mathrm{[Mg/H]}= -0.16$ corresponds to the abundances obtained with MARCS models. \cite{1998yCat.3193....0T}, on the other hand, obtained a value of 0.2. We obtain a value of -0.32, which is lower than that of \cite{2010AA...513A..35A} but when considering the uncertainties in our measurement (up to 0.17~dex), both results are in agreement.
    
    We obtain a value of [Mg/H] = 0.012 for \ksiHya. The standard deviation of the line-to-line scatter is relatively large (0.1~dex)  but the errors due to stellar parameters or NLTE corrections are very small. This star is  a relatively metal-rich giant ($ \mathrm{\feh} = +0.16 \pm 0.2$) so its spectrum is more difficult to analyse, explaining why the scatter in the [Mg/H] abundance is quite high.  We could not find a value of [Mg/H] reported in the literature for this star.  This star has a ratio of [Mg/Fe] = -0.15, which is consistent with [Mg/Fe] ratios of metal-rich stars observed in the Galactic disk \citep[see e.g.][for recent studies]{2014AA...562A..71B, 2014A&A...572A..33M, 2014A&A...565A..89B, 2015arXiv150104110H}
    
 Based on our literature search, we provide  new values of [Mg/H] for \alfCet, \gamSge, \cygB\ and \betAra. These are all cool stars for which we analysed between three and four lines, among them the line at $\lambda 6319$~\AA. The line-to-line scatter can be large for these measurements (about 0.1~dex).  The uncertainty due to stellar parameters can also produce large errors, up to 0.2~dex when the error in \teff\ is taken into account for \alfCet. Given the difficulty of analysing such cool spectra, we find it encouraging to be able to measure Mg with our combined method.  The final lines were carefully inspected to ensure reliability of this new measurement.  The [Mg/Fe] ratios of  \alfCet, \gamSge\ and \betAra\ are slightly enhanced with respect to solar (0.16, 0.19, and 0.16, respectively), which is within the typical values found for  disk giants for solar and slightly lower metallicities \citep{2015arXiv150104110H}. The [Mg/Fe] ratio of \cygB\ is more enhanced (+0.3~dex) with respect to its binary companion \cygA\ (+0.15~dex). When considering the errors, especially the metallicity of \cygB, which is of 0.38~dex (see \tab{tab:params}) both ratios come to a better agreement.

  \subsection{Silicon}
  
  % why is this element interesting
Silicon has several clean lines in the optical part of the spectrum, making it an important $\alpha-$capture element, as it plays an important role in testing supernovae and chemical evolution models. It is believed that silicon is created during later evolution of massive stars \citep{1995ApJS..101..181W} and by Type Ia supernovae \citep{1995MNRAS.277..945T}. Some works studying Galactic chemical evolution models that use spectroscopic analyses of stars and measure silicon are \cite{1993AA...275..101E,  1995ApJS...98..617T, 2005ApJS..159..141V, 2003AA...410..527B} and \cite{2014A&A...572A..33M}.

    % how many lines do we have, how many golden lines overlap between groups, any line with gf to be corrected?
    We have a total of 13 neutral and 2 ionised Si lines, all of them having rather high excitation ($> 4.9$ eV).  The 15 lines were classified as golden lines for the group of $FG-$dwarfs, 13 of them as golden for the $FGK-$giants. For extremely metal-poor stars such as  HD140283, lines usually below $\lambda$4100 \AA\ are preferred to determine Si abundances \citep[see e.g.][who has determined Si for many metal-poor GBS]{2003AA...404..187G}, which lie blue wards from our spectral region. We could still determine the abundance of Si for this extreme case based on only one very weak line ($\lambda$5708~\AA), which has an EW of 2.2 m\AA\ as measured by the ULB method. Only the ULB and iSpec (which are based on synthesis) methods could provide an abundance for this line, which was checked with the synthesis performed to a rather stronger line but blended with telluric features. Our  result  for [Si/H] of -2.2 agrees well with previous reported works of -2.3 \citep{2003AA...404..187G} or of -2.1 \citep{1998yCat.3193....0T, 1986AA...160..264F}.  
    
    We found that the line $\lambda8899$~\AA, which is a golden line for the $FG-$dwarf and $FGK-$giant groups,  yields an absolute abundance that agrees between the methods, but the absolute abundance is systematically lower (about 0.4~dex) than when looking at the value obtained for that line with the differential approach. This suggests for  a revision of the atomic data of this line.   
  
  % nlte corrections (our work or literature)
  We performed NLTE corrections for this element, showing that most of the cases the corrections are zero or below 0.03~dex. These corrections are negligible compared to the line-by-line scatter found for Si for all cases.  The analysis of Si abundances in LTE and NLTE of \cite{2012ApJ...755...36S} agrees with us in the sense that the abundances differences are very small in the optical region. { We note however that \cite{2012ApJ...755...36S} derived significantly larger NLTE departures for strong lines in the infrared (between 10200 and 10900~\AA).} 
  
  % uncertainties in general. How accurate?
For the cool stars \alfCet, \cygB\ and \betAra\ we used 2-5 lines only, because most of the lines were blended with molecules or the continuum could not be properly placed.  The line-to-line scatter is for these cases rather large ($> 0.1$~dex) but for the rest of the stars the scatter is in general below 0.1~dex. The cool stars \alfTau, \epsEri\ and \cygA\ have also a rather large line-to-line scatter, but given the difficulty in analysing cool spectra, we find the values acceptable.    The uncertainties in Si abundances due to errors in stellar parameters is for most the stars smaller than the line-to-line scatter.  The case of \betAra\ is an exception presenting large differences in [Si/H] of up to 0.17~dex when considering the errors of \feh\ and \vmic, but this star has parameters that are uncertain and it is very cool, which makes the analysis of spectra in general more difficult. 
  
  % comparison with literature and Sun 
  The abundance of Si we obtain for the Sun agrees very well with \cite{2007SSRv..130..105G} within the errors. When comparing our results of silicon abundances with the literature (see \fig{fig:final_alpha}), we can conclude that our values agree for most of the stars very well with those reported in previous works, with exception of \betAra\ \citep[+0.67,][]{1979ApJ...232..797L}, which considers a value for \teff\ 500~K and for \vmic\ 3~km/s larger than our own.    For \cygA\ our value agrees very well with that one of \cite{2005AA...433..647A} but disagrees with the result of \cite{2008AA...489..923M}. 
   
  % any star in particular? 
  Finally, to our knowledge, we are the first ones giving a value of [Si/H] for the cool giant \alfCet, which is based on the analysis of the two lines $\lambda$5684~\AA and $\lambda$5701~\AA. The line-to-line scatter is  of 0.1~dex but as seen in Table (ALF CET), few methods (those performing synthesis)  were able to analyse these lines, thus the scatter is overestimated due to the small statistic of this measurement. We visually inspected these lines to ensure the reliability of this new measurement.  The [Si/Fe] value of \alfCet\ is rather low (-0.07) which is lower than [Mg/Fe]. This star is very cool and the errors in the measurement of Si abundances are significant in all our sources of uncertainties. When considering the errors, [Si/Fe] and [Mg/Fe] are consistent. It is worth to comment the trend in temperature found for Si abundances by \cite{2015arXiv150104110H} in the large sample of APOGEE giants. The coolest stars (those around 3500~K such as \alfCet) have systematically lower (even negative) [Si/Fe] ratios with respect to the warmer giants.

  \subsection{Calcium}  
    % why is this element interesting
{ Similar to the case of the Mg~Ib lines}, the Ca II triplet around 8500~\AA\ is  a very important feature to derive surface gravities. Furthermore, the Gaia-RVS spectra are centred on this feature, and so the spectra of RAVE and the HR21-Giraffe setup of the Gaia-ESO survey.  Examples of works  determining parameters using this feature are \cite{2011A&A...535A.106K} and \cite{2013A&A...559A..74B}. Furthermore, calcium is an $\alpha-$capture element, and like Si and Mg, a star can have  different values of  [Ca/Fe] depending on the age and star formation history of the gas that formed the star \citep[e.g. ][]{2004AJ....128.1177V, 2010AA...511L..10N}. 

  % how many lines do we have, how many golden lines overlap between groups, any line with gf to be corrected?
  All the initially selected 22 lines were considered for the final abundance determination, where 18 of them are neutral and 4 ionised  lines. For the $M-$giant group, 3  lines are used, for the $FGK-$giant group, 7 lines (all golden), for $K-$dwarfs 6 lines (3 golden), for $FG-$dwarfs 19 lines and for metal-poor stars 18 lines. The lines at $\lambda 5867$ is classified as golden for all groups except for the metal-poor stars. Five lines are golden lines in both the \fgdwarf\ and the \metalpoor\  groups. The overlap between metal-poor and giants or the \kdwarf\ is much smaller,  because these lines become too strong ($\log (\mathrm{EW}/\lambda) > -4.8$) for  cooler stars.    The lowest excitation potential of Ca lines is about 1.9~eV while the highest is 8.4~e.V. 
 
  % nlte corrections (our work or literature)
 We performed NLTE corrections for this element as described in \sect{nlte}, with large negative corrections for $FGK-$giants. Note that this averaged NLTE correction is in this case not representative for the majority of the individual corrections. One can see that the red ionised lines at $\lambda$8912 and $\lambda$8927~\AA\ have corrections of up to $-0.25$~dex, while the other lines have corrections below 0.05~dex.  Solar-like stars have usually smaller NLTE collections than the line-to-line or the errors due to stellar parameters. The aforementioned two lines are also considered in the final Ca abundance and have  large corrections as well, although slightly smaller than for giants  (of  0.1~dex) and since there are more lines used for the final abundance, these two NLTE sensitive lines have less weight in the  final value.
 
   % uncertainties in general. How accurate?
Normally uncertainties in final Ca abundances are below 0.1~dex when considering errors in stellar parameters or the line-to-line scatter.  The uncertainties in \logg\ produce in most of the cases a negligible difference or zero difference in the measured Ca abundance. The uncertainty in \vmic, on the contrary, produces differences in [Ca/H]  more significant although still small except for the cool stars. Uncertainties in temperature produces slightly smaller differences in final Ca abundance than \vmic\, with the warm giants having a larger difference with respect to the warm dwarfs. Metallicity errors do not significantly affect the final Ca abundance, in a similar way as \logg. 
  
  % comparison with literature and Sun 
We were able to determine Ca abundances for all GBS, which is important because this makes Ca a good element for calibration of homogeneous $\alpha-$abundances using all benchmark stars.  The abundance of Ca we obtain for the Sun agrees very well with \cite{2007SSRv..130..105G} (see \fig{fig:sun}). Our results for the rest of the stars also agree well with those found in the literature. One exception is the cool giant \gamSge, for which we found that \cite{1995AZh....72..864B} obtained a value of [Ca/H] = 0.05, which is 0.4~dex higher than our result. They use a temperature that is 100~K higher, a surface gravity that is 0.3 ~dex lower and a metallicity that is 0.2~dex higher. For the same line we used here ($\lambda$6156~\AA) they determined a value of  $\log \epsilon = 6.4$  while we determine 6.0. The $\log gf$ value they employed agrees with ours within 0.1~dex. We attribute the difference on the stellar parameters and probably different methodologies and continuum placement.   
  
      We could not find works reporting abundances of Ca for the cool giants \psiPhe, \alfCet\ and \betAra. Our abundances are determined from one and two lines.   In these stars a difference in \vmic\ causes a difference in [Ca/H] of about $\sim0.1$~dex or more,  while the rest of the uncertainties do not cause a significant change in the final abundance. We visually inspected these lines to ensure reliability of our results. The [Ca/Fe] ratio for these cool giants are consistent with the expectation of chemical evolution models. The [Ca/Fe] ratio of \psiPhe\ of 0.5 agrees with $\alpha-$enhanced stars at low metallicities (-1.25, see \tab{tab:params}). For the rest of the giants, [Ca/Fe] ratios between 0 and -0.15 were obatined, which agrees with the systematic lower  [Ca/Fe] ratios found by \cite{2003AA...404..715B} and \cite{2015arXiv150104110H} with respect to other $\alpha-$elements in their samples of stars. 

  % any star in particular? 

  It is worth commenting  that although we provide a measurement for [Ca/H] for \psiPhe, this is done using the \feh\ value determined by us in Paper~III. We recall that in Paper I we could not use a reliable stellar evolutionary track to measure \logg\ when using the metallicity determined in Paper III, thus a solar metallicity was employed.   The stellar parameters of this star have been recently determined by \cite{2014MNRAS.443..698S}, obtaining a rather metal-rich star of metallicity $+0.1 \pm 0.4$, with temperature and surface gravity consistent with Paper I. This shows how difficult the determination  of abundances of this cool giant is.  Therefore, we recommend to take this value of Ca with caution and we suggest to reanalyse this line using a revisited value of metallicity.

    \subsection{Titanium}
 Titanium is sometimes referred to as an iron-peak element \citep{1995ApJS...98..617T}, but whose overabundance at low metallicities follows the $\alpha $-element  behaviour. Optical spectra have for FGK stars numerous Ti lines, which makes this element a common one in chemical analyses of stars, as one can determine this abundance from spectra at almost every wavelength range.   Due to its ``$\alpha$'' nature, Ti abundances are widely used to study the structure and evolution of our Galaxy \citep[e.g., ][]{2010AA...511L..10N, 2014A&A...572A..33M, 2014A&A...568A..71B}. Titanium lines are in addition sometimes employed in metal-poor stars for determination of parameters, when the amount of iron lines is not enough to evaluate excitation and ionisation balance \citep[e.g., ][]{2000AJ....120.1014P}

  % why is this element interesting
  % how many lines do we have, how many golden lines overlap between groups, any line with gf to be corrected?
    Titanium is the element where we have most of the lines and thus the mean and standard deviation for the derivation of the final abundances allowed us to sigma clip the bad lines quite straightforwardly. We have a final selection of 55 neutral lines and 12 ionised ones, all of them rather with low excitation ($< 3.1$ eV). For the stars belonging to the group of \mgiant\ we used 12 lines, among them 7 classified as golden. Among these 4 lines, one \ion{Ti}{ii} ($\lambda 5005$~\AA) and three \ion{Ti}{i}  ($\lambda  5689$, $\lambda 5702$ and $\lambda 6091$ have been classified as golden for rest of the groups except for the \metalpoor\ one.  As for Mg, we included a line ($\lambda$7819~\AA) which has no quality flag form the line list. This line shows good results and its synthesis and atomic data should be analysed with more detail for future versions of the GES line list. 
  % nlte corrections (our work or literature)
  
  NLTE corrections for Ti were not computed as for the previous cases, so the respective column in \tab{tab:tih_fin} is empty. The corrections should be however small for FGK stars \citep{2007PASJ...59..335T, 2003AA...410..527B}. As extensively discussed by \cite{2003AA...404..715B}, neutral Ti lines are more sensitive to NLTE effects, especially for metal-poor stars. Our abundances are based on mostly \ion{Ti}{i} lines, meaning that uncertainties due to NLTE should be investigated with detail in the future. The large parameter coverage of the GBS, together with the large number of Ti lines makes this task challenging.  The  line-to-line scatter of Ti abundances, in particular for metal-poor stars, is rather large, which could be attributed to NLTE effects, in which abundances obtained from ionised and neutral lines are different. Since we do not have ionised and neutral lines for all the elements in this study, we prefer to take a final averaged one rather than two separate values, but we list abundances for each line in the online material, so it is possible to choose only ionised Ti lines, which should be  less sensitive to NLTE effects \citep{1983ApJ...265L..93B, 2003AA...404..715B}.   
  
  % uncertainties in general. How accurate?
 Uncertainties due to NLTE corrections of iron from Paper III yield very small or zero differences in Ti abundances.   Even smaller are the differences in Ti abundances when uncertainties of \logg\ are considered, where most of these stars, especially the warm stars, have a zero difference.  One exception is the cool giant \gamSge, for which an error of 0.35~dex in \logg\ causes a difference of 0.07~dex in [Ti/H].  Uncertainties in effective temperature do not significantly affect the final values of Ti abundances. Metallicity uncertainties have in general a small impact in the final Ti abundances of less than 0.05~dex. Finally, errors in \vmic\ produce in the cases where the spectral lines are the strongest (cool or metal-rich stars) greater differences, as expected.   The line-to-line scatter is below 0.1~dex except for metal-poor and cool stars. These stars have a slightly larger line-to-line scatter, mostly due to the few lines used to measure the Ti abundance and stronger NLTE effects. 
 
   % comparison with literature and Sun 
   Our abundance of Ti for the Sun agrees perfectly with the solar abundance of \cite{2007SSRv..130..105G}. When comparing our values with the literature in \fig{fig:final_alpha}, we can see that in general there is a very good agreement.  As in the case of Ca, we are able to determine abundances of Ti for all GBS in an homogeneous fashion. For the cool stars \gamSge, \alfTau\ and \cygB, however, we obtain systematically lower [Ti/H] values than the literature.  As above, the abundance of Ti was determined for \gamSge\ by  \cite{1995AZh....72..864B}, who used a hotter and more metal-rich set of atmospheric parameters than us. The star \alfTau\ was analysed by \cite{2010AA...513A..35A} and by \cite{1998yCat.3193....0T} reporting Ti and Mg abundances (see above). These two works in this case agree with a value of [Ti/H] zero or slightly above, which is 0.4~dex above our result.  Note  \cite{2010AA...513A..35A} uses a solar metallicity for \alfTau, while our \feh\ is $-0.4$. For \cygB, a value of [Ti/H] of $-0.05$ was reported by \cite{2005AJ....129.1063L}, which is 0.3~dex below our result. In Paper III we discussed that our measurement of metallicity had a difference of 0.3~dex with respect to this work and discussed its reasons. 
   
  The case of Gmb1830 is worth commenting. We found 10 measurements of Ti abundance in the literature for this star, for which the results varied from $-1.33$ \citep{2003AA...404..187G} to $-0.85$ \citep{1998yCat.3193....0T}. Our result of $\mathrm{[Ti/H]} = -1.64$ is 0.3~dex below the literature range of values. We discussed in Paper~III the effect on abundances of having a significant difference of effective temperature between the fundamental value determined in Paper~I and the spectroscopic value. The metallicity is too low, and as a consequence, the abundances are too low as well. 
  
  The final  case worth mentioning is the star  HD49933, for which \cite{2007PASJ...59..335T} provides two values for the abundance of titanium, which come from the analysis of neutral and ionised Ti lines and have 0.5~dex difference. In their Figure 11 one can see how the difference between [Ti\ion{I}/H] and [Ti\ion{II}/H] behaves as a function of temperature. For hot stars this difference increase, explaining the difference of 0.5~dex they obtain for HD49933. Our value is computed averaging all lines and lies in-between as expected. Note the line-to-line scatter of this star, as well as of Procyon, which are the hottest stars of our sample, are rather large (above 0.07~dex). For these stars, like for the metal-poor stars, the difference of abundances from ionised and neutral lines could be a NLTE effect.  One might consider lines of ionisation stages separately to decrease this scatter.  In this case we aim for an homogeneous analysis of 33 different stars and 10 different chemical elements, so we prefer to have a larger line-to-line scatter in the Ti abundance of hot stars than having two separate [Ti/H] values. Since we provide the individual abundances for all lines in the online tables (STAR\_LINE) one can easily calculate the separate Ti abundances if needed.  
  
  % any star in particular? 
For \betAra\ our Ti abundance is much lower than the only reported one by \cite{1979ApJ...232..797L}  by 0.8~dex. The atmospheric parameters considered by us and by  \cite{1979ApJ...232..797L}  are very different, causing this difference, which goes in the same direction as [Si/H].  The errors of this star, in particular the line-to-line scatter, are very large. We provide a measurement of Ti for the rest of the cool giants  (\psiPhe\ and \alfCet) for which we could not find a value of [Ti/H] in the literature. The [Ti/Fe] ratios obtained for these stars are +0.36 for \psiPhe, which is consistent with the enhancement of this element seen for metal-poor stars. As discussed for Ca, this abundance should be taken with care and a revision for the metallicity is needed. The cool giants \alfCet\ and \alfTau\ have solar [Ti/Fe], which is consistent with the ratios obtained for Ca.  However, \betAra\ has very low [Ti/Fe] abundance (-0.4) but the errors of this value are very large when considering the errors of Ti and Fe together. Determination of Ti abundances in giants is difficult, which can also be seen in the large dispersion of cool giants in the [Ti/Fe] vs [Fe/H] diagram of \cite{2015arXiv150104110H}.

\subsection{Scandium}
  % why is this element interesting
 This  element is  important for studying the structure and chemical evolution of the Milky Way as [Sc/Fe] v/s [Fe/H] seem to relate differently for the different Galactic components \citep{2012AA...545A..32A}. This suggest that although both elements are synthesised through the same process -- core-collapse and thermonuclear supernovae \citep{1997nceg.book.....P} -- the physical environment, (in particular the IMF) in which these supernovae happened are different in different parts of the Galaxy. For example, the Sc yields show very large variations as a function of the mass of the progenitor in the computations of \cite{2002ApJ...577..281C} producing scatter in [Sc/Fe] v/s [Fe/H].
  
  % how many lines do we have, how many golden lines overlap between groups, any line with gf to be corrected?
  For scandium, we have 6 neutral and 7 ionised lines, all of them with low excitation ($< 1.9$~eV). All the lines were used for \fgdwarf, all except two for the \kdwarf. All ionised lines (3 of which are golden) and no neutral line were used for the stars of the \metalpoor\ group, and \fgkgiant\ considered 4 ionised and all the neutral lines.  The group of \mgiant\ used 4 ionised and 3 neutral lines. Three ionised lines ($\lambda 5667, \lambda 5684$ and $\lambda 6604$~\AA) were used in common for all the groups, but are only golden for some of the groups. The metal-poor stars are those that need stronger lines to be visible at low metallicities, which saturate for cooler more metal-rich stars.
  
  % nlte corrections (our work or literature)
We did not calculate NLTE corrections for Sc, but an extensive discussion on NLTE corrections of Sc for the Sun  can be found in \cite{2008A&A...481..489Z} They found large  NLTE departures to abundances from neutral Sc lines (about 0.15~dex) in the Sun. Recently, \cite{2015AA...577A...9B} studied the differences in [Sc/Fe] with and without NLTE corrections for Sc finding that at low metallicities these differences are larger. No particular trend was found in temperature or surface gravity. Unfortunately  there is still little information on NLTE corrections for Sc for stars very different than the Sun, such as most of the stars studied in this work. 

{ Scandium is an odd-Z element that can be affected by hfs. As extensively discussed in \sect{hfs}, hfs total effect of  hfs in the determination of odd-Z element abundances is different for each GBS. Although we could not see a clear systematic difference on the effects of hfs in Sc, to be consistent with the analysis of the other odd-Z elements analysed in this work, we restricted our determination of Sc to only the methods considering hfs. }

  % uncertainties in general. How accurate?
We provide abundances of Sc for all GBS. The  line-to-line scatter can be in many cases quite large, probably due to the imbalance between abundances obtained between \ion{Sc}{i} and \ion{Sc}{ii} (see \tab{tab:sch_fin}). This scatter is normally below 0.1~dex  except for the Sun, which is also seen in \fig{fig:sun}. Note that the abundance of the Sun is not determined with a differential analysis, so it is expected that the scatter at line-by-line basis can increase with respect to the rest of the stars. From \tab{tab:sch_fin} one can also see that the difference in [Sc/H] is normally much smaller than the line-to-line scatter.

  % comparison with literature and Sun 
  Our result of scandium for the Sun agrees very well with the solar abundance of \cite{2007SSRv..130..105G}. For the rest of the stars our results also agree very well with the literature, except for \betAra\ and Gmb1830.  As discussed above for titanium, \betAra\ has been only analysed by  \cite{1979ApJ...232..797L}, who systematically found higher abundances than us because of the reasons mentioned above. We found in the literature five works reporting abundances of Sc for  Gmb1830, which vary from $-1.23$ to  $-0.68$ \citep[both values determined by ][and the rest of the works lie in between]{2007PASJ...59..335T}. As mentioned before, \cite{2007PASJ...59..335T} made a separate analysis of ionised and neutral lines, providing two different values. { Our value agrees with the low value of \cite{2007PASJ...59..335T}. }
  
  % any star in particular? 
  We were able to determine [Sc/H] for \psiPhe\ and \alfCet. For that we used two lines in common ($\lambda$5667 and $\lambda$5684~\AA). The abundances are uncertain, with a  large line-by-line scatter as seen in \tab{tab:sch_fin} for these stars. We checked the line profiles and decided to keep the abundances, even if the different methods gave different results, as we could not find an obvious reason to reject those lines. These lines are used for all group of stars.  The [Sc/Fe] ratios of the cool giants have expected values \citep[see trends of ][for dwarfs]{2015AA...577A...9B} when considering the uncertainties, with $+0.16, -0.05$ and $+0.16$ for  \psiPhe, \alfCet\ and \alfTau. The [Sc/Fe] ratio of \betAra\ has a similar behavior as [Ti/Fe], i.e. it { is rather low ($-0.3$)}.   This abundance is very uncertain but a revision of the stellar parameters  or a 3D-NLTE investigation of its line profiles could  bring this star back to normal chemical evolution level expectations.

\subsection{Vanadium}
  % why is this element interesting
  The nucleosynthesis channel of vanadium is not properly understood, and the supernovae yields lead to largely underestimated values compared to observed abundances \citep[see ][for a recent review]{2013ARA&A..51..457N}. Regarding its Galactic distribution and enrichment history, [V/Fe] shows a very large dispersion, suggesting different trends for different data \citep{2003AA...404..715B, 2015AA...577A...9B, 2015arXiv150104110H}.   The formation channels of this element as well as the usage of V to track the chemical enrichment history of the Milky Way can be better controlled when better and more consistent observed abundances are derived.  
  % how many lines do we have, how many golden lines overlap between groups, any line with gf to be corrected?

{ Vanadium is another odd-Z element sensitive to hfs. Although this effect is hidden in the line-to-line and node-to-node scatter, it can produce significant systematic differences in the determination of V for some of the stars.  In our aim to have an homogeneous analysis for all GBS, we restricted the determination of V of only methods that took hfs into account. }

  We determined  homogeneous abundances of vanadium for all the GBS.  In total 30 neutral V lines were selected, all of them at wavelength below $\lambda$6600~\AA, except for $\lambda 8933$~\AA. The lines have rather low excitation potential, with the highest excitation of 1.2~eV. All the 4 lines used for the \mgiant\ were classified as golden, while all the 8 lines used for the \metalpoor\ were not classified as golden. A large number of lines is used for the dwarfs and the \fgkgiant.      
  % nlte corrections (our work or literature)
  
  As discussed in \cite{2015A&A...573A..26S} and \cite{2015AA...577A...9B}, departures of NLTE for V lines in the Sun have not been reported. In this work we determined abundances from neutral line only, and as known from other iron-peak elements, neutral lines suffer more from NLTE effects than ionised ones. We agree with the claim of \cite{2015A&A...573A..26S} that a study of NLTE of V is urgently needed.  We did not analyse ionised lines to see if a systematic difference would hint at possible NLTE effects.  
  
  % uncertainties in general. How accurate?
  What concerns our internal uncertainties, final V abundances when changing the overall \feh\ -- either considering its uncertainty of the value determined from LTE -- do not change significantly, being usually less than 0.05~dex. The  cool dwarf \cygA\  presents a slightly larger difference of 0.08, which is still very small and can be neglected when looking at the line-by-line scatter obtained for V abundance of that star.  The dependency of V on \logg\ errors is almost zero in all cases, while the dependency on \teff\ or \vmic\  can be significant although always less than 0.1~dex.  This behavior is not surprising as all lines employed here are neutral lines so we see a similar behaviour to what would happen with the determination of iron from \FeI\ lines only, which is a strong dependency on \teff\ and \vmic.  The larger uncertainty in the V abundances is thus the line-to-line scatter, which can be slightly above 0.1~dex for some cases, such as the very hot or very cool stars, which is probably due to the fewer lines employed for these stars. For solar-like stars, the scatter is very small, usually below 0.05~dex. 

    % comparison with literature and Sun 
    The abundance of vanadium obtained by us for the Sun agrees within the errors with the solar abundance of \cite{2007SSRv..130..105G}, with our abundance being slightly lower. As previously discussed, our value agrees well with \cite{2015AA...577A...9B}. For the rest of the stars, there is in general a very good agreement with the literature as can be seen in \fig{fig:final_iron}. Note there are few measurements of V abundance in the metal-poor stars HD122563 \citep{2000AJ....120.1841F, 2000ApJ...530..783W}, Gmb1830 \citep{1998yCat.3193....0T, 2000AJ....120.1841F, 2003AA...404..187G, 2006AA...454..833K, 2007PASJ...59..335T} and HD22879 \citep{2006MNRAS.367.1329R, 2009AA...497..563N, 2003AA...404..187G, 2000AJ....120.1841F, 2006AA...449..127Z}.     \cite{2000AJ....120.1841F} for example, who reported an abundance of V for the three stars mentioned here, analysed 5 V lines, among them only two overlapping with our selection of lines. Our values agree well for HD122563 and HD22879. Regarding Gmb~1830, our value follows the same direction than the rest of the elements, that is, it is underabundant with respect to the literature. The reason is the different \teff\ employed in this work, which needs urgent revision.  Note also that for the hot metal-poor dwarf HD84937  and for the very metal-poor star HD140283, we could not find a measurement of [V/H] in the literature. 
    
We recall that we could not find lines in common between \alfTau\ and the Sun, so  the V abundances were determined with respect to the absolute value for \alfTau. Our result for that star agrees within the errors  with \cite{1998yCat.3193....0T} and our results for \gamSge\ agree well within the errors with \cite{1995AZh....72..864B}. For \betAra\ again our only comparison is the work of \cite{1979ApJ...232..797L}, which is higher than our values.  For the other two cool giants \psiPhe\ and \alfCet\  we could not find a measurement of V in the literature. The abundances are determined from 3-4 lines, obtaining a high line-to-line scatter, especially for \psiPhe. Both stars have a rather high [V/Fe] (about +0.2), but given the high dispersion of [V/Fe] measured on disk stars, especially on giants \citep{2015arXiv150104110H} it is difficult to ensure that these values are expected. 

The cool dwarf \cygB\ is another case where we could not find a reported value in the literature. For this star we used 16 lines, with 5 methods providing abundances for them. The line-to-line scatter is below 0.1~dex, which is encouraging for this cool dwarf that has so many molecular bands blending the atomic lines. The errors due to stellar parameters are negligible except the error due to \vmic, which is expected since the  V lines of \cygB\ are rather strong. The [V/Fe] ratio for this cool dwarf { is around solar}, consistent with the trends found by \cite{2015AA...577A...9B} or \cite{2003AA...404..715B}, although their study involved warmer stars. 
  % any star in particular? 
  Finally, the hottest GBS, HD49933, has no [V/H] reported in the literature. Our value has a large line-to-line scatter (or better said in this case a method-to-method scatter) of { less than 0.1}~dex. Only one  line was used for the determination of V ($\lambda 5627$\AA), which is very weak, having an EW measured only by the Porto method of 8.2 m\AA.   { Note that} the abundance obtained by the Porto method is significantly higher than the abundances obtained by the other 3 synthesis methods analysing that line, {  suggesting a strong hfs effect. These EW values do not contribute to the final result of V abundances, but these} detailed results can be found in the table (HD49933) as part of the online material. The [V/Fe] ratio for HD49933 is high, of +0.3, which still would follow the trend of \cite{2015AA...577A...9B} for metallicities of -0.4.

\subsection{Chromium}
The scatter of Cr is very small since there is not very much variety in the SN II yields and [Cr/Fe] is almost zero for all metallicities \citep{1995ApJS...98..617T, 2011ApJ...729...16K}. However, several works in the literature report that while for \ion{Cr}{ii} this is the case, for \ion{Cr}{i}  the [Cr/Fe] ratio  increases with metallicity \citep{1991A&A...241..501G, 2004A&A...425..671B, 2004A&A...416.1117C, 2008ApJ...681.1524L, 2012AA...545A..32A}. \cite{2010AA...522A...9B} explained this discrepancy by finding that neutral Cr lines are heavily sensitive to NLTE for metal-poor stars. 

  % why is this element interesting
  % how many lines do we have, how many golden lines overlap between groups, any line with gf to be corrected?
 We have a final selection of 23 \ion{Cr}{i} lines, which have rather low excitation ($< 3.8$~eV). Only two lines have been  analysed by the five groups of stars simultaneously ($\lambda 5272$ and $\lambda 5287$~\AA), mainly because of  the fact that lines that are strong enough to be visible in metal-poor stars are usually saturated in more metal-rich stars.  Lines classified as golden in four groups -- all except the \metalpoor\ one -- are the Cr line at $\lambda$5272, $\lambda$5827 and $\lambda$5628~\AA.  For the group of \mgiant, only these 3 lines out of six are classified as golden. 
 
  % nlte corrections (our work or literature)

Chromium is the only iron-peak element for which we calculated the departures of NLTE for each star and line used in this analysis (see \sect{nlte}). In consistency with \cite{2010AA...522A...9B}, the departures  for solar-type stars are of the order of 0.05~dex, while for metal-poor stars they are very large (0.24~dex for HD84937, -- 0.39~dex for HD122563). 
 
  % uncertainties in general. How accurate?
The systematic uncertainty due to strong NLTE effects for \ion{Cr}{i} lines are much larger than uncertainties in Cr abundances due to errors in stellar parameters or the line-to-line scatter. As in the case of vanadium, the major change in Cr abundance is when the error of temperature and at a certain level the micro turbulence velocity are taken into account. The reason is the same: in this analysis we study only neutral lines so measuring abundances of Cr or Fe have the same impact from the different parameters.  Our line-to-line 
abundance determination is in most of the cases very accurate, with a line-to-line scatter of about 0.05~dex or in several cases below. Few exceptions, mostly those stars where few lines were analysed, present  a line-to-line scatter slightly above 0.1~dex. These are few cases, and in general the cool and metal-poor stars. 

% comparison with literature and Sun 
Our solar abundance is slightly lower than the value of \cite{2007SSRv..130..105G} but they agree well within the errors.  The abundance corrected for 0.05~dex NLTE departure  (see \tab{tab:crh_fin} for the Sun) would produce a better agreement with \cite{2007SSRv..130..105G} but we restrict our analysis to LTE for homogeneity.  The rest of the stars show a good agreement with the literature as seen in \fig{fig:final_iron}.  We were able to determine the abundances of Cr for all GBS homogeneously. It is worth to comment on the star Arcturus, for which our value of [Cr/H] = -0.58 is 0.38~dex lower than the result obtained by \cite{1998yCat.3193....0T}. Our result however agrees very well with the other two values  of \cite{2011ApJ...743..135R} and \cite{2005AJ....129.1063L}. 

  % any star in particular? 
  
   {   One can see in \fig{fig:final_iron} that there is a scatter in Cr for \epsFor\ spanning from [Cr/H] = -0.22 \citep{2004AA...425..187T} to [Cr/H] = -0.95 \citep{2003AA...404..187G}.  The main reason for the discrepancy in the literature values is the different consideration of the stellar parameters in each of these works. In the particular example mentioned above, there is a difference in \teff\ of about 400~K, in \logg\ of $\sim$0.8~dex and \feh\ of about 0.4~dex. \cite{2004AA...425..187T} showed that a difference of 150~K can produce up to 0.09~dex difference in the abundance of Cr for \epsFor.  Furthermore, Cr is an element which is sensitive to NLTE for metal-poor stars (see \sect{nlte}) which might cause a scatter in the literature. As explained in Paper~III, the literature is highly inhomogeneous, which is a reason why we re-determine the abundances of all GBS in an homogeneous way. }
   
  As usual, the cool giant \betAra\ has a value of the abundance of Cr from \cite{1979ApJ...232..797L} which is much higher than our value. We provide [Cr/H] for the two very cool giants \psiPhe\ and \alfCet\ and no comparison is available in the literature. The abundances are determined from few lines ($\lambda 5272$, $\lambda 5287$ and $\lambda 5628$~\AA), which have large errors, especially when considering the errors of the measured \feh.  NLTE corrections for these lines are in both cases  of the order of  0.05 to 0.08~dex. The line-to-line scatter is small (0.16~dex) given the difficulty of analysing such cool giants.   The [Cr/Fe] ratio would become +0.4 and solar for \psiPhe\ and \alfCet, respectively.  Chromium is an iron-peak element and thus it is expected to follow a rather flat trend in metallicity \citep{2014AA...562A..71B}, although studies of lower resolution targeting also giant stars have found some higher abundances of Cr like the one of \psiPhe\ \citep{2014A&A...572A..33M}. The abundances of this star, especially the ratios as a function of iron, should be treated with care until a revision of the iron abundance is made. 
  
\subsection{Manganese}
  % why is this element interesting

  Manganese is produced more by SNe Ia than Fe. Thus,  from $\mathrm{[Fe/H]} \sim - 1$, [Mn/Fe] starts showing an increasing trend toward higher metallicity, which is caused by the delayed enrichment of SNe Ia \citep{2011ApJ...729...16K}.  Furthermore, this trend varies within the different Galactic components \citep{2012AA...545A..32A, 2013A&A...559A...5B}, from cluster to field stars \citep{1989A&A...208..171G}, and it is different for the Milky Way and dwarf galaxies \citep{2000ApJ...537L..57P, 2012A&A...541A..45N}. 
   Measuring accurate abundances of Mn are thus important also for studying the structure of our Galaxy because, for example, 
 there is observational evidence of the existence of low [$\alpha$/Fe] stars \citep[e.g.][]{2010AA...511L..10N, 2014A&A...571L...5J}, which are important for discussions of the formation history of the Galactic halo \citep[e.g. ][]{2014MNRAS.445.2575H}. The majority of low [$\alpha$/Fe] stars should also have high [Mn/Fe] because of the SN Ia contribution as explained in \cite{2011ApJ...729...16K}.  Recently, in Hawkins et al 2015 (submitted) Mn has shown to be one of the best candidates to disentangle Galactic components. { While thin disk stars have enhanced [Mn/Fe] ratios, thick disk stars have solar [Mn/Fe]. They explain this difference with the fact that Mn is produced at a higher fraction compared to Fe during SNIa, meaning that at a given metallicity, $\alpha-$poor stars (which have been polluted by more SNIa) will have higher [Mn/Fe] ratios compared their $\alpha$-rich counterpart. }
 
The abundance determination of Mn is however complicated. It has significant hyperfine structure splitting, which broadens the spectral lines. { As discussed in \sect{hfs}, reliable abundances should preferentially not be obtained by only using the EW and one total oscillator strength of a given line.  In  \cite{2012A&A...541A..45N} one can see that a difference of up to 1.7~dex can be obtained depending on whether hfs is or is not taken into account for Mn. The large scatter found for this element at a method-by-method and line-by-line when the absolute abundances are taken into account, could be attributed to the fact that EW methods  did not consider hfs while the synthesis methods (except iSpec) synthesise the line profiles including hfs. We could see in \fig{fig:hfs} the systematic difference of Mn abundances for the EW and iSpec method compared to the other synthesis methods that considered hfs, the latter being 0.13~dex lower in the Sun. Differentially, systematic differences of Mn abundances for methods considering or neglecting hfs were considerable for several stars. For that reason, the determination of Mn was done considering only the synthesis methods that took hfs into account. Furthermore, many Mn lines lie in spectral regions that are crowded making the continuum identification nontrivial, which contributes to a large line-to-line scatter respect to other elements.  }

 % how many lines do we have, how many golden lines overlap between groups, any line with gf to be corrected?
 Our Mn abundance determination was based in the analysis of 10 \ion{Mn}{i} lines, all of them with excitation potential between 0 and $\sim$3~eV. All these lines were used for the \fgdwarf\ group, except the bluest one  $\lambda 4823$~\AA.  For the \mgiant\ stars we used only 2 lines ($\lambda 5004$ and $\lambda 6021$~\AA), none of them classified as golden.  The latter line was used for all groups.  Almost all lines were used for the determination of Mn of the \metalpoor\ stars, none of them golden except $\lambda 4823$~\AA.  For the \fgkgiant, only six lines were used, but all of them were classified as golden.

  % nlte corrections (our work or literature)
 The NLTE effects in \ion{Mn}{i} have been studied for the Sun and for metal-poor stars by \cite{2008AA...492..823B}, who found that NLTE effects on formation of Mn lines can be very large, in fact the abundances of Mn deviated  from a given line by up to 0.4~dex depending on the stellar parameters of the star. Note they did not consider hfs of Mn in that study.  In  the recent study of \cite{2015AA...577A...9B}  Mn NLTE corrections of \cite{2008AA...492..823B} have been applied for a large sample of stars, finding that indeed differences of 0.4~dex are possible, especially for metal-poor stars. In this study, hfs was taken into account.  We applied NLTE corrections of \cite{2008AA...492..823B} for Mn for most of the GBS, finding that in the majority of the cases the corrections are negligible. Metal-poor stars however have large corrections, but of 0.2~dex rather than 0.4~dex.  We stress that these corrections are subject of uncertainties due to EW measured in lines where hfs is very pronounced and should be taken with care.

   % uncertainties in general. How accurate?
In general, Mn abundances have a small dependency on stellar parameters, which behaves as the other iron-peak elements discussed above. That is, its uncertainty due to different \logg\ or \feh\ (errors or the consideration of \feh\ before NLTE corrections of iron abundances) is negligible compared to the uncertainty due to \teff\ or \vmic. The line-to-line scatter can be up to 0.15~dex, especially for giants. We attribute this high scatter to the different ways to deal with hfs  in our different methods.  

  % comparison with literature and Sun 
For the Sun the comparison between our value and that one of \cite{2007SSRv..130..105G} shows { a systematic offset, with our result being less than those of  \cite{2007SSRv..130..105G}. This presents a similar behaviour than V, which we attribute to the LTE analysis used here and NLTE analysis performed by \cite{2007SSRv..130..105G}. As for V, our results for Mn for the Sun agree well with \cite{2015AA...577A...9B}.}   The comparison for the rest of the stars with the literature (\fig{fig:final_iron}) shows for warm stars very good agreement, but for the cool stars HD107328, \muLeo\ and \cygA\ we see systematic higher literature abundances.  Manganese in HD107328 has been studied by \cite{2007AJ....133.2464L} and by \cite{1998yCat.3193....0T}, obtaining a value of [Mn/H] of -0.6 and -0.1, respectively. Our value of -0.68 agrees very well with  \cite{2007AJ....133.2464L}. The case of \muLeo\ has been investigated by \cite{2007AJ....133.2464L} obtaining a value of [Mn/H] of 0.7, which is 0.4 dex higher than our value of 0.33.  Finally, the star \cygA\ has been studied by \cite{2005AJ....129.1063L}, who obtained a value of [Mn/H] = -0.04. As previously discussed, we determined a value of [Fe/H] for this star that was about 0.4~dex lower than this value in Paper III, which is translated to an abundance of 0.4~dex lower, which is what we see here. Furthermore,  \cygB\ has reported a measurement of [Mn/H] by  by \cite{2005AJ....129.1063L}, which is higher than our value. This is because the metallicity is higher by the same amount.

There are several values of [Mn/H] reported for the metal-poor stars in the literature. One example analysing many of them is \cite{2003AA...404..187G}, who used lines lying bluer than our spectral range. We could not detect any Mn line for the main-sequence metal-poor HD84937.    For \betAra, we could not find trustable lines to provide a value of [Mn/H], so \cite{1979ApJ...232..797L} is still the only reference to our knowledge reporting a [Mn/H] value for this star.  For the stars \psiPhe\ and \gamSge\ we were not able to find trustable lines for Mn abundances, and no work in the literature has reported a value for Mn either. 
  % any star in particular? 

In this work we could however provide a new measurement of Mn for \alfCet, \alfTau, \ksiHya\ and HD49933.    For the \mgiant\, the two lines mentioned above were used. For $\lambda 5004$~\AA\ no NLTE corrections were provided, while for $\lambda 6021$~\AA, the corrections were negligible.  The [Mn/Fe] value for both stars is very similar ({ -0.49} and { -0.47} for \alfTau\ and \alfCet, respectively), which is also observed for the APOGEE giants in \cite{2015arXiv150104110H}. The determination of Mn abundances of \ksiHya\ was based on six lines, among them the two ones used for the \mgiant. No NLTE corrections were possible to calculate for this metal-rich giant, in the same way as for \delEri\ and \muLeo, which are other rather metal-rich stars.  The line-to-line scatter of Mn abundances of \ksiHya\ is relatively large (0.1~dex), which is expected for metal-rich giants, whose lines are very strong and with large hfs. The [Mn/Fe] ratio of \ksiHya\ is of { -0.1}, which is { slightly} lower than what is observed in Galactic disk populations, where [Mn/Fe] tends to increase with metallicity.  Although in \cite{2015AA...577A...9B} some metal-rich dwarfs are observed with [Mn/Fe] values of -0.2, in \cite{2015arXiv150104110H} the bulk of giants is rather at higher [Mn/Fe] values.  Although we can not compare both datasets and our results directly because each of them is calibrated differently, we can see that the [Mn/Fe] value obtained for \ksiHya, \alfCet\ and \alfTau\ are normal for disk stars.  
 
 The hot dwarf HD49933 is the last star of our sample for which we provide new Mn abundances. They are based on four rather weak lines of typical EW of 10 m\AA. We were able to perform NLTE corrections, which for all lines are  of the order of 0.03~dex or less. The line-to-line scatter of this abundance determination is relatively high {(0.07~dex)} reflecting the uncertainties of the different methods in measuring the abundance from these weak lines. The line $\lambda 5407$~\AA\ is particularly uncertain between the different methods, as well as particularly weak (13 m\AA). The [Mn/Fe] value for HD49933 is of -0.3, consistent with the negative trend towards lower metallicities seen when LTE abundances are used in dwarfs \citep{2015AA...577A...9B}.
  
\subsection{Cobalt}
  % why is this element interesting
  Co has very similar behaviour as Cr in terms of supernova yields \citep{2011ApJ...729...16K}. It is another odd-Z iron-peak element which is synthesised principally in explosive silicon burning in SNII \citep{1995ApJS..101..181W, 2013ARA&A..51..457N}, but also in SNIa \citep{2012PhRvC..85e5805B}.  The chemical evolution of Co is supposed to follow the same trend with metallicity as Cr, which is Co evolves with Fe and [Co/Fe] remains constant. Observations however show that Co behaves like an $\alpha-$element in the sense that at low metallicities it is enhanced by more or less the same amount than the $\alpha-$elements, decreasing towards solar values at higher metallicities \citep{2004A&A...416.1117C,  2013ARA&A..51..457N, 2013ApJ...771...67I, 2015AA...577A...9B}. 

  % how many lines do we have, how many golden lines overlap between groups, any line with gf to be corrected?
  We employed 21 lines of \ion{Co}{i} for our analysis which have excitation potentials from $\sim 1$ to $\sim 4$~eV.   Only three lines were useful for measuring [Co/H] of the metal-poor stars, but none of them was classified as golden. Lines used for all the groups are $\lambda 5280$ and $\lambda 5352$~\AA\ but are not golden for all groups.  The lines $\lambda 5647$ and $\lambda 5915$~\AA\  are golden lines used for all groups except for the \metalpoor.  For the \fgdwarf\ we used 19 lines, out of which only two were not classified as golden. For the \kdwarf\ 8 out of the 16 lines were classified as golden. \fgkgiant\ have a total of 16 lines (13 golden) analysed; \mgiant\ a total of 11 (7 golden).  
    
  % nlte corrections (our work or literature)
  NLTE analysis for cobalt has been carried out by \cite{2010MNRAS.401.1334B} obtaining corrections of  up to 0.6 dex for neutral Co lines depending on temperature and surface gravity and low metallicities. They claimed that  the main stellar parameter that controls the magnitude of NLTE effects in Co is in fact metallicity. Although their analysis includes some GBS such as HD84937, the corrections are done for   lines lying bluer than our wavelength coverage. We provide Co abundances for more metal-rich stars compared to the sample of  \cite{2010MNRAS.401.1334B},  making it difficult to estimate a value of NLTE effects in our case. Since we only have neutral Co lines, a possible ionisation imbalance due to NLTE can not be source of line-by-line scatter in our abundances. It could however affect systematically the absolute value of Co. 
  
  % uncertainties in general. How accurate?\  
Since Co is an odd-Z element, it is affected by hfs. { As discussed in \sect{hfs}, the splitting affects the net abundances by different amounts for each GBS, even in the differential approach. For that reason, we restricted the determination of abundance of Co to only methods considering hfs}.
  The uncertainties of Co determination behave like the rest of the iron peak elements, namely small uncertainties for surface gravity and metallicity  variations, and slightly larger uncertainties for temperature and \vmic\ variations.  These  uncertainties  are usually smaller than the line-to-line scatter. The latter is normally less than 0.1~dex, except for some giants, in particular the cool ones. Since we base our abundances in neutral lines for this iron-peak element, it is expected that the errors in \teff\ and \vmic\ propagate the most to the uncertainties in the Co abundances.  
  % comparison with literature and Sun 
  
  The abundances of Co for the Sun agree very well with the solar abundances of \cite{2007SSRv..130..105G}. The agreement for the rest of the stars with the literature is also very good, except for \betAra, which we again compare with the abundances of   \cite{1979ApJ...232..797L} having the same behavior as the rest of the elements. i.e. our abundances lower than those of   \cite{1979ApJ...232..797L}  because of a systematic difference in the stellar parameters. We were able to provide abundances for the cool giants \psiPhe\ and \alfCet, which have no Co abundance reported in the literature to our knowledge.  The measurements come from 4 lines for \psiPhe\ and from 11 lines for \alfCet, obtaining a line-to-line scatter lower than for other elements on these stars. The uncertainties of this measurement due to errors in \feh\ and \vmic\ are above 0.1~dex, which is not surprising as the error in \feh\ (see \tab{tab:params})  is very large as well.  When considering the metallicities of these stars, the [Co/Fe] abundance obtained is of 
 { +0.34} for \psiPhe\ and {-0.07} for \alfCet, both being consistent with trends observed for stellar populations of those metallicities \citep{2013ApJ...771...67I, 2015AA...577A...9B}. 
 
On the other hand, we could not detect Co lines for the metal-poor stars HD140283 and HD84937, as well as the hot (and rather metal-poor) HD49933. A value for [Co/H] exists in the literature for HD140283 as determined by  \cite{1998yCat.3193....0T} of -2.3, while a value for HD49933 has been provided by  \cite{2007PASJ...59..335T}. For HD84937, the abundance of Co still remains to be determined. 
  
  % any star in particular? 
  
\subsection{Nickel}
  % why is this element interesting
Nickel is the last iron-peak element analysed in this work. Its production mechanism is similar to iron being produced principally in SNIa  \citep{2013ARA&A..51..457N}. Abundances of Ni scale linearly with Fe \citep{1993AA...275..101E, 2006MNRAS.367.1329R, 2013ARA&A..51..457N, 2015arXiv150104110H}, with a remarkable low dispersion.   The behaviour of Ni abundances at low metallicities is more uncertain, making  the chemical enrichment history of this element difficult to model. It has been shown that low-$\alpha$ stars in the halo, which are believed to have formed in a smaller gas cloud than typical Milky Way stars, have Ni abundances that are much lower than what models predict \citep[][Hawkins et al 2015]{2010AA...511L..10N, 2011ApJ...729...16K}.

    % how many lines do we have, how many golden lines overlap between groups, any line with gf to be corrected?
  
 The fact that the dispersion in the [Ni/Fe] is so small is partly due to the many clean Ni lines available in optical spectra for several spectral type of stars.  We selected 24 \ion{Ni}{i} lines, with excitation potentials between 1.6 and 4.3~eV, approximately.  There are three Ni lines ($\lambda 5846$, $\lambda 6176$ and $\lambda 6327$~\AA) that overlap for all five group of stars, but none of them are golden lines for all groups.   The two first ones are not golden only for the metal-poor stars.  Almost all lines used for the dwarfs and \fgkgiant\ were classified as golden (only the line $\lambda 5137$~\AA\ was not golden, and was not used by any other group). The metal-poor stars used mostly non-golden lines, while the \mgiant\ group had 3 golden and 3 non-golden lines.    
  
  % nlte corrections (our work or literature)
  There is little known on the NLTE departures of Ni in stellar atmospheres. A recent paper which summarises what has been investigated on NLTE departures of nickel is that one of \cite{2013ApJ...769..103V}. Unfortunately they studied lines at the near-UV, finding that the effects can be quite significant in some cases.  We did not calculate NLTE corrections for Ni in our stars, but as discussed in \cite{2015A&A...573A..26S}, they should be small for neutral Ni lines in the optical range. 
  
  % uncertainties in general. How accurate?
  The uncertainties in the determination of Ni are very similar to the rest of the iron-peak elements for which we analysed only neutral lines. Errors due to metallicity (LTE or uncertainty), as well as errors in surface gravity, give negligible changes in Ni abundances.  Uncertainties in \teff\ and \vmic\ give somehow larger differences in the final Ni abundance, although still small  usually of the order of 0.05~dex or less.  The line-to-line scatter is the larger source of uncertainty for most of the cases, although our measurements for each line are quite accurate and the scatter is usually just above 0.05~dex. 
  
  % comparison with literature and Sun 
  For the Sun we have a good agreement of the result obtained for nickel with respect to \cite{2007SSRv..130..105G}, with our value slightly lower than the one of \cite{2007SSRv..130..105G}.  For the rest of the stars there is also a generally very good agreement with the literature. One can see that the dispersion in the literature is for this element particularly low. For \betAra\ we again obtain a lower abundance than \cite{1979ApJ...232..797L}, as expected.   For \alfTau\ and \cygB\ our results are slightly different than those of  \cite{1998yCat.3193....0T} and \cite{2005AJ....129.1063L}, respectively. Differences for \cygB\ are seen in the rest of the iron-peak elements, and the reason is the difference in the value employed for \feh\ by us and by  \cite{2005AJ....129.1063L}.  
 % any star in particular? 
 
We were able to determine Ni abundances for all GBS, providing new values for the coolest stars \psiPhe\ and \alfCet. 
The abundance of Ni  for \psiPhe\ was determined using two clean lines at $\lambda 5587$ and $\lambda 5846$~\AA.  Synthesis and EW methods were able to provide abundances for those lines, however with large differences causing a scatter rather large (0.34~dex). For \alfCet\ [Ni/H] was determined  using five lines, including those used for \psiPhe. Because these lines are clean, synthesis and EW methods were able to derive abundances, having several measurements to calculate a line-to-line scatter with significance. The value of 0.16 is indeed very small for such cool and complicated star. The scatter compares with the error propagated from the \vmic\ uncertainty.  The [Ni/Fe] ratio of these cool giants are very close to solar, with \psiPhe\ having a value of 0.03 but \alfCet\ a bit lower (-0.15~dex). As shown in \cite{2015arXiv150104110H}, the systematic offset of [Ni/Fe] ratios for very cool giants becomes larger for higher metallicities.  

Although a direct comparison of our results with those of \cite{2015arXiv150104110H} should not be taken too seriously, due to the different spectral ranges and calibrations employed, it is interesting to realise that our systematic offsets for very cool giants show the same trend as APOGEE data. The absolute values should not be directly compared but there is a bias in homogeneous determination of abundances towards very cool giants that goes to the same direction in our analysis and that one of APOGEE.

 \section{Summary and conclusions}\label{conclusions}
 The GBS are 34 stars spanning a wide region in the HR diagram. Their atmospheric parameters (\teff, \logg\ and \feh) and spectra are excellent material to evaluate methods to analyse stellar spectra, as well as to cross-calibrate different stellar spectroscopic surveys. 
 Since the on-going and future surveys collect high resolution data, methods analysing these spectra do not only aim at determining the main stellar parameters, but also abundances of individual elements.  In this article, being the fourth of the series of papers on the GBS, we determined abundances of four $\alpha$ and six iron-peak elements.
 
 The abundances were determined using eight different methods, combining different strategies of measuring equivalent widths and computing synthetic spectra. The methods were applied on a spectral library especially created for this project with the tools described in \cite{2014A&A...566A..98B}, which covers the wavelength ranges of  $\sim$480~--~680~nm and  $\sim$848~--~875~nm, the latter overlapping with the Gaia-RVS spectral range.
 
 The analysis was done using the MARCS atmosphere models and the common line list of the Gaia-ESO Survey \citep{1402-4896-90-5-054010}. 
 The abundances were determined fixing the stellar parameters as defined in \cite{2015arXiv150606095H} for \teff\ and \logg\  and \cite{2014A&A...564A.133J} for \feh\ and \vmic. Three runs were performed by all methods: the first one using the above parameters; the second one considering a slightly different \feh\ aiming to quantify the effect of NLTE in the iron abundances; the third one considering the above parameters and their uncertainties, aiming at quantifying the 
 the effect of the uncertainties in stellar parameters on the derived abundances. 
 
 To reduce the sources of scatter among different lines and methods, our analysis was done in a differential mode, by looking at common lines between two stars. For that, we separated the GBS into five groups: \metalpoor, \fgdwarf, \fgkgiant, \mgiant\ and \kdwarf. For each group we chose one reference star, being HD22879, the Sun, Arcturus, \alfTau\ and \cygA, respectively. The differential analysis was done between the reference star and the rest of the stars in that group.  At last, the reference stars were analysed differentially with respect to the Sun, which was set to be our zero point.  Each final line used in our analysis was carefully inspected to ensure trustable abundances. An extensive discussion was done on this subject. 
 
 We performed NLTE corrections at a line-by-line basis of the elements Mn, Cr, Ca, Si and Mg. For most of the cases these corrections were below 0.1~dex, although Cr for metal-poor stars had a more significant departure e.g. up to 0.3~dex for HD122563.  Furthermore, we discussed how our results compare with the literature, showing that in general our results agree very well, except Gmb1830. We explained this difference by claiming that the temperature we employ for this star might be too low.   In the last part of this article we discussed with more detail the results for each individual element, giving a description of the general behaviour and explaining special cases.

In this article we provide homogeneous abundances of 10 elements for the GBS and quantify several sources of uncertainties, such as the line-to-line scatter and the differences obtained in the abundances when the stellar parameter uncertainties are taken into account. Furthermore, we quantify the effects of NLTE departures for iron which is translated into a different \feh\ value. We also perform direct NLTE calculations in four elements at a line-by-line basis. These values for each star and element can be found in  Tables~\ref{tab:mgh_fin}, \ref{tab:sih_fin}, \ref{tab:cah_fin} and \ref{tab:tih_fin} for Mg, Si, Ca and Ti, respectively, and in Tables ~\ref{tab:sch_fin}, \ref{tab:vh_fin}, \ref{tab:crh_fin}, \ref{tab:mnh_fin}, \ref{tab:coh_fin} and \ref{tab:nih_fin}, for Sc, V, Cr, Mn, Co and Ni, respectively. 
 
 In addition to final abundances and their uncertainties, we present in this work all the material we used to derive the final values of \tab{tab:mgh_fin} to \ref{tab:nih_fin}. That is, we provide the atomic data of each line; the final abundance we obtained for each line;  the  abundances derived by each method and each line; the equivalent widths determined by our methods; and the NLTE correction of each line. We believe this material is crucial to calibrate and develop new methods, as well as to understand better FGK stars in general. 
 
The GBS  are bright stars, in fact many of them visible at naked eye in a clear night. They are so well known  that some of them even belong to ancient star catalogs done by our ancestors millennia ago\footnote{MUL.APIN is one of the first stellar catalogs done by the Babylonians 3 thousands of years ago. The GBS Arcturus, Aldebaran, Pollux and Procyon are part of it.}.  Bright stars have always been necessary pillars to guide us in the sky.  Now the Gaia satellite is orbiting in space, collecting data for the largest and most accurate 3D stellar map of our history. The spectra of million of stars yet unknown as observed by Gaia, Gaia-ESO, GALAH, RAVE, APOGEE, 4MOST, or any other future survey, will be analysed and parametrised according to calibration samples. With our dedicated documentary work on their atmospheric properties and spectral line information, the GBS provide fundamental material to connect these surveys and and contribute to a better understanding of our home galaxy.

\begin{acknowledgements}  P.J. is pleased to thank A. Casey for fruitful discussions and suggestions on the subject, as well as  K. Hawkins and T. M\"adler for their support. We thank S. Hubrig for providing us with the yet unpublished reduced spectrum of \alfCenA.  We also acknowledge the constructive report from the referee. This work was partly supported by the European Union FP7 programme through ERC grant number 320360 and also by the Leverhulme Trust through grant RPG-2012-541. UH acknowledges support from the Swedish National Space Board (Rymdstyrelsen).
 This research has made use of the SIMBAD database,
operated at CDS, Strasbourg, France. 
\end{acknowledgements}

\small
\bibliographystyle{aa} % style aa.bst
\bibliography{refs_paperIV} % your references Yourfile.bib

\begin{appendix}
\section{Final abundances}\label{ap:abundances_fin}
In this appendix we write  the results obtained for the abundances of four $\alpha$ elements and six iron-peak elements, which are listed in independent  tables.  The first column indicates the star, which is listed in increasing order of temperature. The second column indicates the final value of [X/H] determined as discussed above. The third column corresponds to the standard deviation at a line-by-line basis for each measurement. The next 4 columns indicate the difference obtained in the abundances when considering the errors of the metallicity, effective temperature, surface gravity and micro turbulence velocity, respectively (see text). The column labeled with $\Delta$LTE corresponds to the difference obtained in the abundance when the metallicity used was the one before NLTE corrections (see Paper III and \sect{runs}).  The column labelled with NLTE lists the averaged NLTE correction, when available. The last column indicates the number of lines used to derive the final abundance. 
 \begin{table*}
 \begin{center}
\caption{Final abundances for magnesium. }
 \label{tab:mgh_fin}
\begin{tabular}{c | c c c c c c c c  c}
\hline
star & [Mg/H] & $\sigma  (\log \epsilon )$ & $\Delta$[Fe/H] & $\Delta$Teff & $\Delta \log g$ & $\Delta$\vmic & $\Delta$LTE & NLTE & NLIN \\
\hline \\
%literature abundances of Mg
%Cambridge, P Jofre, Fri Jun 19 23:37:46 2015
%star, [X/H](LTE), sigma, delta[FE/H], deltaTeff, deltaLOGG, deltaVMIC, deltaLTE, NLTECORR, NLIN
psiPhe& -- & -- & -- & -- & -- & -- & --& -- & -- \\
alfCet & -0.271 &  0.109 &  0.062 &  0.165 &  0.080 &  0.039 &  0.277 &  -0.010 &03 \\
gamSge &  0.042 &  0.097 &  0.005 &  0.005 &  0.017 &  0.047 &  0.021 &   0.000 &04 \\
alfTau & -0.172 &  0.151 &  0.034 &  0.017 &  0.084 &  0.053 &  0.000 &   0.130 &04 \\
61CygB & -0.061 &  0.069 &  0.002 &  0.036 &  0.021 &  0.001 &  0.000 &   0.002 &04 \\
betAra &  0.131 &  0.036 &  0.016 &  0.018 &  0.014 &  0.093 &  0.000 &   0.013 &03 \\
\hline
 \end{tabular}
 \end{center}
 \end{table*}
 
  \begin{table*}
 \begin{center}
\caption{Final abundances for silicon. }
 \label{tab:sih_fin}
\begin{tabular}{c | c c c c c c c c  c}
\hline
star & [Si/H] & $\sigma  (\log \epsilon )$ & $\Delta$[Fe/H] & $\Delta$Teff & $\Delta \log g$ & $\Delta$\vmic & $\Delta$LTE & NLTE  & NLIN\\
\hline \\
%literature abundances of Si
%Cambridge, P Jofre, Fri Jun 19 23:37:46 2015
%star, [X/H](LTE), sigma, delta[FE/H], deltaTeff, deltaLOGG, deltaVMIC, deltaLTE, NLTECORR, NLIN
psiPhe& -- & -- & -- & -- & -- & -- & --& -- & -- \\
alfCet & -0.524 &  0.098 &  0.022 &  0.141 &  0.039 &  0.063 &  0.062 &  -0.040 &02 \\
gamSge &  0.150 &  0.105 &  0.124 &  0.080 &  0.151 &  0.050 &  0.110 &  -0.025 &06 \\
alfTau & -0.179 &  0.172 &  0.033 &  0.045 &  0.084 &  0.073 &  0.001 &  -0.024 &08 \\
61CygB & -0.328 &  0.182 &  0.027 &  0.043 &  0.041 &  0.029 &  0.000 &   0.000 &03 \\
betAra &  0.228 &  0.175 &  0.164 &  0.056 &  0.085 &  0.180 &  0.005 &  -0.036 &05 \\
\hline
 \end{tabular}
 \end{center}
 \end{table*}

 \begin{table*}
 \begin{center}
\caption{Final abundances for calcium.  }
 \label{tab:cah_fin}
\begin{tabular}{c | c c c c c c c c  c}
\hline
star & [Ca/H] & $\sigma  (\log \epsilon )$ & $\Delta$[Fe/H] & $\Delta$Teff & $\Delta \log g$ & $\Delta$\vmic & $\Delta$LTE & NLTE & NLIN \\
\hline \\
%literature abundances of Ca
%Cambridge, P Jofre, Fri Jun 19 23:37:47 2015
%star, [X/H](LTE), sigma, delta[FE/H], deltaTeff, deltaLOGG, deltaVMIC, deltaLTE, NLTECORR, NLIN
psiPhe & -0.694 &  0.291 &  0.241 &  0.016 &  0.002 &  0.084 &  0.147 &  -0.060 &01 \\
alfCet & -0.456 &  0.132 &  0.025 &  0.029 &  0.015 &  0.117 &  0.055 &  -0.015 &02 \\
gamSge & -0.292 &  0.089 &  0.071 &  0.045 &  0.013 &  0.136 &  0.065 &  -0.033 &03 \\
alfTau & -0.340 &  0.064 &  0.015 &  0.043 &  0.009 &  0.125 &  0.009 &  -0.020 &03 \\
61CygB & -0.403 &  0.059 &  0.010 &  0.007 &  0.008 &  0.043 &  0.000 &  -0.003 &03 \\
betAra & -0.136 &  0.081 &  0.064 &  0.063 &  0.003 &  0.117 &  0.110 &   0.010 &02 \\
\hline
 \end{tabular}
 \end{center}
 \end{table*}

 \begin{table*}
 \begin{center}
\caption{Final abundances for titanium.  }
 \label{tab:tih_fin}
\begin{tabular}{c | c c c c c c c c  c}
\hline
star & [Ti/H] & $\sigma  (\log \epsilon )$ & $\Delta$[Fe/H] & $\Delta$Teff & $\Delta \log g$ & $\Delta$\vmic & $\Delta$LTE & NLTE & NLIN \\
\hline \\
%literature abundances of Ti
%Cambridge, P Jofre, Fri Jun 19 23:37:47 2015
%star, [X/H](LTE), sigma, delta[FE/H], deltaTeff, deltaLOGG, deltaVMIC, deltaLTE, NLTECORR, NLIN
psiPhe & -0.886 &  0.332 &  0.026 &  0.075 &  0.018 &  0.165 &  0.135 & -- &10 \\
alfCet & -0.525 &  0.108 &  0.061 &  0.027 &  0.068 &  0.104 &  0.064 & -- &05 \\
gamSge & -0.266 &  0.099 &  0.008 &  0.038 &  0.068 &  0.158 &  0.002 & -- &12 \\
alfTau & -0.338 &  0.147 &  0.006 &  0.037 &  0.031 &  0.138 &  0.091 & -- &12 \\
61CygB & -0.369 &  0.090 &  0.002 &  0.016 &  0.009 &  0.069 &  0.026 & -- &28 \\
betAra & -0.414 &  0.215 &  0.052 &  0.082 &  0.028 &  0.102 &  0.033 & -- &06 \\
\hline
 \end{tabular}
 \end{center}
 \end{table*}

% -------- IRON PEAK -----
 \begin{table*}
 \begin{center}
\caption{Final abundances for scandium.  }
 \label{tab:sch_fin}
\begin{tabular}{c | c c c c c c c c  c}
\hline
star & [Sc/H] & $\sigma  (\log \epsilon )$ & $\Delta$[Fe/H] & $\Delta$Teff & $\Delta \log g$ & $\Delta$\vmic & $\Delta$LTE & NLTE  & NLIN\\
\hline \\
%literature abundances of Sc
%Cambridge, P Jofre, Fri Jun 19 23:37:47 2015
%star, [X/H](LTE), sigma, delta[FE/H], deltaTeff, deltaLOGG, deltaVMIC, deltaLTE, NLTECORR, NLIN
psiPhe & -0.674 &  0.507 &  0.077 &  0.025 &  0.060 &  0.094 &  0.045 & -- &02 \\
alfCet & -0.397 &  0.199 &  0.158 &  0.015 &  0.111 &  0.110 &  0.068 & -- &04 \\
gamSge & -0.319 &  0.290 &  0.025 &  0.029 &  0.108 &  0.154 &  0.005 & -- &07 \\
alfTau & -0.264 &  0.198 &  0.025 &  0.035 &  0.068 &  0.158 &  0.114 & -- &08 \\
61CygB & -0.324 &  0.075 &  0.005 &  0.010 &  0.020 &  0.040 &  0.000 & -- &04 \\
betAra & -0.716 &  0.213 &  0.111 &  0.098 &  0.023 &  0.043 &  0.000 & -- &03 \\
\hline
 \end{tabular}
 \end{center}
 \end{table*}

  \begin{table*}
 \begin{center}
\caption{Final abundances for vanadium.  }
 \label{tab:vh_fin}
\begin{tabular}{c | c c c c c c c c  c}
\hline
star & [V/H] & $\sigma  (\log \epsilon )$ & $\Delta$[Fe/H] & $\Delta$Teff & $\Delta \log g$ & $\Delta$\vmic & $\Delta$LTE & NLTE  & NLIN\\
\hline \\
%literature abundances of V
%Cambridge, P Jofre, Fri Jun 19 23:37:48 2015
%star, [X/H](LTE), sigma, delta[FE/H], deltaTeff, deltaLOGG, deltaVMIC, deltaLTE, NLTECORR, NLIN
psiPhe & -1.178 &  0.353 &  0.035 &  0.136 &  0.043 &  0.146 &  0.138 & -- &04 \\
alfCet & -0.172 &  0.274 &  0.011 &  0.055 &  0.060 &  0.106 &  0.213 & -- &03 \\
gamSge & -0.314 &  0.156 &  0.003 &  0.058 &  0.079 &  0.125 &  0.157 & -- &04 \\
alfTau & -0.204 &  0.240 &  0.005 &  0.047 &  0.040 &  0.101 &  0.225 & -- &04 \\
61CygB & -0.442 &  0.097 &  0.005 &  0.023 &  0.011 &  0.074 &  0.030 & -- &16 \\
betAra & -0.474 &  0.087 &  0.047 &  0.128 &  0.041 &  0.029 &  0.375 & -- &03 \\
\hline
 \end{tabular}
 \end{center}
 \end{table*}
 
  \begin{table*}
 \begin{center}
\caption{Final abundances for chromiun.  }
 \label{tab:crh_fin}
\begin{tabular}{c | c c c c c c c c c }
\hline
star & [Cr/H] & $\sigma  (\log \epsilon )$ & $\Delta$[Fe/H] & $\Delta$Teff & $\Delta \log g$ & $\Delta$\vmic & $\Delta$LTE & NLTE & NLIN \\
\hline \\
%literature abundances of Cr
%Cambridge, P Jofre, Fri Jun 19 23:37:48 2015
%star, [X/H](LTE), sigma, delta[FE/H], deltaTeff, deltaLOGG, deltaVMIC, deltaLTE, NLTECORR, NLIN
psiPhe & -0.835 &  0.103 &  0.138 &  0.070 &  0.023 &  0.070 &  0.195 &   0.025 &02 \\
alfCet & -0.472 &  0.115 &  0.133 &  0.068 &  0.030 &  0.075 &  0.191 &   0.060 &03 \\
gamSge & -0.152 &  0.060 &  0.035 &  0.043 &  0.056 &  0.104 &  0.017 &   0.025 &06 \\
alfTau & -0.295 &  0.214 &  0.019 &  0.033 &  0.020 &  0.097 &  0.190 &   0.060 &06 \\
61CygB & -0.339 &  0.095 &  0.000 &  0.008 &  0.015 &  0.019 &  0.075 &   0.005 &13 \\
betAra & -0.223 &  0.165 &  0.119 &  0.080 &  0.009 &  0.066 &  0.000 &   0.063 &06 \\
\hline
 \end{tabular}
 \end{center}
 \end{table*}

 \begin{table*}
 \begin{center}
\caption{Final abundances for manganese.  }
 \label{tab:mnh_fin}
\begin{tabular}{c | c c c c c c c c c }
\hline
star & [Mn/H] & $\sigma  (\log \epsilon )$ & $\Delta$[Fe/H] & $\Delta$Teff & $\Delta \log g$ & $\Delta$\vmic & $\Delta$LTE & NLTE & NLIN\\
\hline \\
%literature abundances of Mn
%Cambridge, P Jofre, Fri Jun 19 23:37:48 2015
%star, [X/H](LTE), sigma, delta[FE/H], deltaTeff, deltaLOGG, deltaVMIC, deltaLTE, NLTECORR, NLIN
psiPhe& -- & -- & -- & -- & -- & -- & --& -- & -- \\
alfCet & -0.457 &  0.048 &  0.013 &  0.037 &  0.045 &  0.204 &  0.030 &  -0.000 &02 \\
gamSge& -- & -- & -- & -- & -- & -- & --& -- & -- \\
alfTau & -0.487 &  0.096 &  0.005 &  0.058 &  0.033 &  0.176 &  0.000 &  -0.000 &02 \\
61CygB & -0.443 &  0.096 &  0.005 &  0.012 &  0.015 &  0.042 &  0.030 &  -0.000 &06 \\
betAra& -- & -- & -- & -- & -- & -- & --& -- & -- \\
\hline
 \end{tabular}
 \end{center}
 \end{table*}

  \begin{table*}
 \begin{center}
\caption{Final abundances for cobalt.  }
 \label{tab:coh_fin}
\begin{tabular}{c | c c c c c c c c  c}
\hline
star & [Co/H] & $\sigma  (\log \epsilon )$ & $\Delta$[Fe/H] & $\Delta$Teff & $\Delta \log g$ & $\Delta$\vmic & $\Delta$LTE & NLTE & NLIN \\
\hline \\
%literature abundances of Co
%Cambridge, P Jofre, Fri Jun 19 23:37:49 2015
%star, [X/H](LTE), sigma, delta[FE/H], deltaTeff, deltaLOGG, deltaVMIC, deltaLTE, NLTECORR, NLIN
psiPhe & -0.617 &  0.361 &  0.099 &  0.002 &  0.072 &  0.090 &  0.156 & -- &04 \\
alfCet & -0.332 &  0.128 &  0.088 &  0.012 &  0.079 &  0.106 &  0.045 & -- &11 \\
gamSge & -0.101 &  0.088 &  0.067 &  0.014 &  0.104 &  0.146 &  0.013 & -- &11 \\
alfTau & -0.248 &  0.181 &  0.026 &  0.016 &  0.045 &  0.108 &  0.045 & -- &11 \\
61CygB & -0.338 &  0.153 &  0.003 &  0.005 &  0.016 &  0.015 &  0.029 & -- &10 \\
betAra & -0.083 &  0.102 &  0.074 &  0.010 &  0.055 &  0.141 &  0.069 & -- &06 \\
\hline
 \end{tabular}
 \end{center}
 \end{table*}
 
  \begin{table*}
 \begin{center}
\caption{Final abundances for nickel.  }
 \label{tab:nih_fin}
\begin{tabular}{c | c c c c c c c c  c}
\hline
star & [Ni/H] & $\sigma  (\log \epsilon )$ & $\Delta$[Fe/H] & $\Delta$Teff & $\Delta \log g$ & $\Delta$\vmic & $\Delta$LTE & NLTE  & NLIN\\
\hline \\
%literature abundances of Ni
%Cambridge, P Jofre, Fri Jun 19 23:37:49 2015
%star, [X/H](LTE), sigma, delta[FE/H], deltaTeff, deltaLOGG, deltaVMIC, deltaLTE, NLTECORR, NLIN
psiPhe & -1.220 &  0.393 &  0.109 &  0.126 &  0.069 &  0.138 &  0.436 & -- &02 \\
alfCet & -0.637 &  0.161 &  0.062 &  0.012 &  0.128 &  0.161 &  0.082 & -- &05 \\
gamSge & -0.174 &  0.131 &  0.044 &  0.019 &  0.110 &  0.209 &  0.001 & -- &05 \\
alfTau & -0.397 &  0.214 &  0.017 &  0.014 &  0.048 &  0.181 &  0.026 & -- &06 \\
61CygB & -0.428 &  0.075 &  0.003 &  0.004 &  0.023 &  0.046 &  0.088 & -- &12 \\
betAra & -0.142 &  0.250 &  0.086 &  0.036 &  0.030 &  0.218 &  0.018 & -- &02 \\
\hline
 \end{tabular}
 \end{center}
 \end{table*}

\section{Literature compilation of abundances}\label{lit}
In this appendix we list the individual abundances that we found in the literature for the benchmark stars. Each element is in a different table for all stars, with the abundance in [X/H] and the value obtained from the reference indicated in the table.  In several cases the value [X/H] had to be calculated using the solar abundances  as indicated in the respective reference. Important part of our compilation comes from the Hypatia catalog \citep{2014AJ....148...54H}.

{ 
 A general comparison of our results (see Appendix~\ref{ap:abundances_fin}) and the literature are shown in  \fig{fig:final_alpha} and \fig{fig:final_iron} in the text. An important feature to note from  both figures is that we could not find for all stars and elements a value in the literature. This is in particular for the cool stars such as \alfCet, \psiPhe\ and  \gamSge.  In addition, we could not find reported abundances of V and  Mn  for our hottest GBS, HD49933, and Mg for \ksiHya.  However, there are cases where although there is a value from the literature, we could not provide an abundance such as Co for HD140283 and HD49933, and Mn for HD84937 and \betAra. Example works in the literature providing several of these abundances for our metal-poor  GBS are  \cite{2011ApJ...742...54} and \cite{2003AA...404..187G}. They analysed spectral lines lying below 4500 \AA\ which is outside the wavelength range of our spectra. We were still able to provide most of the abundances  based on very few lines, but since we have several methods, we can be confident that our abundances are robust. We can see how our results, which are mostly determined from very few lines, agree well with the literature for metal-poor stars, which provide abundances of more lines located at the blue part of the spectrum.  
 
 As discussed in \sect{abundances}, for cool stars  we could not trust any of the abundances determined from the selected lines after visual inspection for some cases, such as Mg, Si and Mn for \psiPhe\ and Mn for \gamSge\ and \betAra. We preferred to be conservative and have less abundance determinations but ensure that our values are accurate. It still remains a challenge to have abundances for these elements and stars, as we could not find in the literature a value reported either, except [Mn/H] for \betAra\ of 0.36 by \cite{1979ApJ...232..797L}. 

  In general, our newly determined abundances agree very well with the literature, especially for the solar-like stars.  There are few cases where our abundances do not agree so well such as Ti and Ca abundances \gamSge, \alfTau\ and \cygB, where our abundances are slightly  lower. These stars are, however, very uncertain as their low temperatures make their spectra have several molecular lines which might be blending atomic lines.  It is interesting to note that in general the abundances we determine for Gmb1830 are systematically lower than several literature measurements. We recall that the effective temperature of Gmb1830 is about 400~K below the typical adopted spectroscopic temperature of this star. We could see in Paper~III how this temperature gave us a metallicity with large ionisation and excitation imbalance, suggesting that the angular diameter of this star should be measured again.  Therefore this star does not currently have a recommended benchmark \teff\ (see discussion in Paper I). Here we see that with the stellar parameters of Paper~I and III, most of the abundances we obtained do not agree with previous works in the literature. 
  }

\onecolumn
%\begin{landscape}
\small
\begin{longtab}
%_____________________

\begin{longtable}{lcr}
\caption{Literature compilation for magnesium. } \label{tab:mgh_lit}\\
\hline \hline
Star& [Mg/H] & Reference  \\
\hline
\endfirsthead
\caption{continued.}\\
\hline\hline
Star& [Mg/H] & Reference \\
\hline
\endhead
\hline
\endfoot
%literature abundances of Mg
%Cambridge, P Jofre, Mon May 11 10:46:46 2015
psiPhe & -- & -- \\
alfCet & -- & -- \\
gamSge& -- & -- \\
alfTau&-0.16& \cite{2010AA...513A..35A} \\
 &-0.08& \cite{2010AA...513A..35A} \\
 &0.2& \cite{1998yCat.3193....0T} \\
61CygB& -- & -- \\
betAra& -- & -- \\
\end{longtable}

%_____________________
\begin{longtable}{lcr}
\caption{Literature compilation for silicon. }\\
\hline \hline
Star& [Si/H] & Reference  \\
\hline
\endfirsthead
\caption{continued.}\\
\hline\hline
Star& [Si/H] & Reference \\
\hline
\endhead
\hline
\endfoot

%literature abundances of Si
%Cambridge, P Jofre, Mon May 11 10:46:46 2015
psiPhe & -- & -- \\
alfCet & -- & -- \\
gamSge&0.07& \cite{1995AZh....72..864B} \\
alfTau&-0.11& \cite{2010AA...513A..35A} \\
 &-0.02& \cite{2010AA...513A..35A} \\
 &0& \cite{1998yCat.3193....0T} \\
61CygB&-0.17& \cite{2005AJ....129.1063L} \\
betAra&0.67& \cite{1979ApJ...232..797L} \\
\end{longtable}
%_____________________
\begin{longtable}{lcr}
\caption{Literature compilation for calcium. }\\
\hline \hline
Star& [Ca/H] & Reference  \\
\hline
\endfirsthead
\caption{continued.}\\
\hline\hline
Star& [Ca/H] & Reference \\
\hline
\endhead
\hline
\endfoot

%literature abundances of Ca
%Cambridge, P Jofre, Mon May 11 10:46:46 2015
psiPhe & -- & -- \\
alfCet & -- & -- \\
gamSge&0.05& \cite{1995AZh....72..864B} \\
alfTau&-0.36& \cite{2010AA...513A..35A} \\
 &-0.31& \cite{2010AA...513A..35A} \\
 &-0.1& \cite{1998yCat.3193....0T} \\
61CygB&-0.12& \cite{2005AJ....129.1063L} \\
betAra&1.15& \cite{1979ApJ...232..797L} \\
\end{longtable}
%_____________________
\begin{longtable}{lcr}
\caption{Literature compilation for titanium. }\\
\hline \hline
Star& [Ti/H] & Reference  \\
\hline
\endfirsthead
\caption{continued.}\\
\hline\hline
Star& [Ti/H] & Reference \\
\hline
\endhead
\hline
\endfoot

%literature abundances of Ti
%Cambridge, P Jofre, Mon May 11 10:46:47 2015
psiPhe & -- & -- \\
alfCet & -- & -- \\
gamSge&0& \cite{1995AZh....72..864B} \\
alfTau&0.01& \cite{2010AA...513A..35A} \\
 &0.05& \cite{2010AA...513A..35A} \\
 &0& \cite{1998yCat.3193....0T} \\
61CygB&-0.05& \cite{2005AJ....129.1063L} \\
betAra&0.34& \cite{1979ApJ...232..797L} \\
\end{longtable}

%__________ iron peak___________
\begin{longtable}{lcr}
\caption{Literature compilation for scandium. }\\
\hline \hline
Star& [Sc/H] & Reference  \\
\hline
\endfirsthead
\caption{continued.}\\
\hline\hline
Star& [Sc/H] & Reference \\
\hline
\endhead
\hline
\endfoot

%literature abundances of Sc
%Cambridge, P Jofre, Mon May 11 10:46:47 2015
psiPhe & -- & -- \\
alfCet & -- & -- \\
gamSge&-0.13& \cite{1995AZh....72..864B} \\
alfTau&-0.1& \cite{1998yCat.3193....0T} \\
61CygB&-0.3& \cite{2005AJ....129.1063L} \\
 &-0.26& \cite{2005AJ....129.1063L} \\
betAra&0.3& \cite{1979ApJ...232..797L} \\
 &-0.02& \cite{1979ApJ...232..797L} \\
\end{longtable}
%_____________________
\begin{longtable}{lcr}
\caption{Literature compilation for vanadium. }\\
\hline \hline
Star& [V/H] & Reference  \\
\hline
\endfirsthead
\caption{continued.}\\
\hline\hline
Star& [V/H] & Reference \\
\hline
\endhead
\hline
\endfoot

%literature abundances of V
%Cambridge, P Jofre, Mon May 11 10:46:47 2015
psiPhe & -- & -- \\
alfCet & -- & -- \\
gamSge&0& \cite{1995AZh....72..864B} \\
 &0.08& \cite{1995AZh....72..864B} \\
alfTau&0.1& \cite{1998yCat.3193....0T} \\
61CygB& -- & -- \\
betAra&0.3& \cite{1979ApJ...232..797L} \\
 &0.4& \cite{1979ApJ...232..797L} \\
\end{longtable}
%_____________________
\begin{longtable}{lcr}
\caption{Literature compilation for chromium. }\\
\hline \hline
Star& [Cr/H] & Reference  \\
\hline
\endfirsthead
\caption{continued.}\\
\hline\hline
Star& [Cr/H] & Reference \\
\hline
\endhead
\hline
\endfoot

%literature abundances of Cr
%Cambridge, P Jofre, Mon May 11 10:46:47 2015
psiPhe & -- & -- \\
alfCet & -- & -- \\
gamSge&-0.03& \cite{1995AZh....72..864B} \\
alfTau&-0.1& \cite{1998yCat.3193....0T} \\
61CygB&-0.26& \cite{2005AJ....129.1063L} \\
betAra&0.53& \cite{1979ApJ...232..797L} \\
 &0.64& \cite{1979ApJ...232..797L} \\
Arcturus&-0.57& \cite{2011ApJ...743..135R} \\
\end{longtable}
%_____________________
\begin{longtable}{lcr}
\caption{Literature compilation for Manganese. }\\
\hline \hline
Star& [Mn/H] & Reference  \\
\hline
\endfirsthead
\caption{continued.}\\
\hline\hline
Star& [Mn/H] & Reference \\
\hline
\endhead
\hline
\endfoot

%literature abundances of Mn
%Cambridge, P Jofre, Mon May 11 10:46:47 2015
psiPhe & -- & -- \\
alfCet & -- & -- \\
gamSge& -- & -- \\
alfTau& -- & -- \\
61CygB&0& \cite{2005AJ....129.1063L} \\
betAra&0.36& \cite{1979ApJ...232..797L} \\
Arcturus&-0.73& \cite{2011ApJ...743..135R} \\
 &-0.4& \cite{1998yCat.3193....0T} \\
\end{longtable}
%_____________________
\begin{longtable}{lcr}
\caption{Literature compilation for cobaltus. }\\
\hline \hline
Star& [Co/H] & Reference  \\
\hline
\endfirsthead
\caption{continued.}\\
\hline\hline
Star& [Co/H] & Reference \\
\hline
\endhead
\hline
\endfoot

%literature abundances of Co
%Cambridge, P Jofre, Mon May 11 10:46:47 2015
psiPhe & -- & -- \\
alfCet & -- & -- \\
gamSge&-0.05& \cite{1995AZh....72..864B} \\
alfTau&-0.05& \cite{1998yCat.3193....0T} \\
61CygB&-0.14& \cite{2005AJ....129.1063L} \\
betAra&0.37& \cite{1979ApJ...232..797L} \\
Arcturus&-0.43& \cite{2011ApJ...743..135R} \\
 &-0.2& \cite{1998yCat.3193....0T} \\
\end{longtable}
%_____________________
\begin{longtable}{lcr}
\caption{Literature compilation for nickel. }\\
\hline \hline
Star& [Ni/H] & Reference  \\
\hline
\endfirsthead
\caption{continued.}\\
\hline\hline
Star& [Ni/H] & Reference \\
\hline
\endhead
\hline
\endfoot

%literature abundances of Ni
%Cambridge, P Jofre, Mon May 11 10:46:47 2015
psiPhe & -- & -- \\
alfCet & -- & -- \\
gamSge&-0.01& \cite{1995AZh....72..864B} \\
alfTau&-0.1& \cite{1998yCat.3193....0T} \\
61CygB&-0.15& \cite{2005AJ....129.1063L} \\
betAra&0.34& \cite{1979ApJ...232..797L} \\
Arcturus&-0.46& \cite{2011ApJ...743..135R} \\
 &-0.35& \cite{1998yCat.3193....0T} \\
\end{longtable}

\end{longtab}

%\end{landscape}

...
\end{appendix}

\end{document}